\documentclass[aps,floatfix,amsmath,amssymb,twocolumn]{revtex4-1}

\usepackage{graphicx}
\usepackage[bookmarks=true,colorlinks,linkcolor=blue,urlcolor=blue,citecolor=blue]{hyperref} 
\usepackage[percent]{overpic} 
\usepackage{enumitem} 
\usepackage{bm} 
\usepackage{gensymb} 
\newcommand{\mi}{\mathrm{i}} 
\newcommand{\nn}{\nonumber} 
\allowdisplaybreaks 

\usepackage{tikz}
\newcommand*\mycirc[1]{%
   \begin{tikzpicture}
     \node[draw,circle,inner sep=1pt] {#1};
   \end{tikzpicture}}

\usepackage{scalerel,stackengine}  
\newcommand\pig[1]{\scalerel*[5pt]{\big#1}{%
\ensurestackMath{\addstackgap[1.5pt]{\big#1}}}}

\begin{document}

\title{Multiloop pseudofermion functional renormalization for quantum spin systems: Application to the spin-\texorpdfstring{$\frac{1}{2}$}{1/2} kagome Heisenberg model}

\author{Julian Thoenniss}
\author{Marc K. Ritter}
\author{Fabian B.\ Kugler}
\author{Jan von Delft}
\author{Matthias Punk}
\affiliation{Arnold Sommerfeld Center for Theoretical Physics, Center for NanoScience,\looseness=-1\, and Munich Center for \\ Quantum Science and Technology, Ludwig-Maximilians-Universit\"at M\"unchen, 80333 Munich, Germany} 

\date{\today}

\begin{abstract}

We present a multiloop pseudofermion functional renormalization group (pffRG) approach to quantum spin systems. As a test case, we study the spin-$\tfrac{1}{2}$ Heisenberg model on the kagome lattice, a prime example of a geometrically frustrated magnet believed to host a quantum spin liquid. Our main physical result is that, at pure nearest-neighbor coupling, the system shows indications for an algebraic spin liquid through slower-than-exponential decay with distance for the static spin susceptibility, while the pseudofermion self-energy develops intriguing low-energy features. Methodologically, the pseudofermion representation of spin models inherently yields a strongly interacting system, and the quantitative reliability of a truncated fRG flow is \textit{a priori} unclear. Our main technical result is the demonstration of convergence in loop number within multiloop pffRG. Through correspondence with the self-consistent parquet equations, this provides further evidence for the internal consistency of the approach. The loop order required for convergence of the pseudofermion vertices is rather large, but the spin susceptibility is more benign, appearing almost fully converged for loop orders $\ell \geq 5$. The multiloop flow remains stable as the infrared cutoff $\Lambda$ is reduced relative to the microscopic exchange interaction $J$, allowing us to reach values of $\Lambda/J$ on the subpercent level in the spin-liquid phase. By contrast, solving the parquet equations via direct fixed-point iteration becomes increasingly difficult for low $\Lambda/J$. We also scrutinize the pseudofermion constraint of single occupation per site, which is only fulfilled on average in pffRG, by explicitly computing fermion-number fluctuations. Although the latter are not entirely suppressed, we find that they do not affect the qualitative conclusions drawn from the spin susceptibility.
\end{abstract}

\maketitle 
\section{Introduction}
\label{sec:introduction}
 
The possibility of realizing interesting quantum spin-liquid phases in
frustrated magnets continues to attract considerable experimental and
theoretical interest in the condensed-matter community. These states are characterized by long-range entanglement between the constituent spins and feature fractionalized spinon excitations, which carry spin-$\tfrac{1}{2}$ but are electrically neutral \cite{Savary2016}. 

One prominent candidate material is Herbertsmithite (ZnCu${}_3$(OH)${}_6$Cl${}_2$), a structurally perfect spin-$\tfrac{1}{2}$ Heisenberg antiferromagnet on a kagome lattice \cite{Shores2005}, which does not order magnetically down to the lowest accessible temperatures and is believed to be a spin liquid \cite{Helton2007, deVries2009}. Inelastic neutron scattering experiments indeed revealed signatures of fractionalized spinon exitations \cite{Han2012}, and these experiments sparked renewed theoretical interest in the antiferromagnetic spin-$\tfrac{1}{2}$ Heisenberg model on the kagome lattice \cite{Yan2011,Jiang2012,Depenbrock2012,Iqbal2014,He2017a,Liao2017}. While it is clear by now that its ground state does not break any symmetries, the precise nature of this spin liquid is still under debate.

From a theoretical perspective, the computation of phase diagrams and spin correlation functions for models of frustrated quantum magnets is very challenging. 
Standard analytical approaches are restricted to semi-classical large-$S$ limits, while quantum Monte Carlo (QMC) simulations usually suffer from a sign problem due to the frustrated nature of the exchange interactions. Methods based
on matrix product states (MPS) or tensor networks have been successfully
applied to spin models in one and two dimensions. However, the
simulation of three-dimensional spin models, relevant for the description of quantum spin-ice phases in rare-earth pyrochlores \cite{Gingras2014},  
is currently out of reach for state-of-the-art algorithms, due to the unfavorable scaling of computational cost with the tensor bond dimension in three dimensions.
As far as Heisenberg models on the pyrochlore lattice are concerned, most numerical works so far have been restricted either to exact-diagonalization studies of a small number of interacting spins \cite{Canals1998, Chandra2018}, or to classical spin models, where Monte-Carlo simulations are possible \cite{Yan2017}. The lack of quantitatively accurate methods for large systems, in particular in three dimensions, severely inhibits a better understanding of these interesting and highly relevant models of frustrated quantum magnets.

A promising alternative development originates from the application of the functional renormalization group (fRG) to interacting quantum systems. 
After seminal papers by Polchinski in 1984 \cite{Polchinski1984} and Wetterich in 1993 \cite{Wetterich1993} in the context of high-energy physics, fRG has been extensively applied to condensed-matter systems \cite{Metzner2012,Platt2013}. 
In particular, fRG has been shown to yield reliable results for interacting fermionic systems with moderately strong correlations
\cite{Hedden2004,Andergassen2006,Karrasch2008a,Uebelacker2012,Scherer2012a,Bauer2013,Eberlein2014a,Bauer2014,Schubert2014,Rentrop2016,Heyder2015,Eberlein2016,Rentrop2016,Wentzell2016,Weidinger2017,Schimmel2017,Sbierski2017,Vilardi2017,Weidinger2018,Weidinger2019,Hille2020,Ehrlich2020}.
In 2010, Reuther and W\"olfle extended the fRG methodology to quantum spin systems, using a pseudofermion representation of spin operators \cite{Reuther2010}. 
This breakthrough idea proved very fruitful, leading to an ever growing number of pseudofermion fRG (pffRG) treatments of quantum spin systems in two and three dimensions
\cite{Reuther2011,Reuther2011a,Reuther2011b,Singh2012,Goettel2012,Reuther2014,Suttner2014,Rousochatzakis2015,Iqbal2015,Iqbal2016,Iqbal2016a,Iqbal2016b,Buessen2016,Hering2017,Buessen2018,Rueck2018,Iqbal2019,Buessen2019,Ghosh2019a,Ghosh2019,Kiese2020}.

The fermionic fRG in the condensed-matter context is based on an exact hierarchy of coupled flow equations for the $n$-point vertices. 
In practice, this infinite hierarchy must be truncated, thereby rendering the flow approximate. Most works use \textit{one-loop} fRG, which corresponds to neglecting six-point and all higher vertices.
This truncation can be motivated, e.g., from a weak-coupling 
or leading-log perspective \cite{Diekmann2020}.  However, in general, its quantitative validity is difficult to assess.
Indeed, the need for going beyond one-loop truncation was addressed by
Katanin \cite{Katanin2004} and later by Eberlein \cite{Eberlein2014}. They proposed schemes for incorporating some contributions from the six-point vertex as two-loop contributions to the flow of the four-point vertex; the former via (one-particle) self-energy corrections, the latter via additional (two-particle) vertex contributions. Recently, these ideas were generalized to arbitrary loop number in a scheme called multiloop fRG (mfRG) \cite{Kugler2017b,Kugler2018}. 
It includes all contributions of the six-point vertex to the flow of the four-point vertex and self-energy that can be computed with numerical costs proportional to the one-loop flow.

As the pseudofermion Hamiltonian has no quadratic term, there is no simple notion of a weak coupling limit, and it is \textit{a priori} hard to assess the reliability of a truncated fRG flow. An indication for the versatility of the pffRG approach comes from the fact that it reproduces the exact mean-field limits of both large $S$ (spin length) \cite{Baez2017} and large $N$ (SU($N$) spin symmetry) \cite{Buessen2018a}.
Importantly, this fulfillment requires the Katanin truncation.
Already in their pioneering paper from 2010 \cite{Reuther2010}, Reuther and W\"olfle reported that pure one-loop truncation does not yield meaningful results for pffRG. They hence included Katanin-type two-loop corrections, and this has been standard practice in pffRG ever since. 
In 2018, R\"uck and Reuther \cite{Rueck2018} used an Eberlein-type construction to fully incorporate two-loop contributions. These caused changes, relative to previous work, in the location of phase boundaries and in the values of the critical RG scales, indicating that two-loop contributions help to mitigate violation of the Mermin--Wagner theorem.

Here, we analyze the effect of higher loops in a multiloop formulation of pffRG, applied to the spin-$\frac{1}{2}$ kagome Heisenberg model.
Our main technical result is the demonstration of convergence in the number of loops,
providing further evidence for the internal consistency of the approach.
Loop convergence ensures several important properties; for present purposes,
the following three are particularly notable:  
(i) Results at the end of the flow are independent from the choice of regulator used to introduce the infrared (IR) cutoff $\Lambda$ \cite{Kugler2017}.
(ii) Different strategies for computing the spin susceptibility--either by tracking the RG flow of the couplings to external fields or through contraction of the interaction vertex---yield consistent results \cite{Tagliavini2018}.  
(iii) The mutual screening between two-particle channels is incorporated to full extent \cite{Kugler2017b}.
On the one hand, this is important for the interplay between tendencies toward magnetic order and spin-liquid behavior, which can be reproduced from different particle-hole channels in the large-$S$ and large-$N$ limit, respectively \cite{Buessen2018a}.
On the other hand, depending on the underlying phase, it may stabilize the flow by reducing the spin susceptibility, allowing one to reach further into the low-energy regime.

A useful perspective on the properties of an mfRG flow can be drawn from the parquet formalism \cite{DeDominicis1964a,Bickers2004}. Upon loop convergence, mfRG results fulfill the self-consistent parquet equations \cite{Kugler2018}. This self-consistency encompasses the one- and two-particle level and provides a quality criterion independent of the underlying RG approach. 
Remarkably, however, we find that using an RG flow to obtain a solution of the parquet equations is highly advantageous compared to trying to solve the parquet equations by iteration, since the latter approach is impeded by the singular nature of the pseudofermion propagator.
We show that an iterative solution of the parquet equations at sufficiently large $\Lambda$ yields results in agreement with mfRG, but becomes extremely challenging for $\Lambda$
smaller than the microscopic energy scale
set by the spin-exchange coupling $J$.
By contrast, an mfRG flow in the nonmagnetic phase 
remains stable down to the lowest values of $\Lambda$.
Moreover, the central observable, the spin susceptibility, converges quickly in loop number, even for $\Lambda/J$ below one percent,
allowing us to draw robust physical conclusions. 
Loop convergence on the detailed level of pseudofermion vertices is more challenging
and requires an increasing number of loops for decreasing $\Lambda$. 
For $\Lambda \gtrsim J/2$,
loop numbers on the order of 10 appear sufficient 
for obtaining a well-converged four-point vertex and self-energy. 

In this work, we present zero-temperature multiloop pffRG results for two phases of the antiferromagnetic $J_1$-$J_2$ spin-$\frac{1}{2}$ kagome Heisenberg model:
the spin-liquid phase at the Heisenberg point with nearest-neighbor interactions only ($J_2 \!=\! 0$) as well as the $\mathbf{q} \!=\! 0$ magnetically ordered phase for equal couplings between nearest and next-nearest neighbors ($J_2 \!=\! J_1$). 
We characterize these phases by computing the static spin susceptibility in analogy to earlier (one-loop) pffRG studies of the kagome Heisenberg model \cite{Suttner2014, Buessen2016}. 
In agreement with these works, we find clear signatures of magnetic order at $J_1 \!=\! J_2$ and the absence thereof at $J_2 \!=\! 0$. 
However, in contrast to Refs.~\cite{Suttner2014, Buessen2016}, where an exponential decay of spin-spin correlations as a function of distance was reported at the Heisenberg point, we find clear deviations from exponential decay and our results indicate an algebraic decay there. 
In addition, we analyze the pseudofermion self-energy to better characterize the spin-liquid phase. 
While we are not able to extract a clear power-law dependence at small frequencies, we find no indications of an excitation gap in the self-energy down to values of $\Lambda/J_1 \!=\! 0.02$, 
which thus serves as an upper bound for a potential spinon gap. 
Taken together, our results at the Heisenberg point are consistent with an algebraic $U(1)$ Dirac spin liquid rather than a gapped $\mathbb{Z}_2$ spin liquid. 

The remainder of our paper is structured as follows:
In Sec.~\ref{sec:mfRG}, we summarize the mfRG framework. 
In Sec.~\ref{sec:model_and_implementation},
we introduce the Heisenberg model in the pseudofermion representation
and explain our pffRG implementation.
In particular, we review the vertex parametrization in real and spin space following standard practice \cite{Reuther2010}, 
and---following Ref.~\cite{Wentzell2016}---describe our frequency parametrization 
based on high-frequency asymptotics, which has not been used in the pffRG context before. 
Our results are shown in Sec.~\ref{sec:Results}.
We first report on our major technical finding, establishing loop convergence within the mfRG approach.
We then present our physical results, characterizing two phases with and without magnetic order.
Next, we elaborate on the question
of whether the single-occupation constraint is well fulfilled or not in pffRG.
Finally, we compare qualitative aspects of pffRG results obtained by different computational schemes.
Section~\ref{sec:conclusion} contains a discussion of our main results and conclusions. 
Three appendices are devoted to technical aspects, 
discussing the explicit form of central equations,
symmetry properties of the vertices,
and loop convergence in the magnetic phase, respectively.
 
\section{Multiloop functional renormalization group}
\label{sec:mfRG}

The mfRG flow can be derived in two equivalent ways. 
The first  starts from the fRG hierarchy of flow equations and focuses on the influence of the six-point vertex.
Instead of setting it to zero, one tries to incorporate all those 
contributions to the flow of the four-point vertex and self-energy that can be kept efficiently, i.e., at a cost proportional to that of the standard one-loop flow \cite{Kugler2017b}.
The second way starts from the self-consistent parquet equations, targeting one- and two-particle objects exclusively, and introduces a scale ($\Lambda$) dependence through the bare propagator. This induces a flow of the four-point vertex and self-energy from the requirement that the parquet equations be fulfilled for all $\Lambda$ \cite{Kugler2018}.
Since we will later use the parquet equations as a consistency check for our mfRG results and the direct iteration of the parquet equations to estimate the complexity of the theory, we here focus on the second approach. Nevertheless, we start with a brief summary of the first one.

\subsection{mfRG via flow hierarchy} 
\label{sec:mfRG_hierarchy}

In the fRG hierarchy of flow equations,
the six-point vertex appears explicitly in the flow equation of the four-point vertex. 
So, the first step is to categorize the scale-differentiated four-point diagrams that follow from contracting the six-point vertex with the so-called single-scale propagator $S$. A subclass of them can be generated through four-point diagrams that are already contained in the flow. 
The generation of these diagrams can be performed in an iterative manner; by counting how many propagator pairs connecting \textit{full} vertices are involved, one finds a unique classification by loop order \cite{Kugler2017b}.
In a second step, one realizes that the self-energy $\Sigma$ is \textit{indirectly} affected by the six-point vertex. 
In principle, the standard flow of the self-energy, $\dot{\Sigma} = \Gamma \cdot S$,
expressing $\dot{\Sigma}=\partial_\Lambda \Sigma$ solely through the four-point vertex $\Gamma$, is exact.
However, approximating the six-point vertex in the flow 
equation for $\dot \Gamma$ renders $\Gamma$ approximate, and using 
the resulting $\Gamma$ for $\dot \Sigma$ leaves $\Sigma$ approximate. 
There are corrections to the self-energy flow that are not generated by the standard term $\Gamma \cdot S$ even after accounting for the above higher loop terms in the mfRG flow of $\Gamma$. However, they can be included through extra terms for $\dot{\Sigma}$, which build on specific parts of the four-point flow \cite{Kugler2017b}.
As a third and final step, one now faces additions to the four-point flow, involving $\dot{\Sigma}$ (Katanin substitution), as well as additions to the self-energy flow, depending on differentiated four-point parts. Hence, one actually has an algebraic differential equation, with $\dot{\Sigma}$ and $\dot{\Gamma}$ on both sides of the equation. Such an equation is well-suited for an iterative treatment,
which successively refines $\dot{\Sigma}$ entering $\dot{\Gamma}$ \cite{Kugler2017b} and vice versa. 
As these refinements enter in a rather marginal form, one expects fast convergence such that an internal iteration might not even be necessary.
Upon convergence in loops (and internal iterations if necessary), results of the mfRG flow obey a variety of desirable properties as mentioned in Sec.~\ref{sec:introduction}; particularly, results at the end of the flow are independent from the choice of regulator.

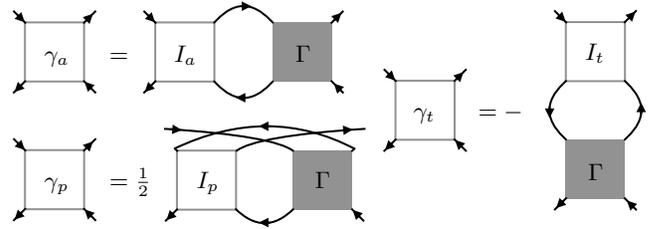
\begin{figure}
\begin{center}
\begin{overpic}[width=.48\textwidth]{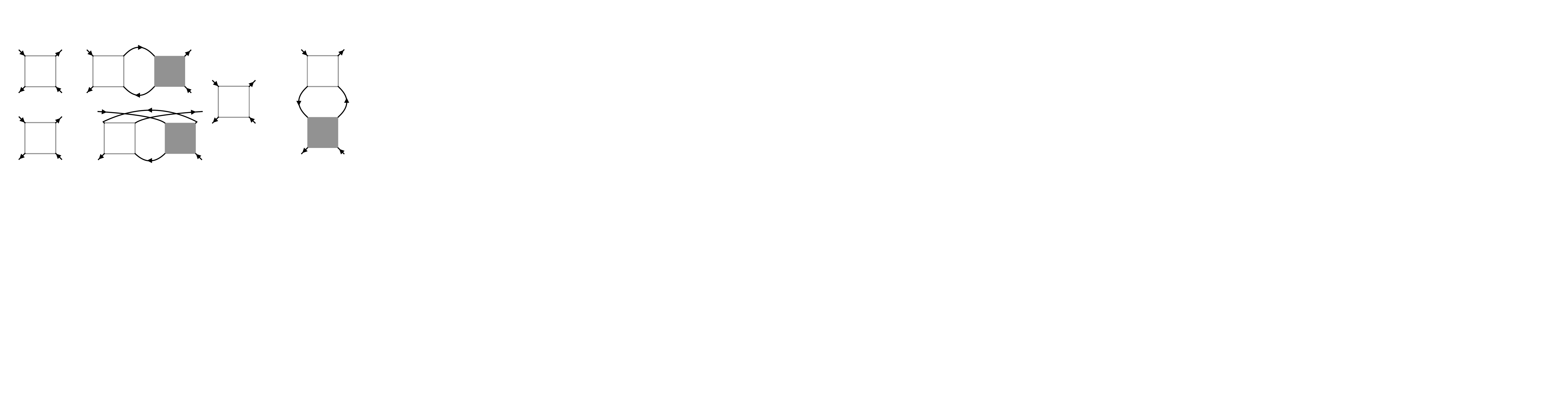}
\put(6,25.7){$\gamma_a$}
\put(16,25.7){$=$}
\put(26,25.7){$I_a$}
\put(44.8,25.7){$\Gamma$}
    
\put(6,6.2){$\gamma_p$}
\put(16,6.2){$=\tfrac 1 2$}
\put(29.5,6.2){$I_p$}
\put(47.8,6.2){$\Gamma$}
    
\put(63,17){$\gamma_t$}
\put(73,17){$=-$}
\put(89.5,26){$I_t$}
\put(90,7.4){$\Gamma$}
\end{overpic}
\caption{Bethe--Salpeter equations for the two-particle reducible vertices $\gamma_r$, $r \in \{a,p,t\}$. }
\label{fig:BetheSalpeter}
\end{center}
\end{figure}

\subsection{mfRG via parquet formalism}

The parquet formalism \cite{DeDominicis1964a,Bickers2004}
is based on categorizing contributions to the four-point vertex by reducibility in propagator pairs.
It assumes a totally irreducible four-point vertex $R$ as input and expresses the full four-point vertex as $\Gamma = R + \gamma_a + \gamma_p + \gamma_t$. The two-particle reducible contributions $\gamma_r$ (with $r=a,p,t$ for antiparallel, parallel and transverse channels) are defined via a set of coupled Bethe--Salpeter equations (BSEs), shown in Fig.~\ref{fig:BetheSalpeter}, where $I_r = R + \sum_{r'\neq r} \gamma_{r'}$ is the vertex that is two-particle irreducible in channel $r$. For a given choice of $R$, the BSEs constitute exact relations. They are formulated with full propagators and thus involve the self-energy, $\Sigma$.
The latter obeys an exact equation of motion, or Schwinger--Dyson equation (SDE), relating $\Sigma$ self-consistently to the full vertex $\Gamma$.
In combination, one also refers to the BSEs and SDE as \textit{parquet equations}. 

For concrete calculations, a choice has to be made for the totally irreducible
vertex, $R$. Here, we adopt the time-honored \textit{parquet approximation} (PA), which sets $R$ equal to the bare vertex $\Gamma_0$.
It can be motivated from the point of view
that it provides all leading contributions in logarithmically divergent perturbation series \cite{Abrikosov1965,Roulet1969} 
as well as from the fact that it yields the simplest possible solution of the full set of parquet equations \cite{DeDominicis1964a,Bickers2004}. 
Including higher-order terms for $R$ would require a drastic increase in numerical effort,
due to multiple nested integrals which \textit{cannot} be computed sequentially.
Such contributions stand in direct correspondence to those contributions of the six-point vertex, entering the fRG flow equation of $\Gamma$, which \textit{cannot} be incorporated efficiently.

The mfRG flow equations can be derived by making the bare propagator $G_0$ scale dependent, such that the low-energy modes below a cutoff $\Lambda$ are suppressed, in precise analogy to the fRG philosophy. One possible choice is a multiplicative regulator where $G_0^\Lambda$ is given by
\begin{equation}
G_0^{\Lambda}(\nu)=\Theta(|\nu|-\Lambda)G_0(\nu)
,
\label{eq:regulator_in_G}
\end{equation}
with the step function $\Theta(x)$. 
This scale-dependent propagator is inserted into the parquet equations.
The requirement that they remain fulfilled for all $\Lambda$ then induces a scale dependence of the two-particle reducible vertices and the self-energy. 
We henceforth consider the propagator, two-particle reducible vertices, and self-energy as scale dependent but omit the superscript $\Lambda$ for brevity.

In symbolic form, the BSEs read $\gamma_r \!=\! I_r \circ \Pi_r \circ \Gamma_r$, where $\Pi_r$ comprises two full propagators $G$ connecting the vertices $I_r$ and $\Gamma$, such that $\gamma_r$ is two-particle reducible in channel $r$.  
With $R \!=\! \Gamma_0$ independent of $\Lambda$, one can differentiate $\gamma_r$ w.r.t.\ $\Lambda$ 
to find the following set of coupled flow equations for the two-particle reducible vertices \cite{Kugler2018}, 
\begin{align}
\label{eq:multiloopgammar}
\dot{\gamma}_r 
& = 
\underbrace{\Gamma \circ \dot{\Pi}_r \circ \Gamma}_{\dot {\gamma}_r^{(1)}} 
\,+\, 
\underbrace{\dot{I}_r \circ \Pi_r \circ \Gamma}_{\dot {\gamma}_r^{\mathrm{(L)}}} 
\,+\, 
\underbrace{\Gamma \circ \Pi_r \circ \dot{I}_r \circ \Pi_r \circ \Gamma}_{\dot {\gamma}_r^{\mathrm{(C)}}} \notag \\
& \ \ + \, 
\underbrace{\Gamma \circ \Pi_r \circ \dot{I}_r}_{\dot {\gamma}_r^{\mathrm{(R)}}} 
,
\end{align}
where 
$\dot I_r =  \sum_{r'\neq r} \dot \gamma_{r'}$.
Moreover, differentiating the SDE (while working in the PA) yields the flow equation \cite{Kugler2018}
\begin{align}
\dot{\Sigma} & = 
\underbrace{[ -\Gamma \cdot S]}_{\dot{\Sigma}_\mathrm{std}} 
\,+\, 
\underbrace{[ - \dot{\gamma}_t^{\mathrm{(C)}} \cdot G ]}_{\dot{\Sigma}_{\bar{t}}} 
\,+\, 
\underbrace{[- \Gamma \cdot (G \cdot \dot{\Sigma}_{\bar{t}} \cdot G)]}_{\dot{\Sigma}_t}
.
\end{align}
Here, $\cdot$ represents a suitable contraction, and
$S = - G \left( \partial_\Lambda [G_0]^{-1} \right) G$
is the single-scale propagator. 

If one simplifies the flow equations for $\dot \gamma_r$ and $\dot \Sigma$ by
retaining only the first term on the right for each, one obtains the flow
equations known from one-loop fRG. The Katanin substitution is already included since $\dot{\Pi}$ in Eq.~\eqref{eq:multiloopgammar} involves the full $\dot{G}$ instead of $S$.  The remaining terms constitute higher-loop contributions. They can be computed systematically by expressing the flow of
$\dot{\gamma}_r \!=\! \sum_{\ell \geq 1} \dot{\gamma}_r^{(\ell)}$ as a sum over
$\ell$-loop contributions, with $\dot \gamma_r^{(2)}$ (which yields Eberlein’s two-loop flow) depending  on $\dot
\gamma_r^{(1)} $, and $\dot \gamma_r^{(\ell \ge 3)}$ depending on both $\dot
\gamma_r^{(\ell-1)}$ and $\dot \gamma_r^{(\ell-2)}$. 
We see that every term of $\dot \gamma_r$ depends on $\dot \Sigma$ through $\dot \Pi_r$. By contrast, $\dot \Sigma$ depends on $\dot \gamma_r$ only through the additions $\dot \Sigma_{\bar t}$ and $\dot \Sigma_{t}$, which build on $\dot \gamma_t^{(\mathrm{C})}$. Hence, one first computes $\dot \gamma_r$ using only $\dot \Sigma_{\mathrm{std}}$. Then, if the self-energy corrections $\dot \Sigma_{\bar t}$, $\dot \Sigma_{t}$ significantly alter $\dot{\Pi}_r$, one can recompute $\dot \gamma_r$ in an additional, internal iteration. 
However, we do not employ such an internal iteration in this work to cap the numerical runtime.

The computational cost of mfRG is proportional to that of one-loop fRG, with a proportionality factor growing linearly with loop order $\ell$. Applications of mfRG to the x-ray--edge problem \cite{Kugler2017}, the two-dimensional Hubbard model \cite{Tagliavini2018,Hille2020}, and the Anderson impurity model \cite{Chalupa2020} have found convergence with increasing loop order for $\ell \lesssim 10$ up to intermediate interaction strengths. Moreover, the converged results agree quantitatively with benchmark QMC data where available. By contrast, no loop convergence is found if the dimensionless interaction strength is tuned too large, signaling a breakdown of the PA. Thus, for these models loop convergence can be understood as a proxy for quantitative accuracy, serving as a useful internal consistency check. This consistency check is particularly valuable for pseudofermion systems where the four-point vertex is generically not small. 

\subsection{Solving the parquet equations by iteration}
\label{sec:parquet}

For a given totally irreducible vertex $R$ in the
decomposition $\Gamma \!=\! R \!+\! \sum_r\! \gamma_r$,
the parquet equations form a closed set of self-consistent equations for the reducible vertices $\gamma_r$
and the self-energy $\Sigma$.
The BSEs for $\gamma_r$, already given above, and the SDE for $\Sigma$ can be written as \cite{Kugler2018}
\begin{subequations}
\label{eq:parquet}
\begin{align}
    \gamma_r 
    & = 
    (\Gamma - \gamma_r) \circ \Pi_r \circ \Gamma, 
    \\
    \Sigma 
    & = 
    - \Gamma_0 \cdot G
    - (\Gamma_0 \circ \Pi_p \circ \Gamma) \cdot G
    .
\end{align}
\end{subequations}
As evident from Eq.~\eqref{eq:parquet}, the parquet equations depend on the combined object $\Psi=(\gamma_a,\gamma_p,\gamma_t,\Sigma)$ and,
after applying them, yield a result $P \Psi$ for it.
They can thus be used for two purposes:
(i) to check whether a solution $\Psi$ fulfills the parquet equations, i.e., is a fixed point of Eq.~\eqref{eq:parquet}, $\Psi = P \Psi$;
(ii) to try finding a fixed-point solution $\Psi = P \Psi$ by iteration of Eq.~\eqref{eq:parquet}.

There are standard techniques for finding fixed-point solutions.
For docile systems, a direct iteration, or \textit{full} update
$\Psi \to P \Psi \to P^2 \Psi \to \dots$,
of Eq.~\eqref{eq:parquet} should converge.
For more challenging systems, Eq.~\eqref{eq:parquet} can become unstable,
such that a \textit{full} update yields divergent results.
One can then introduce a damping factor $0 < z < 1$ and use the 
\textit{reduced} update rule 
$\Psi \to z P \Psi + (1-z) \Psi$.
In the pseudofermion context, we must introduce an IR cutoff $\Lambda$ in the singular bare propagator
$1/(\mi\nu)$ to be able to numerically apply the parquet equations at all.
Then, for $\Lambda$ much greater than the microscopic energy scale $J$, Eq.~\eqref{eq:parquet} converges in a \textit{full} update. 
For lower values of $\Lambda$, one needs a damping $z$,
which in our experience has to be chosen relative to $\Lambda/J$.
For $\Lambda/J$ much smaller than $1$, converging the fixed-point iteration becomes extremely challenging. It requires extremely small values of $z$ (even $z \!\lesssim\! 0.1\Lambda/J$ did not suffice for all $\Lambda$) and soon becomes impractical. 
Possibly, refined techniques like Anderson acceleration \cite{Anderson1965,Walker2011} may help to remedy this. 
However, applying such techniques in the present 
context exceeds the scope of this work.

\section{Model and implementation}
\label{sec:model_and_implementation}

This section is devoted to technical issues.
We define the model to be studied and its pseudofermion representation, and 
then present technical details on the parametrization of the
vertex w.r.t.\ to its real-space, spin, and frequency variables.
 
\subsection{Model}
\label{sec:Model_subsection}

We use the mfRG framework to study the spin-$\tfrac{1}{2}$ Heisenberg model on the kagome lattice,
\begin{equation}
\hat{H} =  \sum_{i<j } J_{ij} \, \hat{\mathbf{S}}_i \cdot \hat{\mathbf{S}}_j 
,
\label{eq:Heisenberg}
\end{equation}
where $\hat{\mathbf{S}}_i$ is the spin operator on lattice site $i$ and $J_{ij}$ are the exchange couplings. 
We choose the pseudofermion representation, 
$\hat{S}^\mu_i=\frac{1}{2} \hat{c}^\dag_{i\alpha}\sigma^\mu_{\alpha\beta}\hat{c}_{i\beta}$,
where  $\hat{c}_{i\alpha}$ annihilates a fermion with spin $\alpha$ at site $i$, 
$\sigma^\mu$ with $\mu \in \{1,2,3\}$ are the Pauli matrices,
and summation over repeated Greek indices is understood implicitly.
The Hamiltonian can then be expressed as a two-particle interaction term: 
\begin{equation}
\label{eq:initialhamiltonian}
\hat{H} 
= 
\frac{1}{4} \sum_{i<j} J_{ij} 
\,
\sigma^\mu_{\alpha\beta}\sigma^\mu_{\gamma\delta} 
\, 
\hat{c}^\dag_{i \alpha}\hat{c}_{i \beta}\hat{c}^\dag_{j \gamma}\hat{c}_{j \delta}
.
\end{equation}

The pseudofermion representation of spin operators requires enforcing a constraint of having precisely one fermion per lattice site. 
Following previous pffRG work, we enforce this constraint only on average through a chemical potential for the fermions, which is fixed at zero by particle-hole symmetry. 
Even though the particle number constraint is not satisfied exactly in this approach, 
it has been argued
(see e.g.\ \cite{Baez2017,Reuther2014a,Reuther2011a}) 
that this is \textit{un}problematic for the computation of ground-state properties,
as long as the ground state of $\hat{H}$ lies in the sector with one fermion 
on each lattice site (or, equivalently, in the sector with maximal spin per lattice site).
We discuss this assumption further in Sec.~\ref{sec:constraint}.

The central observable for describing the model is the spin-spin correlation function,
or spin susceptibility,
\begin{equation}
\chi_{ij}^\omega
= 
\int_{0}^{\beta} \mathrm{d} \tau \,
e^{\mi\omega\tau} 
\langle \mathcal{T}_\tau \, \hat{S}_i^z(\tau) \hat{S}_j^z(0) \rangle
,
\label{eq:static_susc}
\end{equation}
where $\mathcal{T}_\tau$ ensures time ordering and $\beta=1/T$ denotes inverse temperature.
Expressing the spin operators in the pseudofermion representation results in a four-fermion correlator which can be obtained as a sum of a disconnected part involving two propagators and a contraction of the full four-point vertex [see Eq.~\eqref{eq:susceptibility_final}].

The momentum-space resolved static susceptibility is of
particular interest. It is given by the Fourier transform of its real-space counterpart evaluated at the frequency $\omega_0=0$ \cite{sachdev_QPT},
\begin{equation}
\chi_{\omega_0}(\mathbf{q}) = \frac{1}{3}\sum_{i \in \mathrm{u.c.}}\sum_j e^{\mi\mathbf{q} \cdot (\mathbf{R}_{i}-\mathbf{R}_{j})}\chi_{ij}^{\omega_0}.
\label{eq:chi_q}
\end{equation}
Here, $\mathbf{R}_i$, $\mathbf{R}_j$ are vectors pointing from the origin to the lattice sites $i$ and $j$, respectively.
The first sum runs over lattice sites $i$ restricted to one, arbitrary unit cell (u.c.); the second sum runs over all lattice sites $j$.
By symmetry (see Apps.~\ref{sec:appendix_bubbles}, \ref{sec:appendix_symmetries}), $\chi_{ij}^\omega$ is real and symmetric in $i$ and $j$; hence, $\chi_\omega(\mathbf{q})$ is real, too.
In analogy to $\chi_{\omega_0}(\mathbf{q})$, we abbreviate $\chi_{\omega_0}(\mathbf{R}_j) \!=\! \chi^{\omega_0}_{0,j}$, after setting $\mathbf{R}_i \!=\! 0$, for later use.

\subsection{Real-space and spin parametrization}
\label{sec:site_spin_parametrization_subsection}

\begin{figure}
\center
\vspace{0.12cm}
\begin{overpic}[width=0.35\textwidth]{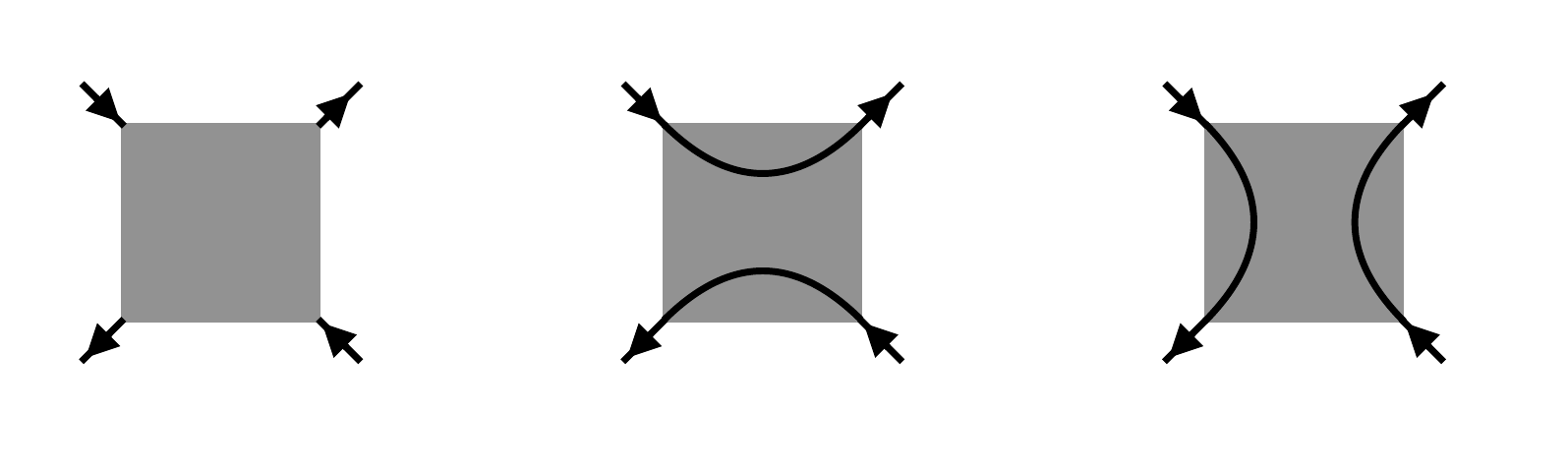}
\put(1,2){$\hat{1}'$}
\put(1,26){$\hat{2}$}
\put(23,2){$\hat{1}$}
\put(23,26){$\hat{2}'$}      
\put(29,14){$=$}          
\put(35,2){$1',i_1$}
\put(35,26){$2,i_2$}
\put(54,2){$1,i_1$}
\put(54,26){$2',i_2$}     
\put(64,14){$-$}                    
\put(70,2){$1',i_2$}
\put(70,26){$2,i_2$}
\put(90,2){$1,i_1$}
\put(90,26){$2',i_1$}
\end{overpic}
\caption{Real-space parametrization of a four-point vertex.}
\label{fig:param1}
\end{figure}
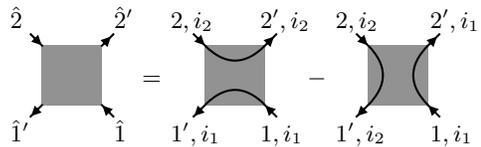

A central property of the pseudofermion representation of spin models is the absence of a quadratic term in the Hamiltonian.
This leads to purely local propagators (and generally purely local one-particle objects) and a restricted real-space dependence of vertices (two-particle objects).
In fact, in a vertex diagram involving purely local propagators,
an incoming leg from site $i_1$ is connected through a 
continuous line of propagators, all carrying the \textit{same} site index,
to an outgoing leg, which thus must have the index $i_1$, too.
Since there are two in- and out-going legs,
only two different lattice sites $i_1$ and $i_2$ can occur.
There remain two possibilities to distribute $i_1$ and $i_2$ among the external legs.
Using the real-space parametrization of Ref.~\cite{Reuther2010}, we can then write
\begin{align}
\nn \Gamma(\hat{1}'\hat{2}';\hat{1}\hat{2}) 
& = 
\Gamma^{ \mathbin{\rotatebox[origin=c]{90}{\tiny)(}}}_{i_1i_2}(1'2';12) 
\,
\delta_{i_1i_{1'}}\delta_{i_2i_{2'}} 
\\
& \ - 
\Gamma^{\mathbin{\rotatebox[origin=c]{0}{\tiny)(}}}_{i_1i_2}(1'2';12) 
\,
\delta_{i_1i_{2'}}\delta_{i_2i_{1'}}
,
\label{eq:real_param}
\end{align} 
where multi-indices with a hat combine frequency, spin, and lattice-site indices, whereas multi-indices without a hat comprise only frequency and spin. This parametrization is represented diagramatically in Fig.~\ref{fig:param1}.

In the pseudofermion representation, a four-point correlator, and thus the full four-point vertex,
is antisymmetric under the exchange of two external arguments: 
$\Gamma(\hat{1}'\hat{2}';\hat{1}\hat{2}) = -\Gamma(\hat{2}'\hat{1}';\hat{1}\hat{2}) = -\Gamma(\hat{1}'\hat{2}';\hat{2}\hat{1})$.
This \textit{crossing symmetry} relates the components of Eq.~\eqref{eq:real_param} as
$\Gamma^{ \mathbin{\rotatebox[origin=c]{90}{\tiny)(}}}_{i_1i_2}(1'2';12) 
= 
\Gamma^{ \mathbin{\rotatebox[origin=c]{0}{\tiny)(}}}_{i_1i_2}(2'1';12)$.
Hence, it suffices to consider only one of the two.
Further, by translation and inversion symmetry, one can always find a map between equivalent pairs of lattice sites, $(i_1,i_2)\rightarrow(\tilde{i}_1 ,\tilde{i}_2)$, such that $\tilde{i}_1$ is always the same reference site that we define as the origin, abbreviated as $0$. This reduces the real-space dependence of the vertices to a single site argument:
$\Gamma^{ \mathbin{\rotatebox[origin=c]{90}{\tiny)(}}}_{i_1i_2}
=
\Gamma^{ \mathbin{\rotatebox[origin=c]{90}{\tiny)(}}}_{0 \; \tilde{i}_2}$.

For a model with SU(2) spin symmetry, one-particle objects are independent of the spin label,
and there are only two independent spin components of a vertex.
Following Ref.~\cite{Reuther2010}, we decompose the vertex in terms of a spin ($s$) and a density ($d$) part,
\begin{align}
\nn \Gamma_{i_1i_2}(1'2';12) 
& = 
\Gamma_{i_1i_2}^s(\nu_{1'}\nu_{2'};\nu_1\nu_2)
\,
\sigma^\mu_{\sigma_{1'} \sigma_{1}}\sigma^\mu_{\sigma_{2'}\sigma_{2}} 
\\
& \ +  
\Gamma_{i_1i_2}^d(\nu_{1'}\nu_{2'};\nu_1\nu_2)
\,
\delta_{\sigma_{1'}\sigma_{1}} \delta_{\sigma_{2'}\sigma_{2}}
. 
\end{align}

So far, this parametrization is standard practice in pffRG.
However, using mfRG, one deals not only with the full vertex $\Gamma$ but also
with the two-particle reducible vertices $\gamma_r$.
While $\gamma_p$ fulfills the same crossing symmetry as $\Gamma$,
the particle-hole vertices are related by
$\gamma_a(\hat{1}'\hat{2}';\hat{1}\hat{2}) = - \gamma_t(\hat{2}'\hat{1}';\hat{1}\hat{2})$ \cite{Kugler2017b,Kugler2018}.
Hence, regarding the real-space parametrization \eqref{eq:real_param}, 
we have the identities
\begin{subequations}
\label{eq:channel_mapping}
\begin{align}
\Gamma^{ \mathbin{\rotatebox[origin=c]{90}{\tiny)(}}}_{i_1i_2}(1'2';12) 
&= 
\Gamma^{ \mathbin{\rotatebox[origin=c]{0}{\tiny)(}}}_{i_1i_2}(2'1';12)
,   
\label{eq:channel_mapping_full}
\\
\gamma^{ \mathbin{\rotatebox[origin=c]{90}{\tiny)(}}}_{p;i_1i_2}(1'2';12) 
&= 
\gamma^{ \mathbin{\rotatebox[origin=c]{0}{\tiny)(}}}_{p;i_1i_2}(2'1';12)
,
\\
\gamma^{ \mathbin{\rotatebox[origin=c]{90}{\tiny)(}}}_{a;i_1i_2}(1'2';12) 
&= 
\gamma^{ \mathbin{\rotatebox[origin=c]{0}{\tiny)(}}}_{t;i_1i_2}(2'1';12) 
,
\label{eq:cr1}
\\
\gamma^{ \mathbin{\rotatebox[origin=c]{90}{\tiny)(}}}_{t;i_1i_2}(1'2';12)
& = 
\gamma^{ \mathbin{\rotatebox[origin=c]{0}{\tiny)(}}}_{a;i_1i_2}(2'1';12) 
\label{eq:cr2}
.
\end{align}
\end{subequations}

The real-space and spin parametrization can then be inserted into general bubble functions,
$B_r(\Gamma,\Gamma') = \Gamma \circ \Pi_r \circ \Gamma',$ as well loop functions,
$L(\Gamma,G) = - \Gamma \cdot G$.
The entire multiloop flow can be formulated in terms of these two types of contractions \cite{Kugler2017b,Kugler2018}.
In terms of general site, spin, and frequency arguments, the bubble functions, illustrated in Fig.~\ref{fig:bubble_param_gen}, read
\begin{align}
\nn
B_a(\Gamma,\Gamma')_{\hat{1}'\hat{2}';\hat{1}\hat{2}} 
&= 
\sum_{\hat{3}\hat{4}} \Gamma(\hat{1}'\hat{4};\hat{3}\hat{2}) 
G(\nu_3) G(\nu_4) \Gamma'(\hat{3}\hat{2}';\hat{1}\hat{4} )
,
\\ 
\nn
B_p(\Gamma,\Gamma')_{\hat{1}'\hat{2}';\hat{1}\hat{2}} 
&= 
\tfrac{1}{2}\sum_{\hat{3}\hat{4}}\Gamma(\hat{1}'\hat{2}';\hat{3}\hat{4}) 
G(\nu_3) G(\nu_4) \Gamma'(\hat{3}\hat{4},\hat{1}\hat{2} )
,
\\
B_t(\Gamma,\Gamma')_{\hat{1}'\hat{2}';\hat{1}\hat{2}} 
&=
-\sum_{\hat{3}\hat{4}} \Gamma(\hat{4}\hat{2}';\hat{3}\hat{2})
G(\nu_3) G(\nu_4) \Gamma'(\hat{1}'\hat{3},\hat{1}\hat{4} )
,
\label{eq:general_bubbles}
\end{align}
and the loop functions in general notation are
\begin{equation} 
\label{eq:general_loops}
L(\Gamma,G)_{\hat{1}}  = - \sum_{\hat{2}} \Gamma(\hat{1}\hat{2};\hat{1}\hat{2}) G(\nu_2).
\end{equation}
The fully parametrized bubble and loop functions are given in App.~\ref{sec:appendix_bubbles}. 
By inserting full vertices, these can be combined to reproduce the 
well-known one-loop flow equations, as in Ref.~\cite{Reuther2010}.
By further inserting reducible vertices, while respecting their channel mapping \eqref{eq:channel_mapping},
they can also be used for all higher loop orders, as in the general mfRG formulation \cite{Kugler2017b,Kugler2018}.

\subsection{Frequency parametrization}

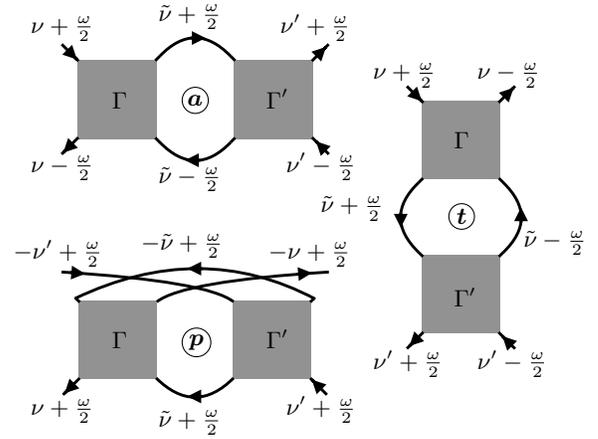
\begin{figure}
\center
\begin{overpic}[width=0.43\textwidth]{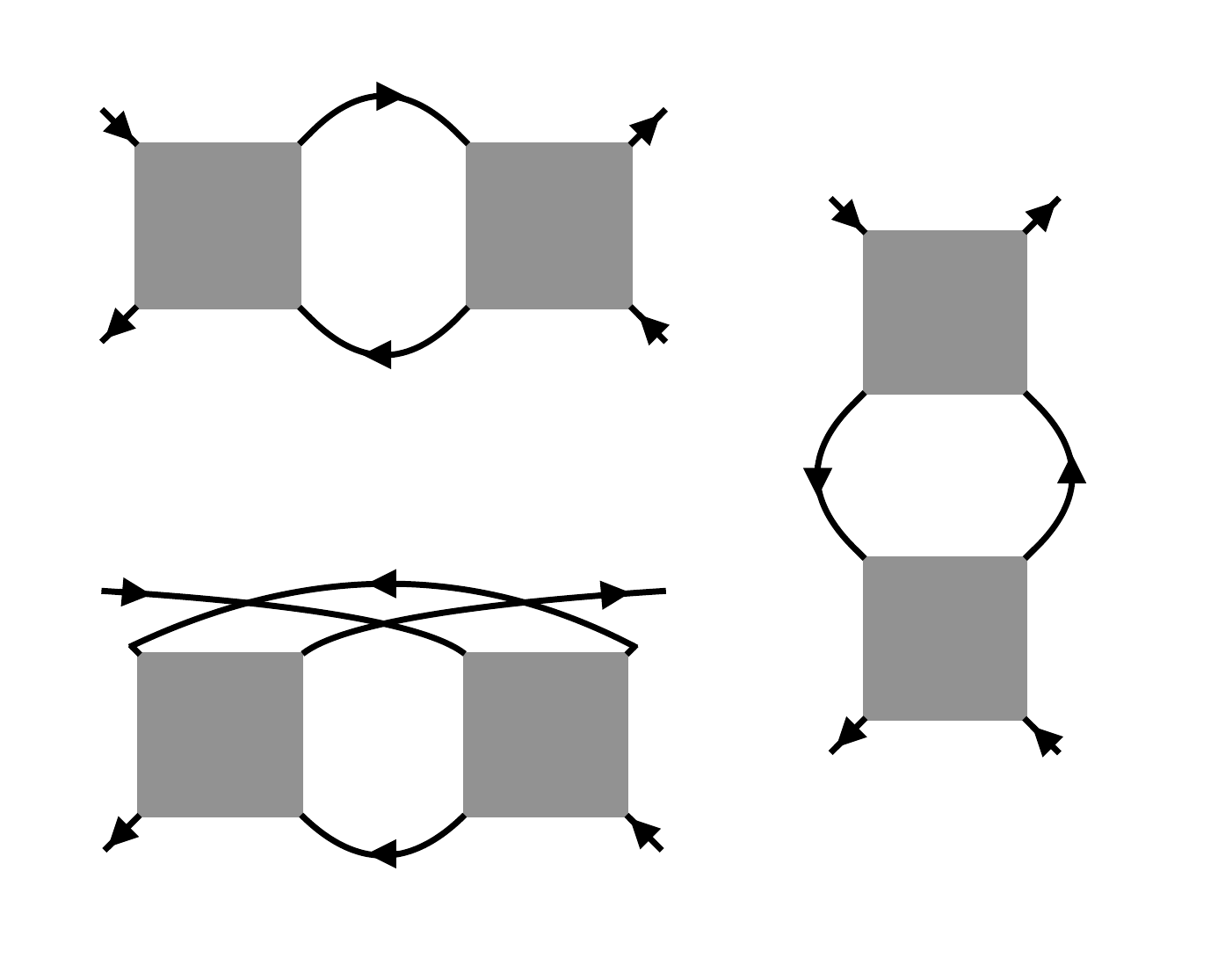}
\put(29,59){\mycirc{$\bm{a}$}}
\put(29,17){\mycirc{$\bm{p}$}}
\put(75,39){\mycirc{$\bm{t}$}}

\put(17,18.5){$\Gamma$}
\put(43.6,18.5){$\Gamma'$}
\put(47,7){\small$\nu'+\tfrac{\omega}{2}$}
\put(44,34){\small$-\nu+\tfrac{\omega}{2}$}
\put(0,34){\small$-\nu'+\tfrac{\omega}{2}$}
\put(3,7){\small$\nu+\tfrac{\omega}{2}$}
\put(25,5){\small$\tilde{\nu}+\tfrac{\omega}{2}$}
\put(22,35.5){\small$-\tilde{\nu}+\tfrac{\omega}{2}$}

\put(17,60){$\Gamma$}
\put(43.6,60){$\Gamma'$}
\put(47,49){\small$\nu'-\tfrac{\omega}{2}$}
\put(46,73){\small$\nu'+\tfrac{\omega}{2}$}
\put(3,49){\small$\nu-\tfrac{\omega}{2}$}
\put(3,73){\small$\nu+\tfrac{\omega}{2}$}
\put(25,47){\small$\tilde{\nu}-\tfrac{\omega}{2}$}
\put(25,75){\small$\tilde{\nu}+\tfrac{\omega}{2}$}

\put(76,53){$\Gamma$}
\put(76,25.7){$\Gamma'$}
\put(80,15){\small$\nu'-\tfrac{\omega}{2}$}
\put(80,65){\small$\nu-\tfrac{\omega}{2}$}
\put(62,65){\small$\nu+\tfrac{\omega}{2}$}
\put(62,15){\small$\nu'+\tfrac{\omega}{2}$}
\put(88,36){\small$\tilde{\nu}-\tfrac{\omega}{2}$}
\put(53,42){\small$\tilde{\nu}+\tfrac{\omega}{2}$}

\end{overpic}
\caption{Natural frequency labels for bubbles in the three channels: $a,p,t$. 
The external legs are amputated and only drawn to fix the arguments.}
\label{fig:bubble_param_gen}
\end{figure}

One of the most important aspects in implementing the mfRG flow is an efficient parametrization of the frequency dependence of vertex functions. A good resolution is clearly important for any fRG flow but becomes crucial in mfRG, for two reasons: 
(i) In one-loop fRG, the bubble functions always contain a differentiated propagator, $S$ or $\dot{G}$, which has a faster large-frequency decay than a normal propagator $G$,
and---in the case of $S$---is often strongly suppressed away from $\nu \!\sim\! |\Lambda|$.
In mfRG, one needs bubble functions with two propagators $G$ as well.
These demand accurate vertex data even for large frequencies.
(ii) Such bubble functions are evaluated and summed up in an iterative fashion. Hence, numerical inaccuracies can accumulate and lead to poor convergence when increasing the number of loops.

Due to energy conservation, the frequency dependence of a vertex can be reduced from four to three independent arguments.
Ever since the invention of pffRG \cite{Reuther2010}, it has been customary to use three bosonic frequencies to parametrize the full vertex.
In the meantime, parquet and (m)fRG studies of interacting electrons in other contexts have revealed that an efficient parametrization of the frequency dependence (and, in appropriate cases, also the momentum dependence) should be adapted to the respective two-particle channel in the form of one (bosonic) transfer argument and two additional (fermionic) arguments \cite{Wentzell2016,Lichtenstein2017a}.
Additionally, Ref.~\cite{Wentzell2016} introduced an efficient frequency parametrization for each two-particle reducible vertex based on high-frequency asymptotics. The underlying idea can be summarized as follows: 
Due to energy conservation, 
a fermionic frequency (say $\nu$) associated with one external leg must also appear on another external leg; 
if both of them are attached to the \textit{same} bare vertex, 
the latter, being frequency-independent, ``deletes the information'' about $\nu$.
This leaves a whole class of diagrams independent of $\nu$,
which can thus be parametrized with less memory requirement and higher resolution.
Such a class is called an ``asymptotic class'', as its contribution remains finite even in the limit $|\nu| \!\to\! \infty$.

Our multiloop pffRG implementation employs the frequency parametrization of Ref.~\cite{Wentzell2016}. 
Each reducible vertex $\gamma_r$ carries its own set of natural frequencies $(\omega_r,\nu_r,\nu'_r)$,
with $\omega_r$ being the bosonic transfer frequency and $\nu_r$, $\nu_r'$ fermionic. 
Two external legs depend on $\nu_r$, the other two on $\nu_r'$; 
we will call them $\nu_r$ legs and $\nu'_r$ legs, respectively. 
Expressing the natural frequencies in terms of the general (fermionic) ones $\nu_i$
for, say, the $a$ channel, one could use 
$(\omega_a,\nu_a,\nu'_a) = (\nu_{2'}-\nu_1,\nu_1,\nu_{1'})$.
In order to center the asymptotics of the vertex functions around the origin, 
we further shift the fermionic frequencies by $\omega_{\mathrm{shift}}$, depending on $\omega_r$.
At zero temperature, where all frequencies are continuous, we can simply set $\omega_{\mathrm{shift}}=\omega_r/2$;
at finite temperature, one would have to ensure that $\omega_{\mathrm{shift}}$ is a bosonic frequency. 
The complete parametrization is shown in Fig.~\ref{fig:bubble_param_gen}, where the channel label $r$ of the frequencies was dropped for brevity.

\begin{figure}
\begin{overpic}[width=0.42\textwidth]{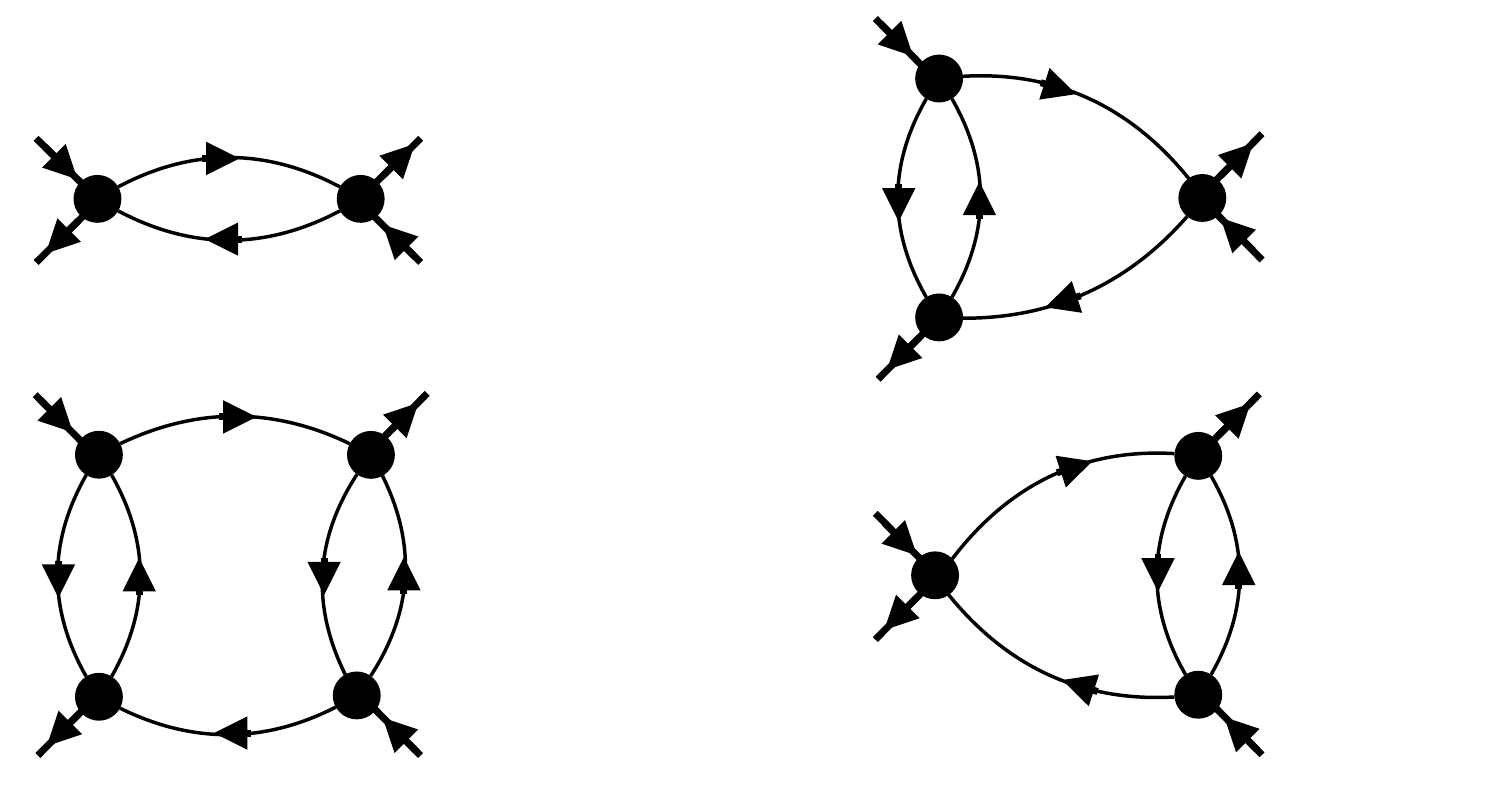}
\put(21,31){\small$\nu'-\tfrac{\omega}{2}$}
\put(21,45){\small$\nu'+\tfrac{\omega}{2}$}
\put(-2,31){\small$\nu-\tfrac{\omega}{2}$}
\put(-2,45){\small$\nu+\tfrac{\omega}{2}$}
\put(31,38){$ \in K_{1,a}^\omega$} 
\put(87,38){$\in K_{2,a}^{\omega,\nu}$}
\put(87,13){$\in K_{2',a}^{\omega,\nu'}$}
\put(31,13){$\in K_{3,a}^{\omega,\nu,\nu'}$}
\end{overpic}
\caption{Four exemplary diagrams in the $a$ channel that each belong to one of the four asymptotic classes.}
\label{fig:asymptotic_classification}
\end{figure}
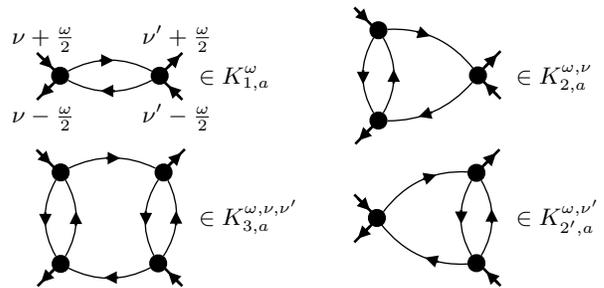

The parametrization based on high-frequency asymptotics classifies two-particle reducible contributions into four classes with an increasing number of arguments,
$\gamma_r^{\omega_r,\nu_r,\nu_r'} 
= K_{1,r}^{\omega_r} \!+\! 
K_{2,r}^{\omega_r,\nu_r} \!+\!
K_{2',r}^{\omega_r,\nu_r'} \!+\!
K_{3,r}^{\omega_r,\nu_r,\nu_r'}$.
The criteria for the asymptotic classification 
(cf.\ Fig.~\ref{fig:asymptotic_classification}) are
\cite{Wentzell2016}:
\begin{enumerate}
\item $K_{1;ij}^{\omega}$: 
The two $\nu$ legs are  connected to one bare 
vertex, and the two $\nu'$ legs are connected to another bare vertex.
Hence, neither $\nu$ nor $\nu'$ 
enter the internal integrals of the diagram, and the diagram depends only on the bosonic frequency $\omega$. 
\item $K_{2;ij}^{\omega,\nu}$:
The two $\nu'$ legs are connected to the same bare
vertex, but the two $\nu$ legs are connected to two different ones. 
Hence, the diagram depends only on $\omega$ and $\nu$. 
\item $K_{2';ij}^{\omega,\nu'}$: As in $2.$ with $\nu \leftrightarrow \nu'$.
\item $K_{3;ij}^{\omega,\nu,\nu'}$: Each external leg
is connected to a different bare vertex. Consequently, the diagram depends on all three frequencies, $\omega$, $\nu$, and $\nu'$.
\end{enumerate}

When evaluating a bubble function $B_r(\Gamma,\Gamma')$, 
one expresses it as sum of $K_n$ contributions and computes these separately.
This is most efficiently done by evoking only those constituents of $\Gamma$ and $\Gamma'$ that contribute to a specific $K_n$, while suppressing all others. 
In the natural frequency parametrization, the dependence on $\nu$ and $\nu'$ enters via $\Gamma$ and $\Gamma'$, respectively (as shown in Fig.~\ref{fig:bubble_param_gen}).
Then, relevant parts of $\Gamma$ and $\Gamma'$ can be determined by a simple rule: 
If a desired class $K_n$ of $B_r$ does not (or does) depend on $\nu$,
then it should be computed using only those contributions to $\Gamma$ that likewise do not (or do) depend on $\nu$; otherwise one would generate diagrams belonging to a different subclass than the one being computed. 
Similarly for $\nu'$ and $\Gamma'$.
We remark that, with this technique, we achieved higher numerical accuracy than with the scheme originally proposed in Ref.~\cite{Wentzell2016}, which uses large frequencies to distill the asymptotic classes.

While the channel-adaptive parametrization through $(\omega_r,\nu_r,\nu_r')$ ensures that the classes $K_n$ of every $\gamma_r$ decay in each of their arguments, 
the frequency dependence can still extend over a wide range.
Empirically, we found that each $K_n$ asymptotically shows a quadratic decay in all its frequency arguments. 
Hence, we can avoid any finite-size effects in frequency space by fitting the decay within the outer parts of the frequency grid to a quadratic function, and analytically extending the latter to infinity.

We choose the number of frequency points in each asymptotic class as:
$n_1 \!=\! n_{1,\omega} \!=\! 1000$ for $K_1^{\omega}$, $n_2 \!=\! n_{2,\omega} n_{2,\nu} \!=\! 250 \times  150$ for $K_2^{\omega,\nu}$ and $K_{2'}^{\omega,\nu'}$, and $n_3 \!=\! n_{3,\omega} n_{3,\nu} n_{3,\nu'} \!=\! 60 \times  40 \times 40$ for $K_3^{\omega,\nu,\nu'}$. 
Further, we use frequency grids symmetric around zero, which are constructed according to the following rules: 
For a (fermionic or bosonic) frequency $\omega$ in the range $0 \!<\! \omega \!<\! \omega_\mathrm{lin}$, 
take a grid with linear spacing $\delta_\text{lin}$ that consists of $n_\text{lin}^\leq$ frequency points and explicitly contains $\omega \!=\! 0$. 
For positive frequencies beyond that range, take an algebraic grid consisting of $n_\text{alg}^>$ points where the $i$-th frequency on the algebraic part is given by 
$\omega_i \!=\! \omega_\mathrm{lin} + \delta_\text{lin} \cdot i ^\alpha$ 
($\alpha > 1$ and $i\in \mathbb{N}$). \\
\includegraphics[width=\linewidth]{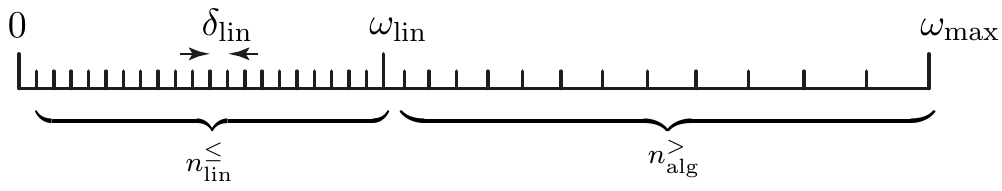}
The linear part serves to properly resolve sharp structures, the algebraic part to efficiently bridge the crossover to the large-frequency regime of pure quadratic decay.
The negative part of the grid is obtained by reflection about $\omega \!=\! 0$.
We use separate grids for each frequency type (fermionic/bosonic) in each asymptotic class and one for the self-energy, i.e., a total of six grids.
Frequency integrals are solved by the adaptive 21-point Gauss--Kronrod routine from the GSL library \cite{gnu}, where we set the relative error to $10^{-3}$. For the computation of vertices, we use an additional absolute error $ \sim\! 10^{-8} \, \text{max}(\Gamma),$ where $\text{max}(\Gamma)$ is roughly the expected maximal value of the vertex class that we compute.

\begin{figure}
    \center
    \includegraphics[scale=1]{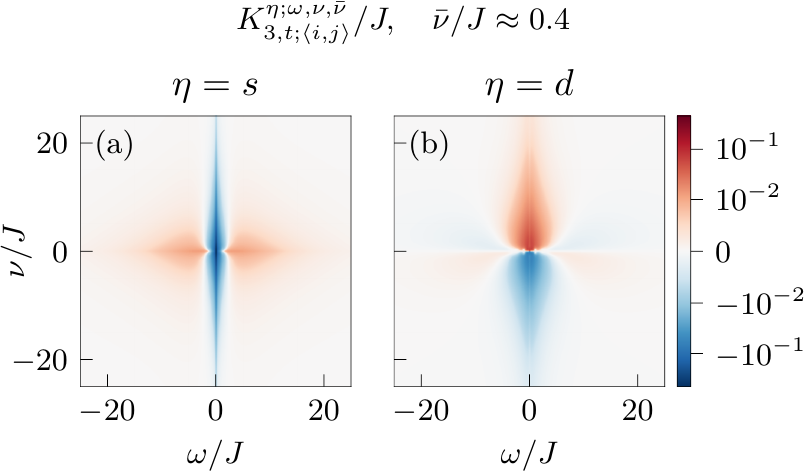}
    \caption{Typical frequency dependence of a vertex, on $\omega$ (bosonic) and $\nu$ (fermionic) at fixed $\nu' \!=\! \bar{\nu}$, as obtained from an mfRG flow ($\ell \!=\! 13$) to $\Lambda/J \!=\! 0.1$ at the Heisenberg point. 
    Shown is the $K_3$ class of the vertex reducible in the $t$ channel, evaluated at nearest-neighbor sites,
    with its (a) spin and (b) density part.
    The former is symmetric, the latter antisymmetric in the fermionic frequencies; 
    all vertices are symmetric in the bosonic frequency (see App.~\ref{sec:appendix_symmetries}).
    } 
    \label{fig:vertex_colorplot}
\end{figure}

It is crucial to adjust the frequency grid dynamically during the flow.
For large values of the IR cutoff, $\Lambda \!\gg\! \Gamma_0$, 
interaction effects are still small and we expect the position of
characteristic features in the frequency dependence to scale directly with $\Lambda$.
Furthermore, even for $\Lambda$ below the exchange coupling, we observe an approximate scaling with $\Lambda$, see e.g.\ Fig.~\ref{fig:loop_conv_and_parquet_check}.
Hence, we choose the maximal frequency $\omega_\text{max}$ on each frequency grid to be roughly proportional to $\Lambda$, but large enough that the frequency dependence of the vertex has decayed sufficiently at the boundaries of the grid that its quadratic decay can be accurately extracted. 
The parameter $\delta_\text{lin}$, mentioned above, is determined by a rather sophisticated routine that scans each vertex and chooses $\delta_\text{lin}$ such that sharp peaks, which typically appear at small transfer frequencies, are well resolved. This value together with $n_\text{lin}^\leq, n_\text{alg}^>$, and $\omega_\text{max}$, fixes the remaining grid parameters $\omega_\mathrm{lin}$
and $\alpha$. 
After each step of the RG flow, we update all six frequency grids and use a multilinear interpolation scheme to transfer the vertices to their new frequency grids. These are then used for the next step of the flow.
In Fig.~\ref{fig:vertex_colorplot}, we illustrate the typical frequency dependence of a vertex,
considering one of the most challenging objects: the $K_3$ class of the two-particle reducible vertex in the $t$ channel.

\subsection{RG flow}

As we investigate the ground state at $T=0$, all Matsubara sums become integrals,
$\frac{1}{\beta} \sum_\nu \to \frac{1}{2\pi} \int \mathrm{d} \nu$.
This makes it convenient to use a smooth version of the regulator introduced in Eq.~\eqref{eq:regulator_in_G}, such as
\cite{Kugler2017}
\begin{equation}
\Theta(|\nu|-\Lambda)
\quad \rightarrow \quad 
\big[ 1-e^{-(|\nu|/\Lambda)^a} \big]
,
\end{equation}
with, e.g., $a=2$.
In fact, the requirements for a scale-dependent bare propagator that
(i) is smooth,
(ii) cures the infrared divergence at $\nu \!=\! 0$,
(iii) obeys particle-hole symmetry $G_0(-\nu) \!=\! -G_0(\nu)$
[implying $G_0(0) \!=\! 0$],
and (iv) recovers the original bare propagator for large frequencies $G_0(|\nu| \!\gg\! \Lambda) \!=\! 1/(\mi\nu)$,
strongly restrict the available choices.
Hence, it is all the more important that mfRG ensures that results at the end of the flow are independent from the particular choice of regulator.

It is essential to start the RG flow from proper initial values of the vertex and self-energy. 
In principle, one could use $\Lambda_\mathrm{ini} \!=\! \infty$ with 
$\Sigma^{\Lambda_\mathrm{ini}} \!=\! 0$ and 
$\Gamma^{\Lambda_\mathrm{ini}} \!=\! \Gamma_0$. 
In practice, however, the flow starts from some finite $\Lambda_\mathrm{ini} \!\gg\! \Gamma_0$;
the smaller $\Lambda_\mathrm{ini}$, the more important it is to use a nontrivial initial condition.
Since the mfRG flow equations yield the solution of the parquet equations at every step of the flow, 
the proper initial values at $\Lambda_\mathrm{ini}$ should similarly fulfill the parquet equations. 
We thus determine the vertex and self-energy at $\Lambda_\mathrm{ini}$ by an
iterative fixed-point solution of the parquet equations \eqref{eq:parquet}. 
(The iteration is initialized with $\Sigma \!=\! 0$ and $\Gamma \!=\! \Gamma_0$.) 
As further discussed below, such a fixed-point iteration converges well for $\Lambda$ much larger than the microscopic energy scale, but becomes extremely challenging, and at some point unfeasible, for lower $\Lambda$.
After initialization, we start the mfRG flow.
For solving the differential equations in $\Lambda$, we use an adaptive sixth-order Runge--Kutta algorithm with a relative error tolerance of $1 \%$.
In order to avoid interpolation errors during the updates of the frequency grids, we limit the Runge--Kutta step size to a maximal value of $h_{\text{max}} \!=\! 0.3 \, \Lambda$. Moreover, we set a \textit{pro forma} lower limit of $h_{\text{min}} \!=\! 10^{-4} \, J$, which is, however, never actually reached except in flows where $\chi$ diverges.

\section{Results}
\label{sec:Results}

In the following, we present our numerical results for the Heisenberg model on the kagome lattice.
We consider nearest- and next-nearest-neighbor exchange interactions $J_1$ and $J_2$;
longer-ranged couplings $J_{ij}$ in the Hamiltonian are set to zero.
It is customary to parametrize $J_1 \!=\! J \cos \alpha$ and $J_2 \!=\! J \sin \alpha$,
such that the phase diagram is covered by the angle $0\degree \!\leq\! \alpha \!\leq\! 360\degree$.
We focus on two different points in the phase diagram, 
(i) the Heisenberg point $\alpha=0\degree$ ($J_1=J$, $J_2=0$), 
where the ground state is expected to be a spin-liquid; and
(ii) $\alpha=45\degree$ ($J_1=J_2$), where the ground state exhibits translationally invariant $\mathbf{q}=0$ coplanar magnetic order, with $120\degree$ angles between spins on the three different sublattice sites.

We divide the presentation of our results into four parts.
First, we analyze the effect of increasing loop order in the multiloop formulation of pffRG.
We demonstrate loop convergence both for the spin susceptibility $\chi$
and on the more detailed level of the pseudofermion self-energy and vertex functions used to compute $\chi$.
For comparison, we also explore an iterative solution of the parquet equations, showing that it becomes extremely challenging if the IR cutoff $\Lambda$ is lower than the exchange coupling $J$.
The second part is devoted to the physically most relevant results:
we discuss the momentum- and real-space dependence of $\chi$ as well as
the frequency dependence of the pseudofermion, or spinon, self-energy.
Third, we take a closer look at the single-occupation constraint inherent in the pseudofermion representation of spin systems. We show that an exact fulfillment of the constraint is neither given in the conventional one-loop pffRG nor in its multiloop extension. We briefly discuss attempts to remedy this, leaving a more thorough investigation for future work.
Fourth, we qualitatively compare results of the pffRG flow obtained by different computational variants.

\subsection{Loop convergence in multiloop pffRG}

\begin{figure}
    \centering%
    \includegraphics[scale=1]{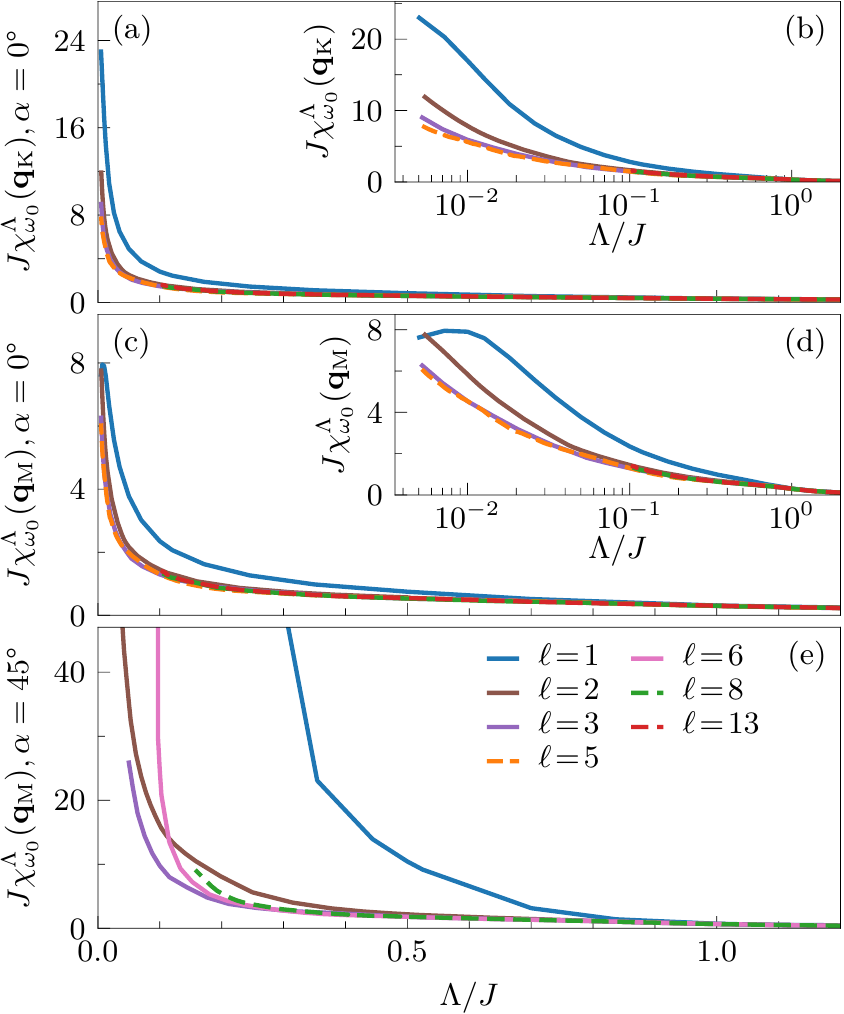}
    \caption{Loop convergence of the mfRG flow for the static spin susceptibility at selected momentum points,
    plotted as function of the IR cutoff $\Lambda$ for different loop orders $\ell$. 
    We consider the spin-liquid phase at the Heisenberg point $\alpha \!=\! 0\degree$ (a--d)
    and (e) $\alpha \!=\! 45\degree$, where the ground state is in the $\mathbf{q} \!=\! 0$ magnetically ordered phase. 
    For the former, we show $\chi_{\omega_0}(\mathbf{q})$ both at the
    K point $\mathbf{q}_\mathrm{K} \!=\! (\frac{8 \pi}{3a},0)$
    and
    M point $\mathbf{q}_\mathrm{M} \!=\! (0,\frac{2 \pi}{\sqrt{3}a})$
    of the extended Brillouin zone;
    for the latter, we consider only the M point.
    Panels (b) and (d) show the same data as (a) and (c), respectively, on a logarithmic $\Lambda/J$ scale,
    demonstrating loop convergence of $\chi$ for $\ell \!\geq\! 5$ 
    down to $\Lambda/J \!=\! 0.01$. 
    Except for $\ell \!=\! 1$ at $\alpha \!=\! 45\degree$, the limiting factor for the lowest $\Lambda/J$ value shown for each curve (here and for similar figures below) was not numerical stability but our choice of numerical runtime.
    Runs with $\ell \!=\! 5$, $13$ ($6$) were only performed for $\alpha \!=\! 0\degree$ ($45 \degree$).
    }
    \label{fig:flow}
\end{figure}

We begin by examining the RG flow of the static spin susceptibility.
To this end, we track $\chi$, computed at zero Matsubara frequency and characteristic momentum points, as a function of $\Lambda/J$, where $J=\sqrt{J_1^2+J_2^2}$. 
This quantity already yields important insights into the underlying phase: although we perform a zero-temperature calculation, $\Lambda$ can be interpreted as an effective temperature \cite{Iqbal2016},
and the flow of $\chi$ with decreasing $\Lambda$ corresponds to its evolution with decreasing temperature.
In an ordered phase, $\chi$, evaluated at a suitable momentum, increases to arbitrarily large values with decreasing $\Lambda$ and eventually diverges. 
In two dimensions and the thermodynamic limit, the critical value of $\Lambda$ should lie at zero, due to the Mermin--Wagner theorem \cite{Mermin1966}.
By contrast, in the absence of order, $\chi$ increases only mildly and remains finite throughout the flow.

\begin{figure*}
    \centering
    \includegraphics[scale=1]{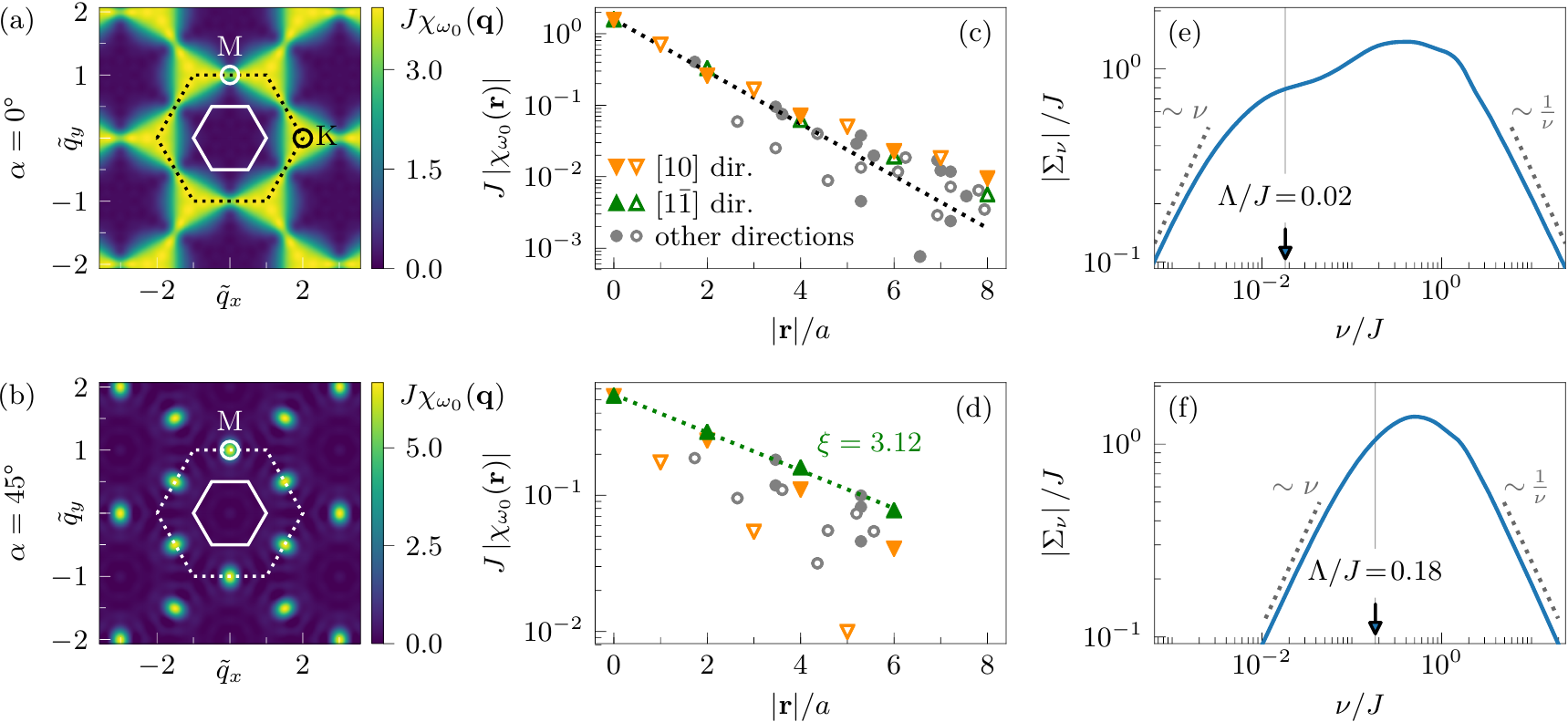}
    \caption{
    Spin-spin correlations and spinon self-energy obtained from the multiloop pffRG flow. 
    Top row: results at the Heisenberg point ($\alpha \!=\! 0\degree$) with $\ell \!=\! 5$ and $\Lambda/J \!=\! 0.02$.
    Bottom row: results in the $\mathbf{q} \!=\! 0$ magnetically ordered phase ($\alpha \!=\!45 \degree$) with $\ell \!=\! 8$ and $\Lambda/J \!=\! 0.18$. 
    (a,b) Static spin susceptibility $\chi_{\omega_0}(\mathbf{q})$ as a function of momentum, where $\tilde q_x \!=\! \frac{3a}{4\pi} q_x$ and $\tilde q_y \!=\! \frac{\sqrt{3} a}{2\pi} q_y$. 
    The solid and dashed lines show the first and extended Brillouin zones,
    with the K and M points marked in reference to Fig.~\ref{fig:flow}.
    (c,d) Absolute value of the static spin susceptibility $\chi^{\omega_0}_{ij}$ as function of distance $|\mathbf{r}| \!=\! |\mathbf{R}_i \!-\! \mathbf{R}_j|$ in real space. 
    Correlations along the  direction where $\chi^{\omega_0}_{ij}$ exhibits the slowest decay for $\alpha \!=\! 0\degree$ ($45\degree$) are highlighted as orange (green) triangles.
    The labels $[10]$ and $[1\bar{1}]$ characterize these directions in terms of Miller indices (see Fig.~\ref{fig:real_space_2d}). 
    Dotted lines indicate exponential decay. 
    In (d), $\xi$ is the correlation length extracted from an exponential fit $\chi_{\omega_0}(\mathbf{r}) \!\sim\! e^{-|\mathbf{r}|/\xi}$ to the data along the $[1\bar{1}]$ direction.
    Full (empty) symbols denote correlations with positive (negative) sign between spins.
    In (d), these occur between sites residing on the same sublattice (on different sublattices). 
    (e,f) Spinon (i.e.~pseudofermion) self-energy $\Sigma$ as a function of frequency $\nu$. 
    Dotted lines serve as guide to the eye for power-law behavior.
    Overall, at the Heisenberg point, the correlations in (c) show slower-than-exponential decay, and the self-energy in (e) reveals a shoulder-type structure in the frequency window $\Lambda \!<\! \nu \!\ll\! J$.
    }
    \label{fig:susc}
\end{figure*}

In the two phases we consider, $\alpha \!=\! 0\degree$ and $\alpha \!=\! 45\degree$, the static susceptibility is maximal at the K point and at the M point, respectively.
Furthermore, for $\alpha \!=\! 0\degree$, the M point constitutes an important `pinch point', as we will see in the contour plots of Fig.~\ref{fig:susc} (there, we also mark the K and M points in the Brillouin zone).
Regarding the RG flow of $\chi_{\omega_0}(\mathbf{q})$ at these momentum points, we indeed observe the previously explained behavior as shown in Fig.~\ref{fig:flow}:
at $\alpha \!=\! 0\degree$, $\chi$ remains finite down to the lowest values of $\Lambda$;
at $\alpha \!=\! 45\degree$, by contrast, it diverges. 
In the comparison of different loop orders $\ell$,
we observe a strong change going from $\ell \!=\! 1$ to $\ell \!=\! 2$.
For loop orders $\ell \!\geq\! 5$,
the value of $\chi$ appears well-converged on the scale 
used for Figs.~\ref{fig:flow}(a--d). 
This highly nontrivial finding is one of our main results:
although the pseudofermion system inherently is in the strong-coupling regime,
convergence of $\chi$ in loop number can be achieved relatively quickly.

Importantly, we find that higher loops tend to reduce the value of $\chi$,
most notably when comparing $\ell \!=\! 1$, $2$, and $3$. 
From previous mfRG studies involving \textit{physical} fermions \cite{Kugler2017,Tagliavini2018,Hille2020,Chalupa2020}
such a reduction is expected:
there, different two-particle channels typically come with different signs for their dominant contributions.
Hence, higher loops, which fully incorporate the feedback between different two-particle channels, lead to a mutual screening.
In the present \textit{pseudo}fermion system, such a screening of two-particle channels with higher loops, leading to a reduction of $\chi$, is found in the spin-liquid phase at $\alpha \!=\! 0\degree$ and, except for very low $\Lambda$, also in the ordered phase at $\alpha \!=\! 45 \degree$.
The one-loop flows, which incorporate inter-channel feedback only partially, have the largest values of $\chi$. Correspondingly, they may overestimate long-range correlations and thus induce stronger finite-size effects, compared to higher loop orders.

Indeed, focusing on $\alpha \!=\! 0\degree$, we see in Fig.~\ref{fig:flow}(d) that the one-loop curve for $\chi$ at the M point, which is very sensitive to finite-size effects, exhibits an unexpected maximum and then decreases for lower $\Lambda$.
The position of an inflection point, marking the onset of this unphysical behavior, may be presumed around $\Lambda/J \!\simeq\! 0.05$
(at this value, we found oscillations in contour plots of $\chi_{\omega_0}(\mathbf{q})$ at $\ell \!=\! 1$ as discussed in the beginning of Sec.~\ref{sec:physical-results}).
The one-loop flow of $\chi_{\omega_0}(\mathbf{q}_\mathrm{K})$ in Fig.~\ref{fig:flow}(b) remains monotonous in $\Lambda$,
but a closer look reveals a notable reduction of the slope for $\Lambda / J \!\simeq\! 0.02$, where $\chi_{\omega_0}(\mathbf{q}_\mathrm{M})$ begins to level off due to finite-size effects.
Higher loops reduce $\chi$ through inter-channel screening and allow us to reach lower values of $\Lambda$ for a given system size
(we ran the $\ell \!=\! 5$ calculation down to $\Lambda/J \!=\! 0.01$). 
Results at the end of the flow are used to predict ground-state properties.

In the ordered phase at $\alpha \!=\! 45\degree$, the effect of higher loops is even more dramatic.
There, the one-loop curve shows a strong upturn already for $\Lambda \!\sim\!  0.3 \, J$.
It thus appears to yield a true divergence at finite $\Lambda$, 
corresponding to a finite-temperature phase transition in violation of the Mermin--Wagner theorem.
By contrast, at higher loop orders, the susceptibility is reduced by orders of magnitude.
The two- and three-loop flows can be followed to much lower values of $\Lambda$, 
suggesting that a true divergence occurs only at $\Lambda \!=\! 0$ 
(in computational practice, though, numerical finite-size effects might set in beforehand). 
For $\Lambda/J \!\lesssim\! 0.2$, $\chi$ at $\ell \!=\! 6$ gets larger than at $\ell \!<\! 6$. We expect that, there, finite-size effects already play a dominant role.
All in all, our results provide confidence that higher loops help to mitigate violation of the Mermin--Wagner theorem in multiloop pffRG. 

\begin{figure*}
    \centering%
    \includegraphics[scale=1]{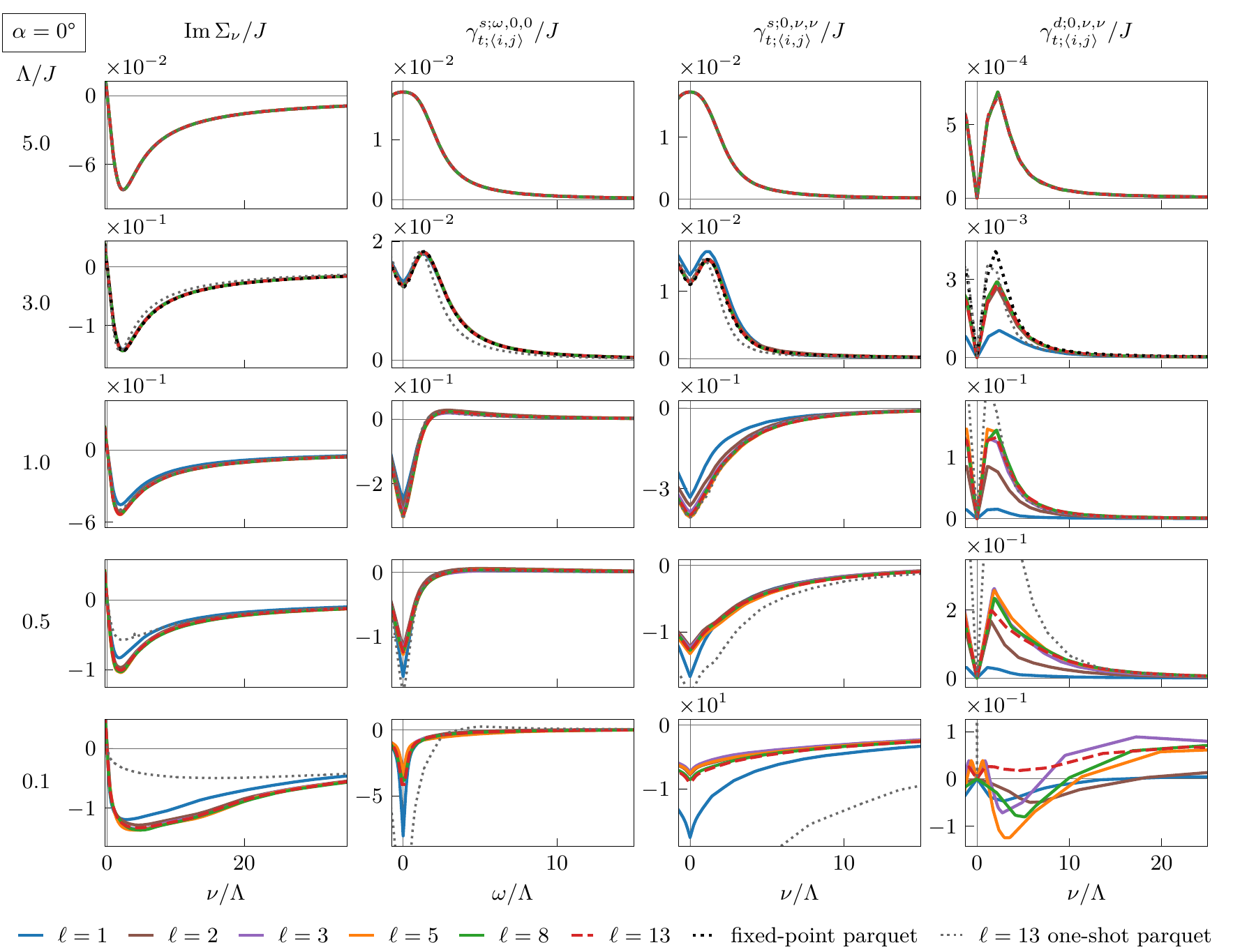}
    \caption{%
        Frequency dependence of the pseudofermion self-energy and representative vertex components, at decreasing values of the IR cutoff $\Lambda$ at $\alpha \!=\! 0\degree$.
        We consider the spin ($s$) and density ($d$) parts of the two-particle reducible vertex in the $t$ channel, evaluated at nearest-neighbor sites and different frequency cuts. 
        Note that $\gamma_{t;i,j}^{d;\omega,0,0}$ vanishes by symmetry.
        Lines show mfRG results at different loop order $\ell$ as well as parquet checks (see main text).
        The first row constitutes the initial condition, obtained from a self-consistency iteration of the parquet equations, and is thus identical for all results. 
        While the one-loop results differ notably from the others already at $\Lambda/J \!\lesssim\! 3$,
        we find convergence with $\ell$ on the order of $10$ up to $\Lambda \!\gtrsim\! J/2$.
        $\Sigma$ and $\gamma_t^s$ are well converged even at $\Lambda/J \!=\! 0.1$.
        By contrast, the density part, $\gamma_t^d$, is harder to converge and fails to converge at $\Lambda/J \!=\! 0.1$, but it is also orders of magnitude smaller than its spin counterpart.
        See Fig.~\ref{fig:loop_conv_and_parquet_check_45deg} for the analogous plot at $\alpha \!=\! 45\degree$. 
    }
    \label{fig:loop_conv_and_parquet_check}
\end{figure*}

A more thorough check of loop convergence can be carried out on the level of the frequency-dependent self-energy and vertex, required to compute $\chi$.
The self-energy is local and purely imaginary, and thus constitutes a single function $\mathrm{Im} \, \Sigma_\nu$.
By contrast, the vertex, albeit purely real, contains an extensive load of information.
Hence, we focus on three representative components for the purpose of this test.
We consider the two-particle reducible vertex in the $t$ channel, as it often is the dominant one signalling magnetic instabilities, evaluated at nearest-neighbor sites.
Regarding its spin  part, we examine (i) its dependence on the bosonic frequency $\omega$ at $\nu \!=\! \nu' \!=\! 0$ as well as (ii) a cut at $\omega \!=\! 0$ and equal fermionic frequencies $\nu \!=\! \nu'$.
The density part at frequencies $(\omega,0,0)$ is zero by symmetry;
hence, we show (iii) its cut at $(0,\nu,\nu)$.

Figure~\ref{fig:loop_conv_and_parquet_check} shows these representative quantities at decreasing values of $\Lambda/J$, computed at different loop order.
Let us first look at the qualitative aspects, common to all results:
as $\Lambda/J$ decreases, characteristic features such as peaks or kinks become much sharper.
Indeed, even though the frequency axes are already rescaled by $1/\Lambda$,
the features still become narrower with decreasing $\Lambda$. 
The self-energy develops intriguing low-energy behavior, as further discussed in the next section.
The vertices exhibit narrow peaks with a cusp at zero frequency.
This cusp occurs at a finite ordinate (or at zero) in the spin (or density) part.
It is interesting to observe that $\gamma_t^s$ changes qualitatively from large to small $\Lambda/J$,
attested by a sign change of the zero-frequency value, which occurs already at a rather large value of $\Lambda/J$ between $3$ and $1$.

On the quantitative level, we see deviations between the results computed at different loop order.
The first row in Fig.~\ref{fig:loop_conv_and_parquet_check}, computed at $\Lambda/J \!=\! 5$, constitutes our initial point of the flow. It is obtained by an iterative solution of the parquet equations and thus identical for all calculations.
Dotted lines represent additional consistency checks via the parquet equations, further discussed below.
The second row shows results of an mfRG flow from $\Lambda/J \!=\! 5$ to $3$.
Although this value of $\Lambda$ is still significantly larger than the exchange coupling $J$, 
we can already observe deviations between the results at $\ell \!=\! 1$ and $\ell \!>\! 1$.
Upon decreasing $\Lambda$, this trend is further enhanced: 
one-loop results deviate strongly from the rest;
starting at $\Lambda/J \!\leq\! 1$, we also observe deviations between $\ell \!=\! 2$ and $\ell \!>\! 2$.
For $\Lambda/J \!\leq\! 0.5$,
we find that loop convergence in the density part of the $t$ channel, $\gamma_t^d$, (right most column in Fig.~\ref{fig:loop_conv_and_parquet_check})
becomes increasingly difficult and is not achieved at $\Lambda/J \!=\! 0.1$.
However, $\gamma_t^d$ is orders of magnitude smaller than its spin counterpart.
Comparing all vertex components, a remaining discrepancy from loop convergence on the percent level is acceptable for our numerical accuracy.
In fact, the dominant spin part of the reducible vertex in the $t$ channel, $\gamma_t^s$, and the self-energy are well converged for $\ell \!\gtrsim\! 10$ even at $\Lambda/J \!=\! 0.1$.

The fast convergence of the spin susceptibility in loop number can be traced back to the influence of different vertex components in the general expression \eqref{eq:susceptibility_final} for $\chi^\omega_{ij}$. 
There, the density part of the vertex, which exhibits the slowest loop convergence in Fig.~\ref{fig:loop_conv_and_parquet_check}, 
enters as a purely local contribution ($\sim\! \delta_{ij}$).
Hence, it yields a $\mathbf{q}$-independent contribution to $\chi_{\omega_0}(\mathbf{q})$
and has no structural effect in contour plots such as Fig.~\ref{fig:susc}(a,b).
Moreover, it 
can actually \textit{not} lead to finite offset for $\chi_{\omega_0}(\mathbf{q} \!=\! \mathbf{0})$ at $\Lambda \!=\! 0$, as its contribution must cancel with other terms from the spin part of the vertex.
The reason is that the imaginary-time spin susceptibility at $\mathbf{q} \!=\! 0$
is proportional to $\langle \hat{S}^z_\text{tot}(\tau) \hat{S}^z_\text{tot} \rangle$ (with $\tau \!\geq\! 0$).
This expectation value vanishes for all $\tau$ since the ground state is a total spin singlet.

Upon loop convergence, mfRG results fulfill the self-consistent parquet equations.
It is for this reason that we initialize the mfRG flow with a converged parquet solution.
Yet, already at a large value of $\Lambda/J \!=\! 5$, it is nontrivial to obtain a solution of the parquet equations by iteration.
We find it to be impossible with a \textit{full} update
and need to incorporate a damping factor of $0.5$ (see Sec.~\ref{sec:parquet}).
The difficulty of converging the parquet equations is further illustrated in the second row of Fig.~\ref{fig:loop_conv_and_parquet_check}, computed at $\Lambda/J \!=\! 3$.
The thick dotted black shows the parquet solution obtained by fixed-point iteration with a damping factor of $0.3$.
In the first three columns of this row, the converged parquet solution lies perfectly on top of the mfRG curves. 
However, in the fourth column, showing the density part, discrepancies between the parquet and mfRG results remain.
The reason for these discrepancies may be that our convergence criterion in the iterative parquet solution is rather insensitive to $\gamma_t^d$, as it is an order of magnitude smaller $\gamma_t^s$.
Another possibility is that the mfRG results for $\gamma_t^d$ would still change notably if one included the internal iterations in the mfRG flow (Sec.~\ref{sec:mfRG_hierarchy}, ``third step''), omitted in this work.
In fact, these incorporate changes to the vertex flow induced by the mfRG corrections to the self-energy flow (Sec.~\ref{sec:mfRG_hierarchy}, ``second step''),
and we find the latter necessary to obtain a perfect match between fixed-point parquet and mfRG results for $\Sigma$, shown in the first column.
All in all, we take the agreement between $\ell \!=\! 13$ and the converged parquet results as confirmation that mfRG provides a parquet solution.
Nevertheless, at any finite loop number (and without internal iterations), 
we expect marginal differences between mfRG results and the \textit{exact} fixed point of the parquet equations.
Indeed, inserting the $\ell \!=\! 13$ results into the parquet equations yields the gray dotted curve. This \textit{one-shot} iteration of the parquet equations, starting from an mfRG result close to the parquet fixed point, leads to notable deviations from the converged solution. 
This corresponds to the fact that iterating the parquet equations without damping is unstable.
For these systems, a \textit{one-shot} insertion into the parquet equations cannot be used to check consistency with the parquet solution.
It rather serves as an estimator of how difficult it is to obtain said parquet solution.

Accordingly, for lower values $\Lambda/J \!<\! 2$, we find increased differences between the mfRG result at $\ell \!=\! 13$ and a \textit{one-shot} parquet iteration subsequently performed on it.
We tried finding a parquet solution by fixed-point iteration for $\Lambda/J \!<\! 1$. However, this required extremely small values of the damping factor and soon became unpractical.
Progress on that front may be possible using more refined fixed-point iteration techniques, as mentioned in Sec.~\ref{sec:parquet}, but we leave that as a topic for future study.

From the fact that our mfRG results match the direct parquet solution at large $\Lambda$ and show convergence in loop number for $\ell \!\gtrsim\! 10$ up to $\Lambda \!\gtrsim\! J/2$, we infer that mfRG can provide a solution of the parquet equations throughout the flow.
It is highly remarkable that the flow yields stable results, even though solving the parquet equations by iteration becomes increasingly challenging with decreasing $\Lambda$.
Pseudofermion systems thus provide an example where the mfRG reformulation of the parquet equations as an RG flow is crucial to the success of the method.

\subsection{Spin susceptibility and spinon self-energy}
\label{sec:physical-results}

In this section, we present our physically most relevant results. 
They are summarized in Fig.~\ref{fig:susc}, showing the spin susceptibility and the pseudofermion, or spinon, self-energy for two parameter sets. 
The upper row shows data for the spin-liquid phase at the Heisenberg point, $\alpha \!=\! \arctan(J_2/J_1) \!=\! 0\degree$, 
whereas the lower row shows data at $\alpha \!=\! 45\degree$, 
where the ground state exhibits $\mathbf{q} \!=\! 0$ magnetic order. 
These results were obtained for loop number $\ell \!\geq\! 5$, where $\chi$ is well converged, ($\ell \!=\! 5$ for $\alpha \!=\! 0\degree$ and $\ell \!=\! 8$ for $\alpha \!=\! 45\degree$) and low values of the IR cutoff $\Lambda$ compared to $J \!=\! \sqrt{J_1^2+J_2^2}$.
How low $\Lambda/J$ can be taken depends on the 
size of the real-space grid for the pseudofermion vertex.
For $\alpha \!=\! 0\degree$, we take $\Lambda/J \!=\! 0.02$.
We also obtained data for $\Lambda/J \!\leq\! 0.1$,
but, there, finite-size effects are discernible [e.g.\ as superposed oscillations in $\chi_{\omega_0}(\mathbf{q})$ in plots such as Fig.~\ref{fig:susc}(a)], 
similarly as they are for $\ell \!=\! 1$ and $\Lambda/J \!\leq\! 0.05$.
For $\alpha \!=\! 45\degree$, we take $\Lambda/J \!=\! 0.18$, which is significantly smaller than $1$ but still large enough that finite-size effects [e.g.\ ring-like oscillations around the maxima in Fig.~\ref{fig:susc}(b)] do not prevail.

Figures~\ref{fig:susc}(a,b) show contour plots of the static spin susceptibility as a function of momentum, $\chi_{\omega_0}(\mathbf{q})$. At the Heisenberg point [Fig.~\ref{fig:susc}(a)], the susceptibility exhibits a broad continuum without sharp Bragg peaks, consistent with a quantum spin-liquid phase. By contrast, results for $\alpha \!=\! 45\degree$ [Fig.~\ref{fig:susc}(b)] show well developed Bragg peaks at the M points of the extended Brillouin zone, as expected in the $\mathbf{q} \!=\! 0$ magnetically ordered phase.

Figures~\ref{fig:susc}(c,d) show static spin correlations $\chi_{ij}^{\omega_0}$ in real space as a function of distance $|\mathbf{r}| \!=\! |\mathbf{R}_i \!-\! \mathbf{R}_j|$. 
Data along the $[10]$ and $[1\bar{1}]$ direction, where correlations decay slowest for $\alpha \!=\! 0\degree$ and $45\degree$, respectively, are highlighted as colored triangles. 
These slowest decaying components dominate the asymptotics of the spin susceptibility at large distances and are therefore of prime interest.
At the Heisenberg point [Fig.~\ref{fig:susc}(c)], the decay of correlations is clearly slower than exponential, 
and our data indicates an algebraic decay, consistent with a gapless $U(1)$ Dirac spin liquid. 
We find this slower-than-exponential decay to be very robust
and already present in our one-loop data (i.e.\ it is not an effect of higher loops)
and also for larger values of the IR cutoff, $\Lambda/J \!\lesssim\! 0.5$ (see Sec.~\ref{sec:chi_qualitative}). 
By contrast, previous (one-loop) pffRG studies observed an exponential decay, suggesting the presence of an energy gap at the Heisenberg point \cite{Suttner2014, Buessen2016}. 
However, in Ref.~\cite{Buessen2016}, the exponential fit
was not restricted to points along the slowest-decaying direction, as done here, but included all data points.
If all data points in Fig.~\ref{fig:susc}(c) were similarly considered together, the presence of slower-than-exponential decay would likewise not be as clear as when focusing on the slowest-decaying direction. 
Moreover, we speculate that our results may differ from those of Refs.~\cite{Suttner2014, Buessen2016} due to our improved treatment of the vertices' frequency dependence (channel-adaptive parametrization, high resolution of low-energy features) 
and the fact the we use a smooth rather than sharp IR cutoff,
so that our RG flow is more susceptible to low-energy features for a given, finite value of $\Lambda/J$.

\begin{figure}
    \centering%
    \includegraphics[scale=1]{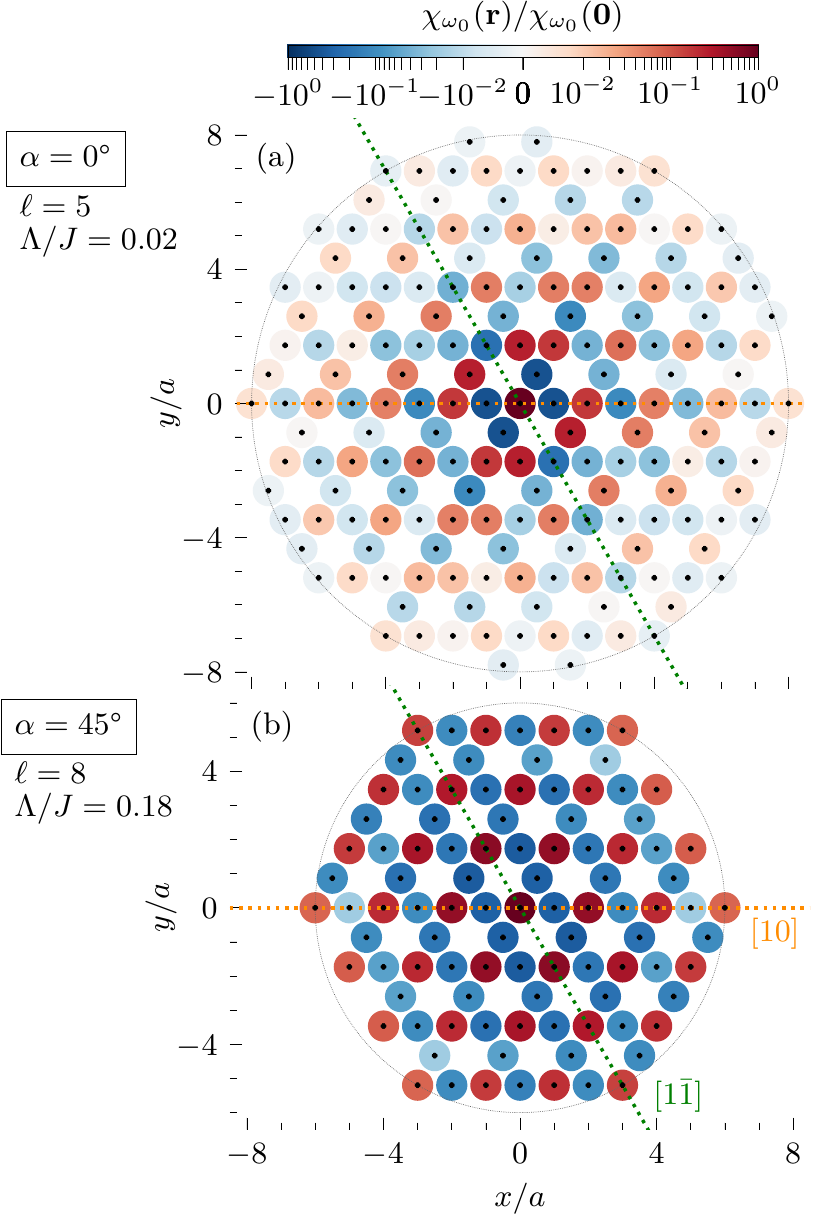}
    \caption{%
    Real-space dependence of the static spin susceptibility, $\chi_{ij}^{\omega_0}$, obtained from the mfRG flow,
    (a) for $\alpha \!=\! 0\degree$ ($\ell \!=\! 5$, $\Lambda/J \!=\! 0.02$, $J\chi_{\omega_0}(\mathbf{0}) \!\approx\! 1.57$) and
    (b) for $\alpha \!=\! 45\degree$ ($\ell \!=\! 8$, $\Lambda/J \!=\! 0.18$, $J\chi_{\omega_0}(\mathbf{0}) \!\approx\! 0.54$).
    The color code represents the value of $\chi_{ij}^{\omega_0}$ on lattice site $j$ with $i$ fixed at the origin. 
    The pictured grid with (a) 181 and (b) 97 lattice sites comprises the finite correlation area in the real-space parametrization of the pseudofermion vertex.
    From the decay of $|\chi_{\omega_0}(\mathbf{r})|$ at the boundaries of the grid in (a), we infer that finite-size effects are still small.
    By contrast, in (b), correlations become long-ranged as one approaches the magnetically ordered phase.
    The orange (green) dashed line shows the $[10]$ ($[1\bar 1]$) direction, where the decay of correlations is slowest (see Fig.~\ref{fig:susc}).
    }
    \label{fig:real_space_2d}
\end{figure}

For $\alpha \!=\! 45\degree$, our data shows an exponential decay of spin-spin correlations [Fig.~\ref{fig:susc}(d)]. 
There is an even--odd effect for correlations along the $[10]$ direction:
correlations between sites on the same sublattice (full symbols) are positive and larger in magnitude than correlations between sites on different sublattices (empty symbols), which have negative sign. 
Correlations of this type are indeed expected upon approaching the $\mathbf{q} \!=\! 0$ magnetically ordered phase, which exhibits $120\degree$ angles between spins at the three different sublattice sites.

To make this more transparent, we plot in Fig.~\ref{fig:real_space_2d} the two-dimensional real-space dependence of the static spin susceptibility for both phases \cite{Buessen2016}.
For $\alpha \!=\! 0\degree$ [Fig.~\ref{fig:real_space_2d}(a)],
one observes antiferromagnetic correlations in the directions across nearest neighbors that decay over a few lattice constants $a$. 
As seen in both Figs.~\ref{fig:susc}(c) and \ref{fig:real_space_2d}(a),
the absolute value $|\chi_{\omega_0}(\mathbf{r})|$ has decayed by over two orders of magnitude at the edge of our finite correlation area. Hence, we can infer that, for $\Lambda/J \!\gtrsim\! 0.02$, finite-size effects from the real-space parametrization of the vertex are small and the real-space truncation does not affect our results significantly.
For $\alpha \!=\! 45\degree$, correlations naturally become more and more long-ranged as one lowers $\Lambda/J$ and approaches the ordered phase.
Since we prioritized our computational resources for studying the more interesting Heisenberg point,
we used a slightly smaller lattice for $\alpha \!=\! 45\degree$, and finite-size effects are still under control for $\Lambda/J \!\gtrsim\! 0.18$.
The corresponding color plot of $\chi_{\omega_0}(\mathbf{r})$ at this value of $\Lambda/J$ [Fig.~\ref{fig:real_space_2d}(b)] already reveals the succinct pattern of the $\mathbf{q} \!=\! 0$ phase, with two spins being parallel or at $120\degree$ to each other if they reside on the same or two different sublattices, respectively.

Lastly, Figs.~\ref{fig:susc}(e,f) show data for the (purely imaginary) spinon self-energy $\Sigma$ as a function of frequency $\nu$. 
For large frequencies $\nu \!\gg\! J$, its magnitude $|\Sigma|$ decays as $1/\nu$; 
for frequencies $\nu \ll \Lambda$ 
(devoid of physical significance, since below the IR cutoff), 
it exhibits trivial linear behavior. At the Heisenberg point,
we would expect a power-law dependence at low frequencies above the IR cutoff if the ground state is indeed an algebraic spin liquid, as suggested by our results for the real-space correlations. 
In Fig.~\ref{fig:susc}(e), our data does show a small frequency regime where such a power law seems to emerge out of a shoulder-like feature. 
However, the value of $\Lambda/J = 0.02$ is not small enough to find a conclusive power law over more than one decade in frequencies. 
(Our one-loop results actually show an extended power law without the shoulder (see Fig.~\ref{fig:self-energy-comparison}), but they are affected by the finite maximum range of correlations in our simulation.)
In contrast to such algebraic behavior, for a gapped spin liquid with a spinon excitation gap $\Delta$, we would expect a diverging self-energy of the form $\Sigma_\nu \!=\! \Delta^2/(\mi\nu)$.
This is clearly not consistent with our data down to $\Lambda/J \!=\! 0.02$, which thus serves as an upper bound for a potential spinon gap, $\Delta \!<\! 0.02 \, J$. 
For $\alpha \!=\! 45\degree$ [Fig.~\ref{fig:susc}(f)], 
we do not find such rich behavior in the self-energy.
Indeed, the energy window between $|\Sigma| \!\sim\! \nu$ and $|\Sigma| \!\sim\! 1/\nu$ is rather small
due to the finite IR cutoff $\Lambda/J \!=\! 0.18$,
which is necessary to ensure finite results for $\chi$ with controllable finite-size effects.

\subsection{On the occupation constraint in pffRG}
\label{sec:constraint}

The pseudofermion representation introduces unphysical states in the Hilbert space.
Physical states are one-fermion, spin-$\tfrac{1}{2}$ degrees of freedom per site, 
unphysical ones have zero or two fermions per site.
In principle, one must therefore enforce a constraint of single fermionic occupation on each lattice site.
In practice, targeting zero-temperature properties, this constraint is only enforced on average through an appropriate chemical potential.
The argument is that unphysical states correspond to vacancies in the lattice and are therefore gapped out;
the ground state should lie in the sector with one fermion, i.e.\ maximal spin magnitude $\tfrac{1}{2}$, per lattice site \cite{Baez2017,Reuther2014a,Reuther2011a}.
The validity of such an argument is, however, less clear for frustrated systems such as the kagome lattice, where vacancies could affect the degree of frustration. 
In this section, we investigate the fulfillment of the constraint in pffRG in more detail.

By particle-hole symmetry, the constraint is indeed fulfilled \textit{on average},
$\langle \hat{n}_i \rangle \!=\! 1$ (where $\hat{n}_i \!=\! \sum_\alpha \hat{c}^\dag_{i\alpha}\hat{c}_{i\alpha}$),
and $\langle \hat{S}^z_i \rangle \!=\! 0$. 
Evidently, these measures are oblivious to fluctuations around their mean, and the relations are still fulfilled if empty and doubly occupied sites contribute equally.
However, \textit{exact} fulfillment of the constraint can be checked through particle-number fluctuations or, equivalently, the spin magnitude per site:
$\langle \hat{n}_i^2 \rangle \!=\! \langle \hat{n}_i \rangle^2 \!\Leftrightarrow\! \langle \hat{S}^z_i \hat{S}^z_i \rangle \!=\! \tfrac{1}{4}$ 
(using $\langle \hat{n}_i \rangle \!=\! 1$).
Indeed, since $\langle \hat{S}^z_i \hat{S}^z_i \rangle$ is bounded by $0$ and $\tfrac{1}{4}$,
finding the maximal value excludes contributions of unphysical states.
Now, with a method capable of computing the spin susceptibility $\chi_{ij}^\omega$,
one can evaluate the spin magnitude per site as a local equal-time correlator:
summing over $\omega$ in Eq.~\eqref{eq:static_susc} yields
\begin{equation}
\frac{1}{\beta} \sum_\omega \chi_{ii}^\omega
=
\chi_{ii}^{\tau=0} 
=
\langle \hat{S}_i^z \hat{S}_i^z \rangle
\overset{!}{=}
\tfrac{1}{4}
. 
\label{eq:chisum}
\end{equation}

In pffRG, the spin susceptibility stems from a four-fermion correlator with two distinct contributions.
The disconnected part [first term in Eq.~\eqref{eq:susceptibility_final}] gives
(no summation over $\alpha$, $\beta$)
\begin{flalign}
-\tfrac{1}{2} G(\tau=0^+) G(\tau=0^-)
& =
\tfrac{1}{2} 
\langle \hat{c}_{i\alpha} \hat{c}_{i\alpha}^\dag \rangle
\langle \hat{c}_{i\beta}^\dag \hat{c}_{i\beta} \rangle 
=
\tfrac{1}{8}
\hspace{-0.5cm} &
\end{flalign}
for the sum rule \eqref{eq:chisum}.
Indeed, this result holds for any finite $\Lambda$ by particle-hole symmetry.
The vertex correction [remainder in Eq.~\eqref{eq:susceptibility_final}] 
is suppressed for large $\Lambda$.
Thus, for very large (but finite) $\Lambda$, one finds
$\chi_{ii}^{\tau=0}=\langle {\hat{S}_i^z} {\hat{S}_i^z} \rangle = 1/8$,
corresponding to the \textit{fermionic} infinite-temperature limit, where
an empty, singly, or doubly occupied site is equally probable.
Decreasing $\Lambda$, the vertex corrections to $\chi$ become relevant.
They should suppress unphysical contributions and increase the value of $\chi_{ii}^{\tau=0}$.
In fact, to realize the spin-maximal value of Eq.~\eqref{eq:chisum} in the ground state,
they should similarly give $1/8$ at $\Lambda=0$.

\begin{figure}
    \centering
    \includegraphics[scale=1]{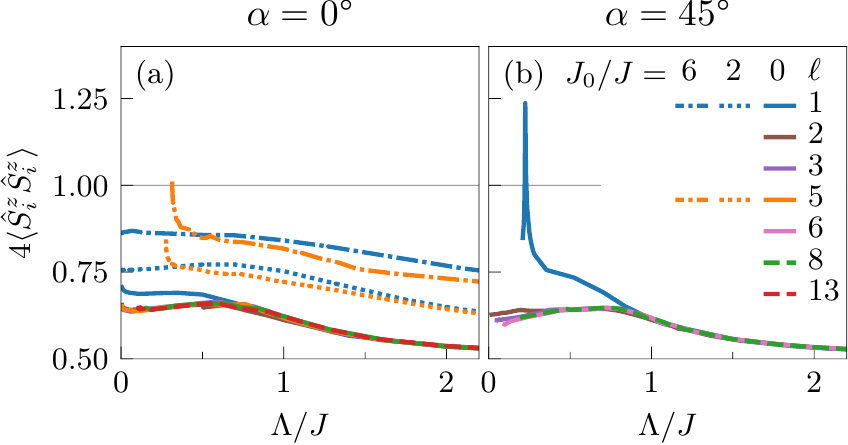}
    \caption{
        Flow of the on-site equal-time 
        correlator $\langle\hat S^z_i \hat S^z_i\rangle$, used to check exclusion of unphysical states in the pseudo\-fermion representation, 
        (a) at $\alpha \!=\! 0\degree$
        and (b) at $\alpha \!=\! 45\degree$.
        Solid, dotted, and dash-dotted lines represent data with an auxiliary on-site interaction $J_0/J \!=\! 0$, $2$, and $6$, respectively.
    }
    \label{fig:spinsq-constraint}
\end{figure}

Figure~\ref{fig:spinsq-constraint} shows our mfRG results for 
$4\chi_{ii}^{\tau = 0} \!=\! 4\langle \hat{S}^z_i \hat{S}^z_i\rangle$ 
along the flow, for both $\alpha \!=\! 0 \degree$ and $45\degree$
(solid lines). 
At the beginning of the flow, for $\Lambda \gg  J$, 
we find the expected value of $\tfrac{1}{2}$. 
This demonstrates that vertex corrections are still suppressed and that physical and unphysical states contribute to similar extent.
As we lower $\Lambda$, we find that the vertex corrections increase the value of $\langle \hat S^z_i \hat S^z_i\rangle$ and therefore do tend to reduce the contributions from unphysical states.
However, already on the level of pure one-loop calculations, 
there are two important observations:
(i) Even for our lowest values of $\Lambda/J$ on the subpercent level, 
the constraint $4\langle \hat S^z_i \hat S^z_i\rangle \!=\! 1$ is far from being exactly fulfilled. 
Instead, at $\alpha \!=\! 0\degree$ [Fig.~\ref{fig:spinsq-constraint}(a)], 
the final value of $4\langle \hat S^z_i \hat S^z_i\rangle$ is around $0.7$.
(ii) The operator bound $4\langle \hat S^z_i \hat S^z_i\rangle \leq 1$ is not guaranteed to be obeyed in a diagram-based approach such as pffRG. 
Indeed, at $\alpha \!=\! 45\degree$ [Fig.~\ref{fig:spinsq-constraint}(b)], 
where the one-loop result signals a divergence of $\chi$ at finite $\Lambda$, 
$4\langle \hat S^z_i \hat S^z_i\rangle$ shows a similar upturn, 
reaching values significantly larger than $1$.

We see in Fig.~\ref{fig:spinsq-constraint} (solid lines) that higher loops appreciably change $\chi_{ii}^{\tau=0}$ only for $\Lambda/J < 1$, similar to $\chi_{\omega_0}(\mathbf{q}_{\mathrm{M}})$ in Fig.~\ref{fig:flow}.
The effect of higher loops is also similar to Fig.~\ref{fig:flow} in that they reduce the values of 
$\chi_{ii}^{\tau=0}$
as they do for
$\chi_{\omega_0}(\mathbf{q}_{\mathrm{M}})$.
For $\alpha \!=\! 0\degree$, this leads to values of $4\langle \hat S^z_i \hat S^z_i\rangle$ around $0.65$.
For $\alpha \!=\! 45\degree$, where the higher loops successfully suppress the divergence of $\chi$ at finite $\Lambda$, they also suppress $4\langle \hat S^z_i \hat S^z_i\rangle$, such that it does not exceed the operator bound $1$. 

An intriguing idea for checking the single-occupation constraint was presented in Ref.~\cite{Baez2017},
where an auxiliary on-site term in the Hamiltonian, say, $\hat{H}_0 \!=\! -J_0 \hat{\mathbf{S}}^2$, was added.
If the constraint is exactly fulfilled, then $\hat{H}_0$ reduces to a constant $-3J_0/4$ and has no effect. By contrast, unphysical states would receive a large energy penalty and not contribute at low energies, i.e., $\Lambda \!\ll\! J_0$.
Of course, the converse statement is not true:
if some, particularly small, values of $J_0$ have no significant effect on $\chi$, it does not guarantee that the constraint is exactly fulfilled.
In Ref.~\cite{Baez2017}, it was shown for the honeycomb lattice that small values $J_0 \!\leq\! 0.6 J$ have no significant effect, and curves of $\chi_{\omega_0}(\mathbf{q})$ as a function of $\Lambda$ collapse for different $J_0$ upon rescaling.
We can confirm this finding on the kagome lattice for similarly small values of $J_0$.
In particular, small values $J_0 \!<\! J$ did not change the maximal value of $4\langle \hat S^z_i \hat S^z_i\rangle$ compared to the results shown in Fig.~\ref{fig:spinsq-constraint}.
Importantly, this does \textit{not} imply that the constraint is exactly fulfilled,
as we find that $4\langle \hat S^z_i \hat S^z_i\rangle$ stays well below $1$ for all $\Lambda/J$ included in the flow.

We also considered values of $J_0$ much larger than $J$.
The dotted and dash-dotted blue lines for $\alpha \!=\! 0\degree$ 
in Fig.~\ref{fig:spinsq-constraint}(a) represent one-loop data
for $J_0/J \!=\! 2$ and $6$, respectively.
(Note that $J$ is still defined as $\sqrt{J_1^2+J_2^2}$.)
The values of $4\langle \hat S^z_i \hat S^z_i\rangle$ increase substantially and monotonically with $J_0$. 
Regarding our main observable $\chi_{\omega_0}(\mathbf{q})$, we find that
these values of $J_0$ lead to quantitative but not qualitative changes (see Sec.~\ref{sec:chi_qualitative}). 
We did not perform calculations for even larger $J_0$.
However, having realized that the bound $4\langle \hat S^z_i \hat S^z_i\rangle \!\leq\! 1$ is easily violated in the one-loop results at $\alpha \!=\! 45\degree$,
we expect that one can similarly produce $4\langle \hat S^z_i \hat S^z_i\rangle \!>\! 1$ at $\alpha \!=\! 0\degree$ for sufficiently large $J_0$.

The crux of the matter is that large $J_0$ enters as a large bare vertex and thereby increases the strong-coupling challenges inherent to pffRG.
To make physical statements, we must reach $\Lambda \!\ll\! J$, ideally independent of $J_0$.
From the dotted lines in Fig.~\ref{fig:spinsq-constraint}(a), however, we see that, for $J_0 /J\!=\! 2$, deviations between loop orders $1$ and $5$ appear much earlier, already for $\Lambda/J \!\gtrsim\! 1$.
Similar deviations occur in the dash-dotted lines computed at $J_0/J \!=\! 6$ already for $\Lambda/J \!\gtrsim\! 2$.
For lower values of $\Lambda/J$, the five-loop curves exhibit a rapid upturn, signaling the onset of numerical instabilities in our computation.
Values for $4\langle \hat S^z_i \hat S^z_i\rangle$ closer to $1$
through a large on-site term $J_0$
thus come with increased challenges in the pffRG methodology, 
regarding convergence in loop number as well as other numerical parameters (accuracy of frequency integrals, solution of the RG flow, etc.).
The limit of $J_0 \!\to\! \infty$ is \textit{not} controlled in a diagram-based approach. 
Therefore, the strategy of using an energy penalty $J_0$ to suppress 
unphysical states cannot be evoked without caveats,
as it fundamentally relies on convergence for large $J_0$.

\subsection{Qualitative effects of \texorpdfstring{$\ell$}{l} and \texorpdfstring{$J_0$}{J0}}
\label{sec:chi_qualitative}

\begin{figure}
    \centering
    \includegraphics[scale=1]{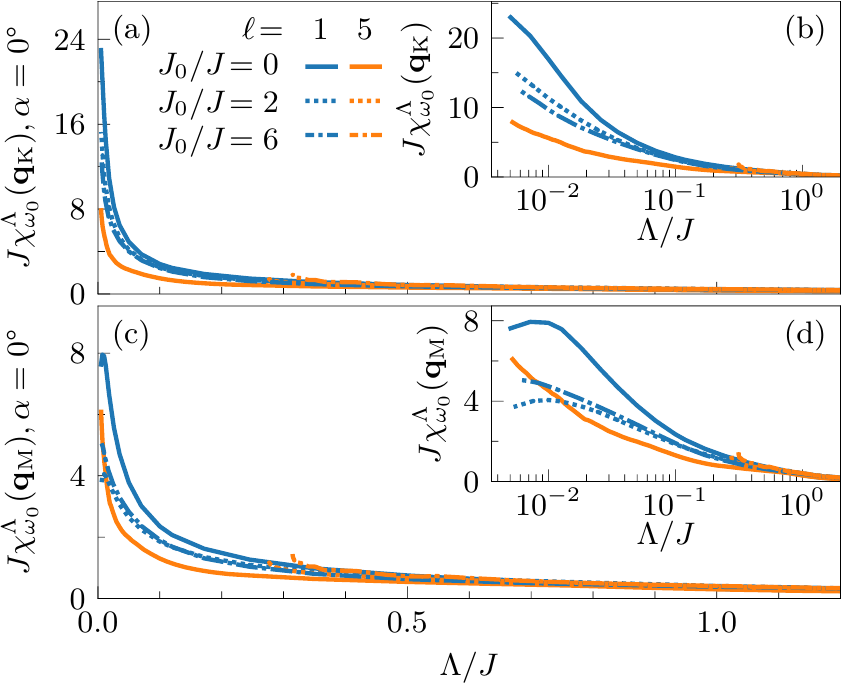}
    \caption{%
    Flow of the static spin susceptibility for $\alpha \!=\! 0\degree$ 
    at the K point (a,b) and the M point (c,d), with insets (b,d) using a logarithmic $\Lambda$ scale, analogous to Fig.~\ref{fig:flow}.
    Lines show six variants of the pffRG flow, obtained 
    by using a loop number of $\ell \!=\! 1$ or $5$,
    and an on-site interaction of $J_0/J \!=\! 0$, $2$, or $6$.
    The one-loop flows for $\chi_{\omega_0}(\mathbf{q}_\mathrm{M})$ level off and show an unphysical downturn for low $\Lambda$. 
    This is not seen for $\ell \!=\! 5$ down to $\Lambda/J \!=\! 0.01$.
    However, a five-loop flow with large on-site interaction $J_0/J \!\geq\! 2$ is numerically unstable for $\Lambda/J \!\lesssim\! 0.3$.
    }
    \label{fig:flow-onsite-comparison}
\end{figure}

In the preceding sections, we first introduced the multiloop pffRG flow, with higher-loop ($\ell$) terms used to incorporate contributions from the six-point vertex,
and then discussed the strategy of adding an on-site interaction proportional to $J_0$ to suppress unphysical states in the pseudofermion representation.
To make contact with previous work, it is worthwhile to analyze the effect of these variations of pffRG on a \textit{qualitative} level. 
We will focus on the Heisenberg point ($\alpha \!=\! 0\degree$),
where, for the given real-space parametrization,
the one- and five-loop flows show finite-size effects 
for $\Lambda/J \!\leq\! 0.05$ and $0.01$, respectively.
Regarding the spin susceptibility for $\Lambda/J \!\gtrsim\! 0.1$, we will find that the momentum dependence of the static spin susceptibility is actually structurally the same in \textit{all} calculations.
By contrast, we will see that the frequency dependence of the pseudofermion self-energy changes significantly with loop number and develops novel features if $\Lambda/J$ is decreased to the percent level.

To get an overview, Fig.~\ref{fig:flow-onsite-comparison} shows the flow of the static spin susceptibility evaluated at the K point (a,b) and the M point (c,d) for the different setups.
The solid curves with $\ell \!=\! 1$ and $\ell \!=\! 5$ are computed at $J_0 \!=\! 0$ and thus are the same as in Fig.~\ref{fig:flow}.
Recall that $\chi$ is well converged for $\ell \!\geq\! 5$ and $J_0 \!=\! 0$.
Additionally, the dotted and dash-dotted lines give one- and five-loop results for $J_0/J \!=\! 2$ and $6$, respectively.
Regarding the one-loop data (blue) in Fig.~\ref{fig:flow-onsite-comparison}(a,b), it is interesting to note that $\chi_{\omega_0}(\mathbf{q}_\mathrm{K})$ decreases with $J_0$,
although $\langle \hat{S}_i^z \hat{S}_i^z \rangle \!=\! \chi_{ii}^{\tau=0}$ increases with $J_0$
(cf.\ Fig.~\ref{fig:spinsq-constraint}).
This is not the case at $\chi_{\omega_0}(\mathbf{q}_\mathrm{M})$ shown in Fig.~\ref{fig:flow-onsite-comparison}(c,d).
However, there, we see that the downturn of $\chi$ for low $\Lambda/J$ persists,
such that we cannot fully trust our one-loop results for $\Lambda/J \!\lesssim\! 0.02$.
While higher loops remove this downturn for $J_0 \!=\! 0$,
a five-loop flow with large $J_0/J \!\geq\! 2$ is numerically unstable,
as already seen in Fig.~\ref{fig:spinsq-constraint}.
Hence, in the following, we compare $\chi$ in more detail among three setups:
(a) $\ell \!=\! 1$ and $J_0 \!=\! 0$,
(b) $\ell \!=\! 5$ and $J_0 \!=\! 0$,
(c) $\ell \!=\! 1$ and $J_0/J \!=\! 6$.

\begin{figure} 
    \centering%
    \includegraphics[scale=1]{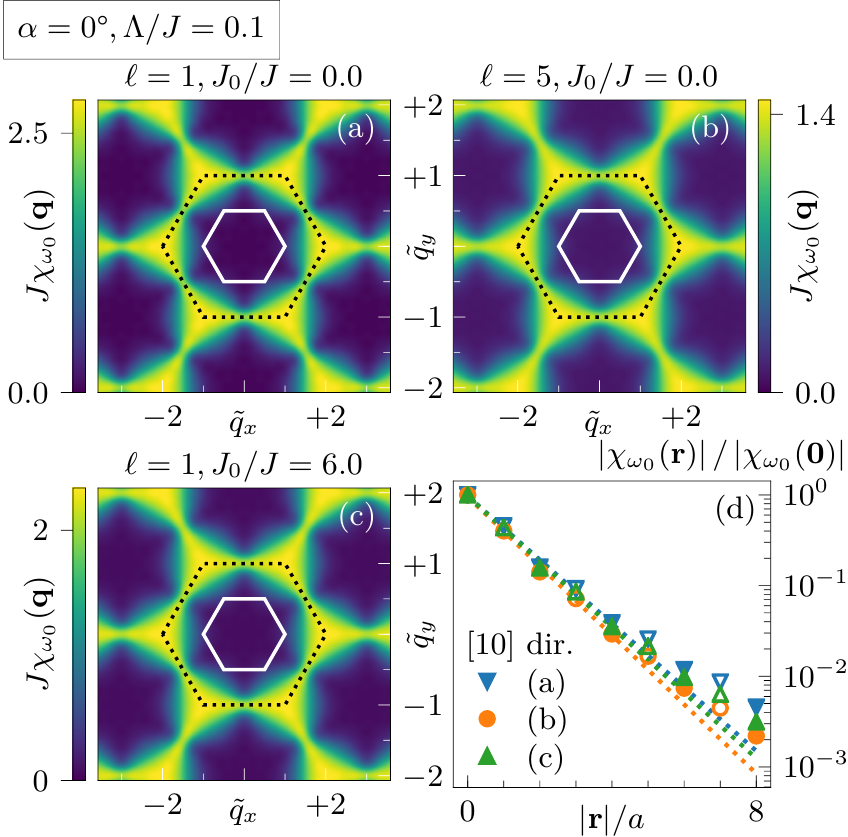}
    \caption{%
    Effect of loop order $\ell$ and onsite interaction $J_0$ on the static spin susceptibility $\chi_{\omega_0}$ at $\alpha \!=\! 0\degree$ and $\Lambda/J \!=\! 0.1$.
    (a--c) Comparison between $\chi_{\omega_0}(\mathbf{q})$ results obtained with
    (a) $\ell \!=\! 1$, $J_0 \!=\! 0$,
    (b) $\ell \!=\! 5$, $J_0 \!=\! 0$, and
    (c) $\ell \!=\! 1$, $J_0/J \!=\! 6$.
    (d) Decay of normalized spin-spin correlations as a function of distance in $[10]$ direction for the same data as in (a--c). Full (empty) symbols denote correlations with positive (negative) sign, analogous to Fig.~\ref{fig:susc}.
    The normalization values $J\chi_{\omega_0}(\mathbf{0})$ for (a), (b), (c) are $1.10$, $0.64$, and $0.96$, respectively.
    }
    \label{fig:onsite-comparison}
\end{figure}

Figure~\ref{fig:onsite-comparison} shows the momentum dependence of the static spin susceptibility for these three cases. 
From the color plots [Fig.~\ref{fig:onsite-comparison}(a,b,c)],
we see that $\chi_{\omega_0}(\mathbf{q})$ changes quantitatively, e.g.\ regarding the maximal value,
but the qualitative features are perfectly analogous. 
This provides strong support for pffRG as a reliable and efficient tool to scan phase diagrams.
Figure~\ref{fig:onsite-comparison}(d) displays the real-space decay of $\chi_{\omega_0}(\mathbf{r})$, normalized to its value at $\mathbf{r} \!=\! \mathbf{0}$.
Again, the maximum value changes among the three computations (see caption), 
but the qualitative behavior of $\chi_{\omega_0}(\mathbf{r})$ is analogous.
In particular, the deviation of $|\chi_{\omega_0}(\mathbf{r})|$ from an exponential decay is common to all calculations and thus a robust outcome of our analysis.

\begin{figure} 
    \centering%
    \includegraphics[scale=1]{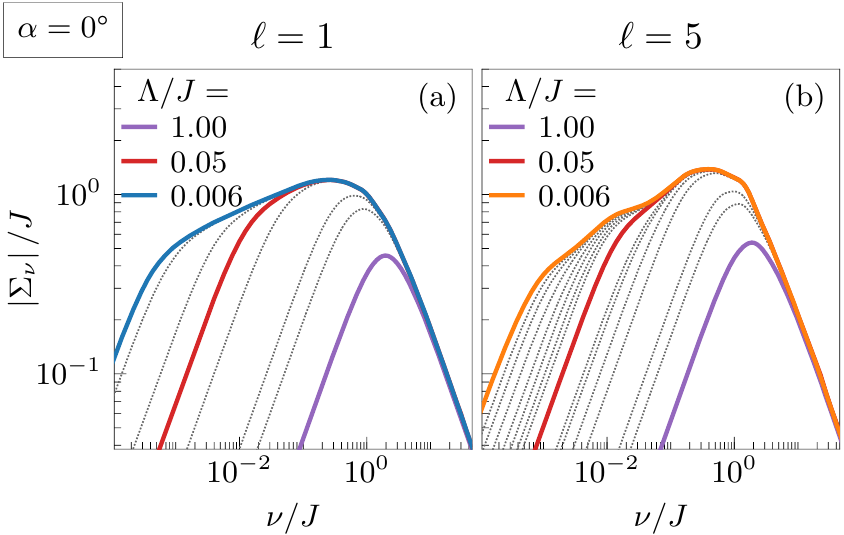}
    \caption{%
    Double logarithmic plots of the pseudofermion self-energy, obtained by (a) a one-loop and (b) a five-loop flow, for various $\Lambda/J$.
    Intriguing low-energy features emerge for $\Lambda/J$ on the percent level.
    For $\Lambda/J \!\geq\! 0.05$, the one-loop curve shows an extended low-energy power law,
    while the five-loop flow shows a shoulder-like low-energy frequency dependence where power-law behavior is less clear.
    However, at $\ell \!=\! 1$, effects of the finite maximum range of correlations in our numerical simulations are seen already for $\Lambda/J \!\leq\! 0.05$ (cf.\ Fig.~\ref{fig:flow}).
    At $\ell \!=\! 5$, they occur later in the flow, appearing for $\Lambda/J \!\leq\! 0.01$.
    }
    \label{fig:self-energy-comparison}
\end{figure}

Lastly, we discuss the low-energy features of the pseudofermion self-energy.
We restrict ourselves to $J_0 \!=\! 0$, since $J_0$ changes the characteristic energy scale of the model to $\sqrt{J^2 \!+\! J_0^2}$ \cite{Baez2017}, 
making it hard to compare results that depend sensitively on the ratio of $\Lambda$ to the characteristic scale.
Figure~\ref{fig:self-energy-comparison} shows $\Sigma_\nu$ in (a) a one-loop and (b) a five-loop calculation, as a set of curves for $\Lambda/J$ ranging from a large value of $1$ down to the subpercent level.
For $\Lambda/J \!>\! 0.1$, 
one detects in Figure~\ref{fig:self-energy-comparison}
mostly the simple $\sim\! \nu$ and $\sim\! \frac{1}{\nu}$ behavior of $\Sigma_\nu$ for $\nu \!<\! \Lambda$ and $\nu \!\gg\! J$, respectively 
(cf.\ Sec.~\ref{sec:physical-results}),
with a maximum in between.
However, if $\Lambda/J$ is reduced to the percent level, intriguing low-energy features emerge.
At $\Lambda/J \!=\! 0.05$ (red curves), the $\nu \!<\! \Lambda$ flank and the maximum are connected by a rather straight line in the double logarithmic plots of both 
Figs.~\ref{fig:self-energy-comparison}(a) and (b).
Quantitatively, we already find notable differences between the two results, with slightly different slopes of the lines for $\ell \!=\! 1$ and $5$.
In the one-loop calculation for even lower $\Lambda/J$, this straight line extends over orders of magnitude, corresponding to a pronounced low-energy power law. 
Yet, this must be taken with a grain of salt, as 
the one-loop flow suffers from finite-size effects already for $\Lambda/J \!\leq\! 0.05$
since it incorporates inter-channel feedback only partially
[see our discussion of Figs.~\ref{fig:flow}(b,d)].
The five-loop flow can be followed to lower energies, with finite-size effects appearing only for $\Lambda/J \!\leq\! 0.01$.
It shows a shoulder-like low-energy frequency dependence, where power-law behavior is less clear.
It is possible that the shoulder widens and flattens to reveal an extended power law upon further decreasing $\Lambda$,
but our data is not conclusive in this regard.
Moreover, convergence in loop number is difficult to achieve in this extremely challenging regime, because finite-size effects become more prominent at small $\Lambda$. 
For $\Lambda/J \!<\! 0.01,$ it would thus be necessary to increase both loop number as well as the the real-space grid of the vertex to achieve loop convergence and obtain conclusive results.

\section{Discussion and conclusions}
\label{sec:conclusion}

While the pseudofermion fRG (pffRG) \cite{Reuther2010} has served as a promising theoretical tool to study quantum spin systems for roughly a decade, 
fundamental questions about its applicability have remained open. 
In this work, we addressed some of these questions by using a multiloop pffRG approach.
We found that both increasing the number of loops within this scheme and adding an energy penalty for unphysical states left the \textit{qualitative} properties of the spin susceptibility $\chi$ unchanged. 
This provides strong support for pffRG as a reliable and efficient tool to scan phase diagrams of quantum spin systems.
The most important \textit{quantitative} effect of the multiloop approach is that it can reduce the values of $\chi$ through inter-channel screening and thus allow one to follow the flow 
further into the low-energy regime. 
For $\Lambda/J$ on the percent level, 
the pseudofermion self-energy in the spin-liquid phase
develops an intriguing low-energy frequency dependence, which still varies between low loop numbers, 
requiring further study with larger computational resources. 
In more detail, our results are summarized as follows. 

First, spin models in the pseudofermion representation are inherently strongly interacting, since the kinetic energy is quenched and no meaningful perturbation theory can be formulated. As such, the applicability of diagram-based approaches like fRG is contentious. 
The truncation in the fRG hierarchy of flow equations is \textit{a priori} uncontrolled and its quantitative reliability difficult to assess.
In this work, we further investigated the effect of truncation
by employing a multiloop fRG (mfRG) scheme. 
In mfRG, higher-loop corrections to the flow of the two-particle vertex are systematically included up to arbitrary loop order,
and converged results agree with those of the parquet approximation.
Despite the fact that the model is strongly interacting, we find loop convergence in mfRG.
For some vertex components, this is relatively slow;
for $\Lambda \!\gtrsim\! J/2$, loop numbers on the order of $10$ seem sufficient.
Fortunately, for the spin susceptibility, loop convergence is much quicker and achievable for values of the infrared (IR) cutoff $\Lambda$ as low as $0.01 \, J$,
\textit{substantially} smaller than the microscopic energy scale $J$.
Hence, loop convergence can be added to the list of arguments
testifying to the internal consistency and heuristically motivating the reliability of pffRG analyses of frustrated quantum magnets. 
By contrast, solving the parquet equations by iteration becomes unfeasible in the regime of low $\Lambda/J$, as fixed-point iterations become unstable and convergence is at best extremely slow.
Thus, mfRG provides a route to circumvent these problems. 

Second, the pseudofermion representation of the spin algebra relies on enforcing a single-occupation constraint of one fermion per lattice site. 
In pffRG, this constraint is typically enforced on average only, 
working on the presumption that the ground state of the fermionic model lies in the physical sector of the Hilbert space with one fermion per site. 
In this work, we explicitly computed fermion-number fluctuations and showed that the single-occupation constraint is generically not fulfilled exactly during the RG flow. This problem can be remedied, in principle, by penalizing unphysical states using an auxiliary on-site interaction, as proposed by Baez and Reuther \cite{Baez2017}. 
Extending their analysis, we find that the energy penalty needs to be substantially larger than the exchange coupling $J$ in order to significantly reduce fermion-number fluctuations. 
Fortunately, qualitative physical conclusions regarding, e.g., the momentum structure of the static spin susceptibility
or its real-space decay, are hardly affected.
However, quantitative predictions
and the detailed behavior of the pseudofermion vertices do change.
Moreover, a large on-site interaction adversely affects the convergence properties of pffRG, particularly w.r.t.\ increasing loop number. Further pursuing this route would thus require more computational resources than were available for this work.
Alternatively, it might be worthwhile to revisit other techniques of enforcing the constraint, e.g., via the Popov--Fedotov trick \cite{Popov1988,Reuther2010,Reuther2011c,Roscher2019}, Majorana fermions \cite{Tsvelik2003,Niggemann2020}, or the recently introduced spin fRG \cite{Krieg2019,Goll2019}. We leave a detailed 
investigation of this problem for future work.

Finally, the discussion of physical results in pffRG studies has been limited to the spin susceptibility in the vast majority of previous works. 
In our approach, the increased frequency resolution in combination with an adaptive frequency grid 
allowed us to follow the RG flow down to very low values of $\Lambda/J$. This enabled us to analyze the frequency dependence of the fermionic self-energy to obtain additional information about the character of potential spin-liquid phases. 
At the Heisenberg point, we find a shoulder-like feature out of which a power-law dependence seems to emerge at small frequencies above the IR cutoff $\Lambda$. A conclusive estimate of this power law would require to flow to even lower values $\Lambda/J \!\ll\! 0.02$. 
At the moment, our computational resources are not sufficient to reach much lower values of $\Lambda$, as this requires larger real-space grids and the RG flow in this regime becomes very slow and computationally costly. In any case, our results at the Heisenberg point at $\Lambda/J \!=\! 0.02$ are \textit{not} consistent with a gapped spin-liquid, for which a diverging self-energy of the form $\Sigma_\nu \!\sim\! |\Delta|^2 / (\mi \nu) $ is expected. The IR cutoff $\Lambda$ thus serves as an upper bound for a potential spinon gap, $\Delta \!<\! 0.02 \, J$.
Furthermore, the real-space decay of spin-spin correlations in our results hints toward a $U(1)$ spin liquid at the Heisenberg point. 
While previous (one-loop) pffRG studies found an exponential decay \cite{Suttner2014,Buessen2016}, 
our results are consistent with an algebraic decay.

As an outlook, we note that our multiloop pffRG approach can be readily applied in higher dimensions as well, 
where powerful numerical tools like QMC or MPS and tensor network methods fail either due to the sign problem, or due to the unfavorable scaling of computational costs. 
Three-dimensional frustrated quantum magnets, such as rare-earth pyrochlores, potentially realizing a quantum spin-ice phase \cite{Gardner1999,Molavian2007,Ross2011,Thompson2011,Fennell2012},
are of particular interest. Minimal models for these materials involve effective spin-$\tfrac{1}{2}$ moments on a pyrochlore lattice, coupled by dominant Heisenberg interactions, typically with an easy-axis exchange anisotropy \cite{Gingras2014}. 
For an extensive discussion of pffRG methodology for system with reduced spin symmetry, see Ref.~\cite{Buessen2019}.

\acknowledgements
Parallel to this work, a separate study on applying mfRG in the pffRG context \cite{Kiese2020b} has been performed by a Cologne--W\"urzburg--Madras collaboration involving Dominik Kiese, Tobias M\"uller, Yasir Iqbal, Ronny Thomale, and Simon Trebst. 
We warmly thank these colleagues for numerous constructive discussions and an open exchange of information throughout the course of these two projects, 
for agreeing to coordinate the submission of our papers, and for postponing the submission date until our computations were complete. 
We also thank Santiago Aguirre, Sabine Andergassen, Patrick Chalupa, Cornelia Hille, Severin Jakobs, Volker Meden, Alessandro Toschi, and Elias Walter for constructive discussions on mfRG in other contexts. 
We acknowledge support from the Deutsche Forschungsgemeinschaft under Germany’s Excellence Strategy—EXC-2111–390814868.

\appendix

\section{Fully parametrized functions}
\label{sec:appendix_bubbles}

We begin by stating the relation between the spin susceptibility and the pseudofermion four-point vertex.
It is obtained by inserting the pseudofermion representation into Eq.~\eqref{eq:static_susc}
and expressing the resulting four-fermion correlator through the full vertex and full propagators. One finds \cite{Reuther2010}
\begin{align}
\nn \chi_{ij}^\omega &
=
\frac{-1}{2\beta} \sum_{\nu} 
G(\nu-\tfrac{\omega}{2}) G(\nu+\tfrac{\omega}{2}) \delta_{ij}
+
\frac{-1}{2\beta^2} \sum_{\nu_{1}, \nu_{2}} 
\\
\nn 
& \times 
G(\nu_{1}-\tfrac{\omega}{2}) G(\nu_{1}+\tfrac{\omega}{2}) 
G(\nu_2-\tfrac{\omega}{2}) G(\nu_2+\tfrac{\omega}{2})
\\
\nn
& \times
\Big[
-
\Gamma^s_{ii}(\nu_2-\tfrac{\omega}{2},\nu_{1}+\tfrac{\omega}{2};\nu_{1}-\tfrac{\omega}{2},\nu_2+\tfrac{\omega}{2}) 
\, \delta_{ij}
\\
\nn
& \ \ \quad 
+
\Gamma^d_{ii}(\nu_2-\tfrac{\omega}{2},\nu_{1}+\tfrac{\omega}{2};\nu_{1}-\tfrac{\omega}{2},\nu_2+\tfrac{\omega}{2}) 
\, \delta_{ij}
\\
& \ \ \quad
- 2 \,
\Gamma^s_{ij}(\nu_{1}+\tfrac{\omega}{2},\nu_2-\tfrac{\omega}{2};\nu_{1}-\tfrac{\omega}{2},\nu_2+\tfrac{\omega}{2})
\Big]
.
\label{eq:susceptibility_final}
\end{align} 
Here, $\Gamma_{ij}$ is short for the vertex with horizontally connected fermion lines, 
$\Gamma_{ij} \!=\! \Gamma^{ \mathbin{\rotatebox[origin=c]{90}{\tiny)(}}}_{ij}$,
which is equal to 
$\Gamma^{\mathbin{\rotatebox[origin=c]{0}{\tiny)(}}}_{ij}$
anyway by the crossing symmetry \eqref{eq:channel_mapping_full} of the full vertex.
We compute the scale-dependent susceptibility during the flow from Eq.~\eqref{eq:susceptibility_final} with scale-dependent propagators and scale-dependent vertices.

Next, we present the fully parametrized form of two-particle bubble functions
(depicted in Fig.~\ref{fig:bubble_param_gen}). 
Upon inserting the real-space parametrization \eqref{eq:real_param} of vertices, 
each bubble function in Eq.~\eqref{eq:general_bubbles} decomposes into two parts, one with horizontally and another one with vertically connected fermion lines.
We give the result for the former case.
Similarly, by insertion of the spin parametrization, the bubble functions decompose into a spin and density part, as determined by the Pauli matrices and Kronecker $\delta$'s in the equations below.

The $a$ bubble with horizontal lines requires only vertices with horizontal lines and reads
\begin{align}
\label{eq:final_a}
&
B_{a;i_1i_2}^{\Gamma,\Gamma'}(1'2';12) =
\frac{1}{\beta} \sum_{\nu_3,\nu_4} G(\nu_3)G(\nu_4) 
\\
\nn
& \, \times
\pig\{
\pig[
2 \, 
\Gamma^s_{i_1i_2}(\nu_{1'}\nu_{4};\nu_{3}\nu_{2})
\,
\Gamma'^s_{i_1i_2}(\nu_{3}\nu_{2'};\nu_{1}\nu_{4} )
\\
\nn
& \quad + 
\Gamma^d_{i_1i_2}(\nu_{1'}\nu_{4};\nu_{3}\nu_{2})
\,
\Gamma'^s_{i_1i_2}(\nu_{3}\nu_{2'};\nu_{1}\nu_{4} )
\\
\nn
& \quad + 
\Gamma^s_{i_1i_2}(\nu_{1'}\nu_{4};\nu_{3}\nu_{2})
\,
\Gamma'^d_{i_1i_2}(\nu_{3}\nu_{2'};\nu_{1}\nu_{4} )
\pig]
\sigma^\mu_{\sigma_{1'}\sigma_{1}} \sigma^\mu_{\sigma_{2'}\sigma_{2}} 
\\
\nn
& \quad + 
\pig[
3 \,
\Gamma^s_{i_1i_2}(\nu_{1'}\nu_{4};\nu_{3}\nu_{2})
\,
\Gamma'^s_{i_1i_2}(\nu_{3}\nu_{2'};\nu_{1}\nu_{4} )
\\
\nn
& \quad + 
\Gamma^d_{i_1i_2}(\nu_{1'}\nu_{4};\nu_{3}\nu_{2})
\,
\Gamma'^d_{i_1i_2}(\nu_{3}\nu_{2'};\nu_{1}\nu_{4} )
\pig]
\delta_{\sigma_{1'}\sigma_{1}} \delta_{\sigma_{2'}\sigma_{2}}
\pig\}
.
\end{align}

As both $\gamma_p$ and $I_p$ are crossing symmetric, the $p$ bubble only acts on crossing-symmetric vertices. Hence, the factor of $1/2$ can be cancelled by choosing only site-resolved vertices with horizontal fermion lines. This yields
\begin{align}
\label{eq:final_p}
&
B_{p;i_1i_2}^{\Gamma,\Gamma'}(1'2';12) 
=
\frac{1}{\beta} \sum_{\nu_3,\nu_4}
G(\nu_3)G(\nu_4)
\\
\nn
& \, \times
\pig\{ 
\pig[ 
-2 \, 
\Gamma^s_{i_1i_2}(\nu_{1'}\nu_{2'};\nu_{3}\nu_{4})
\Gamma'^s_{i_1i_2}(\nu_{3}\nu_{4};\nu_{1}\nu_{2})
\\
\nn
& \quad + 
\Gamma^d_{i_1i_2}(\nu_{1'}\nu_{2'};\nu_{3}\nu_{4})
\Gamma'^s_{i_1i_2}(\nu_{3}\nu_{4};\nu_{1}\nu_{2})
\\
\nn
& \quad + 
\Gamma^s_{i_1i_2}(\nu_{1'}\nu_{2'};\nu_{3}\nu_{4})
\Gamma'^d_{i_1i_2}(\nu_{3}\nu_{4};\nu_{1}\nu_{2})
\pig]
\sigma^\mu_{\sigma_{1'}\sigma_{1}} \sigma^\mu_{\sigma_{2'}\sigma_{2}} 
\\
\nn
& \quad +
\pig[
3 \, 
\Gamma^s_{i_1i_2}(\nu_{1'}\nu_{2'};\nu_{3}\nu_{4})
\Gamma'^s_{i_1i_2}(\nu_{3}\nu_{4};\nu_{1}\nu_{2})
\\
\nn
& \quad + 
\Gamma^d_{i_1i_2}(\nu_{1'}\nu_{2'};\nu_{3}\nu_{4})
\Gamma'^d_{i_1i_2}(\nu_{3}\nu_{4};\nu_{1}\nu_{2}
\pig]
\delta_{\sigma_{1'}\sigma_{1}} \delta_{\sigma_{2'}\sigma_{2}}
\pig\}
.
\end{align}

Finally, the $t$ bubble mixes vertices with horizontal and vertical lines.
We put tilde symbols on site-resolved vertices where the fermion lines were changed [using Eqs.~\eqref{eq:channel_mapping}] from vertical to horizontal ones, 
implying a mapping between the $a$ and the $t$ channel.
The result is
\begin{align}
\label{eq:final_t}
&
B_{t;i_1i_2}^{\Gamma,\Gamma'}(1'2';12) 
=
- \frac{1}{\beta} \sum_{\nu_3,\nu_4} G(\nu_3) G(\nu_4)
\\
\nn
& \, \times 
\pig\{
\pig[
2 
\sum_{i_3} 
\Gamma^s_{i_3i_2}(\nu_{4}\nu_{2'};\nu_{3}\nu_{2})
\Gamma'^s_{i_1i_3}(\nu_{1'}\nu_{3};\nu_{1}\nu_{4} )
\\
\nn
& \quad + 
\tilde{\Gamma}^s_{i_2i_2}(\nu_{2'}\nu_{4};\nu_{3}\nu_{2})
\Gamma'^s_{i_1i_2}(\nu_{1'}\nu_{3};\nu_{1}\nu_{4})
\\
\nn
& \quad - 
\tilde{ \Gamma}^d_{i_2i_2}(\nu_{2'}\nu_{4};\nu_{3}\nu_{2})
\Gamma'^s_{i_1i_2}(\nu_{1'}\nu_{3};\nu_{1}\nu_{4})
\\
\nn
& \quad + 
\Gamma^s_{i_1i_2}(\nu_{4}\nu_{2'};\nu_{3}\nu_{2}) 
\tilde{\Gamma}'^s_{i_1i_1}(\nu_{3}\nu_{1'};\nu_{1}\nu_{4}) 
\\
\nn
& \quad -
\Gamma^s_{i_1i_2}(\nu_{4}\nu_{2'};\nu_{3}\nu_{2}) 
\tilde{\Gamma}'^d_{i_1i_1}(\nu_{3}\nu_{1'};\nu_{1}\nu_{4})
\pig]
\sigma^\mu_{\sigma_{1'}\sigma_{1}} \sigma^\mu_{\sigma_{2'}\sigma_{2}} 
\\
\nn
& \quad +
\pig[
2 
\sum_{i_3} 
\Gamma^d_{i_3i_2}(\nu_{4}\nu_{2'};\nu_{3}\nu_{2})
\Gamma'^d_{i_1i_3}(\nu_{1}'\nu_{3};\nu_{1}\nu_{4} )
\\
\nn
& \quad -
3 \,
\tilde{\Gamma}^s_{i_2i_2}(\nu_{2'}\nu_{4};\nu_{3}\nu_{2})
\Gamma'^d_{i_1i_2}(\nu_{1'}\nu_{3};\nu_{1}\nu_{4})
\\
\nn
& \quad -  
\tilde{\Gamma}'^d_{i_2i_2}(\nu_{2'}\nu_{4};\nu_{3}\nu_{2})
\Gamma^d_{i_1i_2}(\nu_{1'}\nu_{3};\nu_{1}\nu_{4})
\\
\nn
& \quad -
3 \, 
\Gamma^d_{i_1i_2}(\nu_{4}\nu_{2'};\nu_{3}\nu_{2}) 
\tilde{\Gamma}'^s_{i_1i_1}(\nu_{3}\nu_{1'};\nu_{1}\nu_{4})
\\  
\nn
& \quad - 
\Gamma^d_{i_1i_2}(\nu_{4}\nu_{2'};\nu_{3}\nu_{2}) 
\tilde{\Gamma}'^d_{i_1i_1}(\nu_{3}\nu_{1'};\nu_{1}\nu_{4})
\pig]
\delta_{\sigma_{1'}\sigma_{1}} \delta_{\sigma_{2'}\sigma_{2}}
\pig\}
.
\end{align}

As last step, we give the fully parametrized loop function.
Analogously inserting the fully parametrized vertex into Eq.~\eqref{eq:general_loops} 
for loop functions, one obtains
\begin{align*}
L(\Gamma,G)(\nu_1)
& =
- \frac{1}{\beta}\sum_{\nu_2}
G(\nu_2)
\Big[
2 \, \sum_{i_2}\Gamma^d_{i_1i_2}(\nu_1\nu_2;\nu_1\nu_2) 
\\
& \ - 
3 \, \tilde{\Gamma}^s_{i_1i_1}(\nu_2\nu_1;\nu_1\nu_2) 
- \tilde{\Gamma}^d_{i_1i_1}(\nu_2\nu_1;\nu_1\nu_2)  
\Big]
.
\end{align*}

\section{Symmetries of pseudofermion vertices in the asymptotic classification}
\label{sec:appendix_symmetries}

In order to derive symmetry relations for the different diagrammatic classes in all three channels, we apply combinations of the following symmetry operations: 
\begin{enumerate}[label=(\alph*)]
\item exchange of both incoming and both outgoing legs,
\item complex conjugation,
\item time-reversal transformation,
\item local particle-hole transformation.
\end{enumerate}
Operations (a) and (b) can always be used to derive identities for all two-particle quantities such that only a reduced frequency range needs to be computed explicitly. Operation (c), however, is only useful if the system is time-reversal invariant. 
In diagrammatic language, complex conjugation has the effect of switching the direction of the arrows on all fermionic lines and additionally switching the sign of each frequency. 
A time-reversal transformation also switches the arrows but does not change the signs. Exchanging both incoming and outgoing legs amounts to mirroring the diagram at the diagonal. 
Finally, operation (d) is useful if the system 
is invariant under a local particle-hole transformation,
i.e.\ a particle-hole transformation at any arbitrary site,
as is the case for the Heisenberg model in the pseudofermion representation.
Below, we use the index $\eta \in\{s,d\}$ to label the spin and the density part of a vertex class as well as the variable
\begin{equation*}
\xi_\eta = 
\begin{cases}
+1 \quad \mathrm{if} \quad \eta = s \, , 
\\ 
-1 \quad \mathrm{if} \quad \eta = d \, . 
\end{cases}
\end{equation*}

In the pseudofermion system, the bare propagator is purely local and particle-hole symmetric,
$G_0(\nu)^*=G_0(-\nu)=-G_0(\nu)$.
Hence, it is purely imaginary.
By contrast, the bare vertex is real.
In a diagrammatic expansion,
a self-energy diagram with $n$ bare vertices contains $2n-1$ propagators,
and a vertex diagram with $n$ bare vertices contains $2n-2$ propagators.
Hence, the self-energy is purely imaginary while any vertex is real.

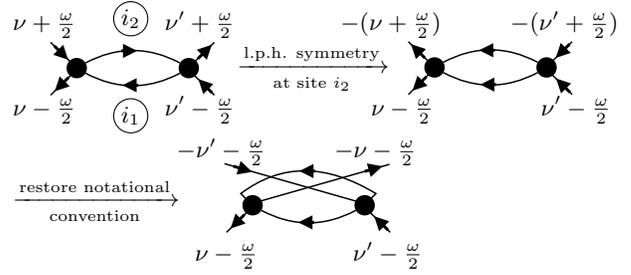
\begin{figure}
\begin{overpic}[width=0.45\textwidth]{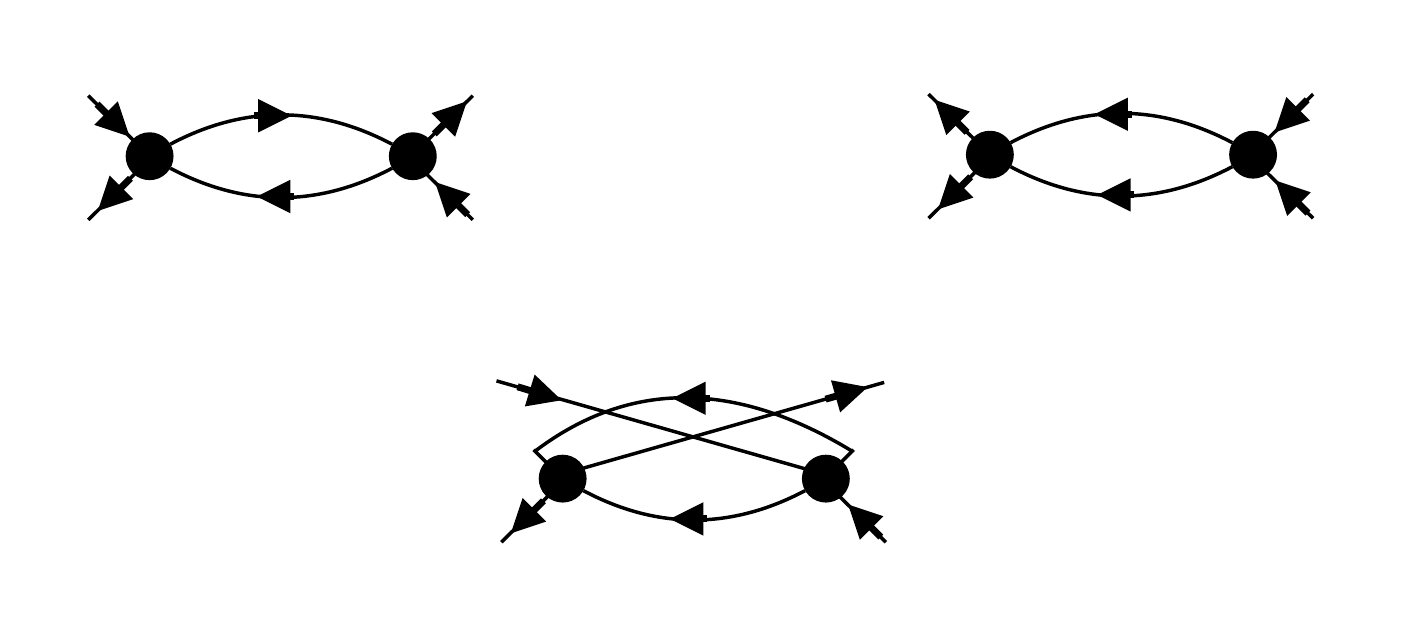}
\put(16.5,22){\small\mycirc{$i_1$}}
\put(16.5,38){\small\mycirc{$i_2$}}
\put(25,25){\small$\nu'-\tfrac{\omega}{2}$}
\put(25,39){\small$\nu'+\tfrac{\omega}{2}$}
\put(0,25){\small$\nu-\tfrac{\omega}{2}$}
\put(0,39){\small$\nu+\tfrac{\omega}{2}$}
 
\put(37,32){$\xrightarrow[\text{at site } i_2]{\text{l.p.h. symmetry}}$}

\put(87,25){\small$\nu'-\tfrac{\omega}{2}$}
\put(82,39){\small$-(\nu'+\tfrac{\omega}{2})$}
\put(59,25){\small$\nu-\tfrac{\omega}{2}$}
\put(54,39){\small$-(\nu+\tfrac{\omega}{2})$}

\put(0,10){$\xrightarrow[\text{convention}]{\text{restore notational}}$}

\put(56,1){\small$\nu'-\tfrac{\omega}{2}$}
\put(53,18){\small$-\nu-\tfrac{\omega}{2}$}
\put(30,1){\small$\nu-\tfrac{\omega}{2}$}
\put(27,18){\small$-\nu'-\tfrac{\omega}{2}$}
\end{overpic}
\caption{Exemplary diagrammatic derivation of the relation $K^{s;\omega}_{1,a;i_1i_2} \!=\! K^{s;-\omega}_{1,p;i_1i_2} $, reflecting local particle-hole symmetry. All lower (upper) legs carry site index $i_1$ ($i_2$).}
\label{fig:p_a_demo}
\end{figure}

We briefly sketch the derivation of vertex symmetry relations following from local particle-hole symmetry. 
All other symmetry relations given thereafter can be found by the use similar arguments.
Under a local particle-hole transformation (d) 
at site $i_2$, the full vertex $\Gamma$ transforms as
\begin{align*}
\Gamma_{i_1i_2}(1',2';1,2)
& = 
\Gamma^s_{i_1i_2}(\nu_{1'},\nu_{2'};\nu_1,\nu_2)\sigma^\mu_{\sigma_{1'}\sigma_1}\sigma^\mu_{\sigma_{2'}\sigma_2}
\\
&
\ +
\Gamma^d_{i_1i_2}(\nu_{1'},\nu_{2'};\nu_1,\nu_2)\delta_{\sigma_{1'}\sigma_{1}}\delta_{\sigma_{2'}\sigma_{2}}
\\
\xrightarrow{\text{(d)}}  
- \sigma_2  \sigma_{2'} 
&
\pig[
\Gamma^s_{i_1 i_2}(\nu_{1'},-\nu_{2};\nu_1,-\nu_{2'})
\underbrace{
\sigma^\mu_{\sigma_{1'}\sigma_1}\sigma^\mu_{\bar{\sigma}_{2}\bar{\sigma}_{2'}}
}_{
- \sigma_2  \sigma_{2'}\sigma^\mu_{\sigma_{1'}\sigma_1}\sigma^\mu_{\sigma_{2'}\sigma_{2}}
}
\\
& \
+
\Gamma^d_{i_1i_2}(\nu_{1'},-\nu_{2};\nu_1,-\nu_{2'})
\underbrace{
\delta_{\sigma_{1'}\sigma_{1}}\delta_{\bar{\sigma}_{2}\bar{\sigma}_{2'}}
}_{
\delta_{\sigma_{1'}\sigma_{1}}\delta_{\sigma_{2'}\sigma_{2}}
}
\pig]
\\
& =
\Gamma^s_{i_1i_2}(\nu_{1'},-\nu_{2};\nu_1,-\nu_{2'})\sigma^\mu_{\sigma_{1'}\sigma_1}\sigma^\mu_{\sigma_{2'}\sigma_2}
\\
& \
-
\Gamma^d_{i_1i_2}(\nu_{1'},-\nu_{2};\nu_1,-\nu_{2'})\delta_{\sigma_{1'}\sigma_{1}}\delta_{\sigma_{2'}\sigma_{2}}
.
\end{align*}
Furthermore, the $a$ and $p$ channel are interchanged in this transformation, i.e., $\gamma^\eta_{a;ij}(\nu_{1'},\nu_{2'};\nu_1,\nu_{2}) = \xi_\eta \gamma^\eta_{p;ij}(\nu_{1'},-\nu_{2};\nu_1,-\nu_{2'})$.
In Fig.~\ref{fig:p_a_demo}, this is demonstrated for a diagram that belongs to the $K_1$ class.
From this, one can deduce the relations (d) for the individual channels and asymptotic vertex classes. 

\begin{figure*} 
    \centering%
    \includegraphics[scale=1]{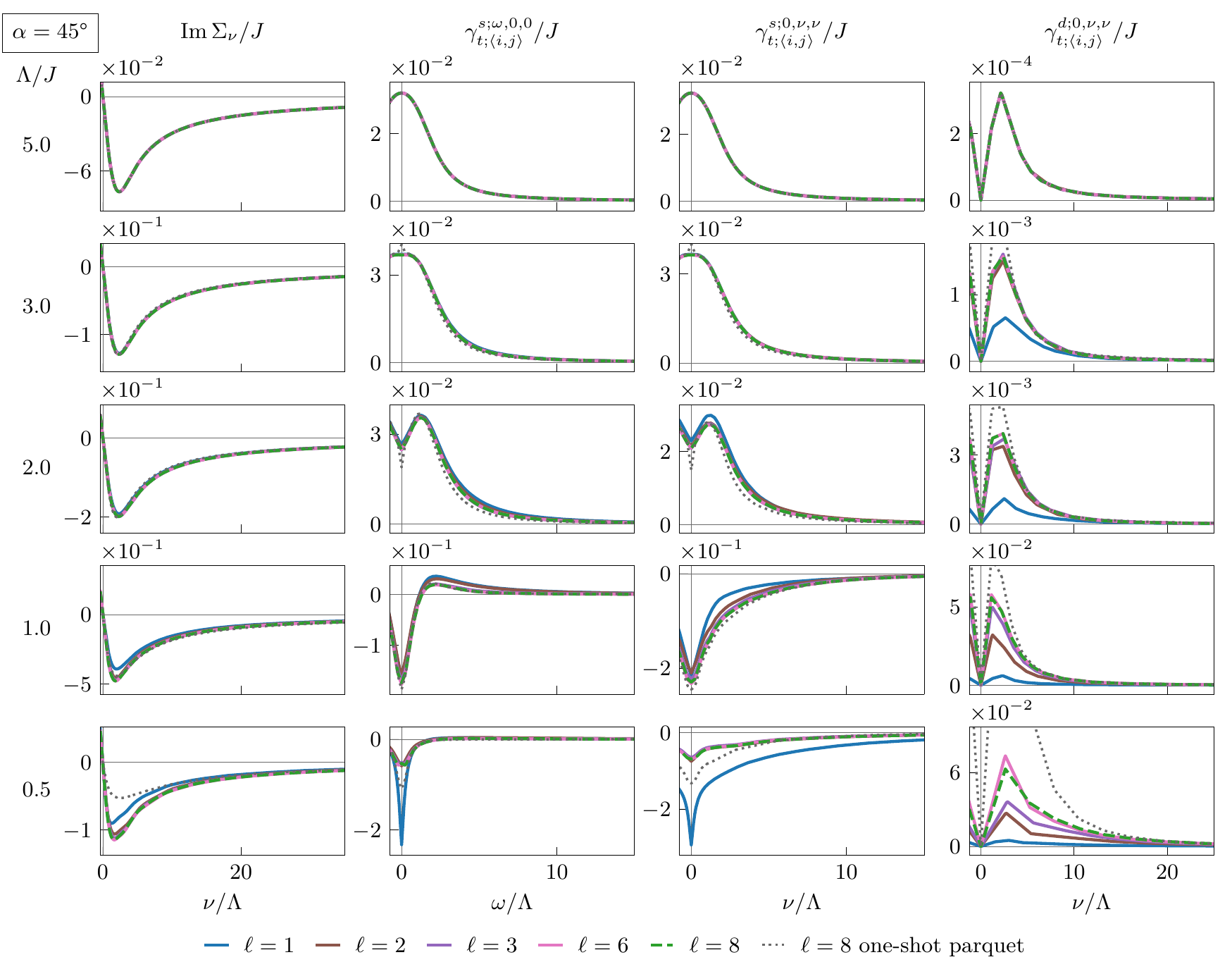}
    \caption{%
        Frequency dependence of the pseudofermion self-energy and vertex at decreasing IR cutoff $\Lambda$,
        analogous to Fig.~\ref{fig:loop_conv_and_parquet_check} 
        but at $\alpha = 45\degree$, corresponding to an ordered phase.
        Compared to the spin-liquid phase (Fig.~\ref{fig:loop_conv_and_parquet_check}),
        loop convergence appears easier to achieve here.
        For $\gamma_t^d$, one observes discrepancies of the one-loop data already at $\Lambda/J \!=\! 3$
        and slight remaining deviations between $\ell \!=\!6$ and $8$ at $\Lambda \!=\! J/2$.
        However, for all other quantities,
        one-loop results start to differ only at $\Lambda/J \!\lesssim\! 2$
        and loop convergence is achieved at $\ell \!=\! 8$ down to $\Lambda \!=\! J/2$.
    }
    \label{fig:loop_conv_and_parquet_check_45deg}
\end{figure*}

In the $K_{1}$ class of diagrams, one obtains the relations:
\begin{alignat*}{2}
K^{\eta;\omega}_{1,a;i_1 i_2} : \qquad
& \overset{\mathrm{(a)}}{=} K^{\eta;-\omega}_{1,a;i_2i_1}
, \qquad
& & \overset{\mathrm{(b)}}{=}(K^{\eta;-\omega}_{1,a;i_1i_2})^* 
, \\
& \overset{\mathrm{(c)}}{=} K^{\eta;\omega}_{1,a;i_1i_2}
, \qquad
& & \overset{\mathrm{(d)}}{=} \xi_\eta K^{\eta;-\omega}_{1,p;i_1i_2}  
;
\\
K^{\eta;\omega}_{1,p;i_1i_2} : \qquad
& \overset{\mathrm{(a)}}{=} K^{\eta;\omega}_{1,p;i_2i_1}
, \qquad
& & \overset{\mathrm{(b)}}{=}(K^{\eta;-\omega}_{1,p;i_2i_1})^*  
, \\
& \overset{\mathrm{(c)}}{=} K^{\eta;\omega}_{1,p;i_2i_1}
, \qquad
& &\overset{\mathrm{(d)}}{=} \xi_\eta K^{\eta;-\omega}_{1,a;i_1i_2} 
;
\\
K^{\eta;\omega}_{1,t;i_1i_2} : \qquad
& \overset{\mathrm{(a)}}{=} K^{\eta;-\omega}_{1,t;i_2i_1}
, \qquad
& & \overset{\mathrm{(b)}}{=}(K^{\eta;\omega}_{1,t;i_1i_2})^*
,  \\
& \overset{\mathrm{(c)}}{=} K^{\eta;-\omega}_{1,t;i_1i_2}
, \qquad
& & \overset{\mathrm{(d)}}{=} \xi_\eta K^{\eta;\omega}_{1,t;i_1i_2}  
.
\end{alignat*}
The last relation directly implies
$K^{d,\omega}_{1,t;i_1i_2}=0$.

The symmetry operations for the classes $K_2$ and $K_{2'}$ yield the relations below. Note that $K_2$ and $K_{2'}$ are interchanged by some operations. In these cases, we combine multiple operations such that we find two equivalent expressions for every object in the class $K_2$.
\begin{alignat*}{2}
K^{\eta;\omega,\nu}_{2,a;i_1i_2} : \quad
& \overset{\mathrm{(a)}}{=} K^{\eta;-\omega,\nu}_{2',a;i_2i_1}
, \quad
& &  \overset{\mathrm{(b)}}{=}(K^{\eta;-\omega,(-\nu)}_{2',a;i_1i_2})^*  
, \\
& \overset{\mathrm{(c)}}{=} K^{\eta;\omega,\nu}_{2',a;i_1i_2}
, \quad
& & \overset{\mathrm{(d)}}{=} \xi_\eta K^{\eta;-\omega,\nu}_{2,p;i_1i_2} 
, \\
& \overset{\mathrm{(a+b)}}{=} (K^{\eta;\omega,(-\nu)}_{2,a;i_2i_1})^* 
, \quad
& & \overset{\mathrm{(a+c)}}{=} K^{\eta;-\omega,\nu}_{2,a;i_2i_1}
; \\
K^{\eta;\omega,\nu}_{2,p;i_1i_2} : \quad
& \overset{\mathrm{(a)}}{=} K^{\eta;\omega,(-\nu)}_{2,p;i_2i_1}
, \quad
& & \overset{\mathrm{(b)}}{=}(K^{\eta;-\omega,\nu}_{2',p;i_2i_1})^*
, \\
& \overset{\mathrm{(c)}}{=}K^{\eta;\omega,(-\nu)}_{2',p;i_2i_1}
, \quad
& & \overset{\mathrm{(d)}}{=} \xi_\eta K^{\eta;-\omega,\nu}_{2,a;i_1i_2} 
, \\
& \overset{\mathrm{(b+c)}}{=} (K^{\eta;-\omega,(-\nu)}_{2,p;i_1i_2})^*
; & &
\\
K^{\eta;\omega,\nu}_{2,t;i_1i_2} : \quad
& \overset{\mathrm{(a)}}{=} K^{\eta;-\omega,\nu}_{2',t;i_2i_1}
, \quad
& & \overset{\mathrm{(b)}}{=}(K^{\eta;\omega,(-\nu)}_{2,t;i_1i_2})^*
, \\
& \overset{\mathrm{(c)}}{=} K^{\eta;-\omega,\nu}_{2,t;i_1i_2}
, \quad
& & \overset{\mathrm{(d)}}{=} \xi_\eta K^{\eta;\omega,-\nu}_{2,t;i_1i_2} 
.
\end{alignat*}
Since all vertices are real, the last relation implies
$K^{d,\omega,\nu}_{2,t;i_1i_2}=0$.

Finally, the symmetry relations for the $K_3$ class are
\begin{alignat*}{2}
K^{\eta;\omega,\nu,\nu'}_{3,a;i_1i_2} : \quad
& \overset{\mathrm{(a)}}{=} K^{\eta;-\omega,\nu',\nu}_{3,a;i_2i_1}
, \quad
& & \overset{\mathrm{(b)}}{=}(K^{\eta;-\omega,(-\nu'),(-\nu)}_{3,a;i_1i_2})^* 
, \\
& \overset{\mathrm{(c)}}{=} K^{\eta;\omega,\nu',\nu}_{3,a;i_1i_2} 
, \quad
& & \overset{\mathrm{(d)}}{=} \xi_\eta K^{\eta;-\omega,\nu,\nu'}_{3,p;i_1i_2}
; \\
K^{\eta;\omega,\nu,\nu'}_{3,p;i_1i_2} : \quad
& \overset{\mathrm{(a)}}{=} K^{\eta;\omega,(-\nu),(-\nu')}_{3,p;i_2i_1}
, \quad
& & \overset{\mathrm{(b)}}{=}(K^{\eta;-\omega,(-\nu'),(-\nu)}_{3,p;i_1i_2})^* 
, \\
& \overset{\mathrm{(c)}}{=} K^{\eta;\omega,\nu',\nu}_{3,p;i_1i_2}
, \quad
& & \overset{\mathrm{(d)}}{=} \xi_\eta K^{\eta;-\omega,\nu,\nu'}_{3,a;i_1i_2}
; \\
K^{\eta;\omega,\nu,\nu'}_{3,t;i_1i_2} : \quad
& \overset{\mathrm{(a)}}{=} K^{\eta;-\omega,\nu',\nu}_{3,t;i_2i_1} 
, \quad
& & \overset{\mathrm{(b)}}{=}(K^{\eta;\omega,(-\nu),(-\nu')}_{3,t;i_1i_2})^* 
, \\
& \overset{\mathrm{(c)}}{=} K^{\eta;-\omega,\nu,\nu'}_{3,t;i_1i_2}
, \quad
& & \overset{\mathrm{(d)}}{=} \xi_\eta K^{\eta;\omega,-\nu,\nu'}_{3,t;i_1i_2}
\\
& & &
= \xi_\eta  K^{\eta;\omega,\nu,-\nu'}_{3,t;i_1i_2}
.
\end{alignat*}

\section{Loop convergence at \texorpdfstring{$\alpha = 45\degree$}{alpha = 45 degrees}}
\label{app:LoopConvergenceAlpha45}

In Fig.~\ref{fig:loop_conv_and_parquet_check_45deg}, we check loop convergence in the pseudofermion self-energy and vertex at $\alpha \!=\! 45\degree$.
We find results reminiscent of those in Fig.~\ref{fig:loop_conv_and_parquet_check}:
After a flow from $\Lambda/J \!=\! 5$ to $3$, one can already observe deviations between the results at $\ell \!=\! 1$ and $\ell \!>\! 1$. 
Down to $\Lambda/J \!=\! 1$, $\ell \!=\! 6$ appears sufficient for loop convergence.
At $\Lambda/J \!=\! 0.5$, one can see minor deviations between $\ell \!=\! 6$ and $8$,
albeit only in the comparatively small density contribution.
Comparing Figs.~\ref{fig:loop_conv_and_parquet_check} and \ref{fig:loop_conv_and_parquet_check_45deg},
it seems that loop convergence is easier to achieve 
in the ordered phase at $\alpha \!=\! 45\degree$ than in the spin liquid phase at $\alpha \!=\! 0\degree$.
For instance, one can check that the sign change of $\gamma_{t;\langle i j \rangle}^{s;0,0,0}$ occurs later in the flow at $\alpha \!=\! 45\degree$ than it does at $\alpha \!=\! 0\degree$.
Moreover, the density part of $\gamma_t$ (which is the hardest to converge in loop number) is even smaller relative to the spin part at $\alpha \!=\! 45\degree$ than at $\alpha \!=\! 0\degree$.

The most striking feature of Fig.~\ref{fig:loop_conv_and_parquet_check_45deg} is the sharp and pronounced peak of $\gamma_{t;\langle i j \rangle}^{s;0,0,0}$ at $\ell \!=\! 1$. It foreshadows a divergence at finite $\Lambda$ at one-loop level. Higher loop corrections successfully dampen this peak through inter-channel feedback, thus helping to mitigate violation of the Mermin--Wagner theorem.

\bibliography{references.bib}

\begin{thebibliography}{96}%
\makeatletter
\providecommand \@ifxundefined [1]{%
 \@ifx{#1\undefined}
}%
\providecommand \@ifnum [1]{%
 \ifnum #1\expandafter \@firstoftwo
 \else \expandafter \@secondoftwo
 \fi
}%
\providecommand \@ifx [1]{%
 \ifx #1\expandafter \@firstoftwo
 \else \expandafter \@secondoftwo
 \fi
}%
\providecommand \natexlab [1]{#1}%
\providecommand \enquote  [1]{``#1''}%
\providecommand \bibnamefont  [1]{#1}%
\providecommand \bibfnamefont [1]{#1}%
\providecommand \citenamefont [1]{#1}%
\providecommand \href@noop [0]{\@secondoftwo}%
\providecommand \href [0]{\begingroup \@sanitize@url \@href}%
\providecommand \@href[1]{\@@startlink{#1}\@@href}%
\providecommand \@@href[1]{\endgroup#1\@@endlink}%
\providecommand \@sanitize@url [0]{\catcode `\\12\catcode `\$12\catcode
  `\&12\catcode `\#12\catcode `\^12\catcode `\_12\catcode `\%12\relax}%
\providecommand \@@startlink[1]{}%
\providecommand \@@endlink[0]{}%
\providecommand \url  [0]{\begingroup\@sanitize@url \@url }%
\providecommand \@url [1]{\endgroup\@href {#1}{\urlprefix }}%
\providecommand \urlprefix  [0]{URL }%
\providecommand \Eprint [0]{\href }%
\providecommand \doibase [0]{http://dx.doi.org/}%
\providecommand \selectlanguage [0]{\@gobble}%
\providecommand \bibinfo  [0]{\@secondoftwo}%
\providecommand \bibfield  [0]{\@secondoftwo}%
\providecommand \translation [1]{[#1]}%
\providecommand \BibitemOpen [0]{}%
\providecommand \bibitemStop [0]{}%
\providecommand \bibitemNoStop [0]{.\EOS\space}%
\providecommand \EOS [0]{\spacefactor3000\relax}%
\providecommand \BibitemShut  [1]{\csname bibitem#1\endcsname}%
\let\auto@bib@innerbib\@empty
\bibitem [{\citenamefont {Savary}\ and\ \citenamefont
  {Balents}(2016)}]{Savary2016}%
  \BibitemOpen
  \bibfield  {author} {\bibinfo {author} {\bibfnamefont {L.}~\bibnamefont
  {Savary}}\ and\ \bibinfo {author} {\bibfnamefont {L.}~\bibnamefont
  {Balents}},\ }\href {\doibase 10.1088/0034-4885/80/1/016502} {\bibfield
  {journal} {\bibinfo  {journal} {Rep. Prog. Phys.}\ }\textbf {\bibinfo
  {volume} {80}},\ \bibinfo {pages} {016502} (\bibinfo {year}
  {2016})}\BibitemShut {NoStop}%
\bibitem [{\citenamefont {Shores}\ \emph {et~al.}(2005)\citenamefont {Shores},
  \citenamefont {Nytko}, \citenamefont {Bartlett},\ and\ \citenamefont
  {Nocera}}]{Shores2005}%
  \BibitemOpen
  \bibfield  {author} {\bibinfo {author} {\bibfnamefont {M.~P.}\ \bibnamefont
  {Shores}}, \bibinfo {author} {\bibfnamefont {E.~A.}\ \bibnamefont {Nytko}},
  \bibinfo {author} {\bibfnamefont {B.~M.}\ \bibnamefont {Bartlett}}, \ and\
  \bibinfo {author} {\bibfnamefont {D.~G.}\ \bibnamefont {Nocera}},\ }\href
  {\doibase 10.1021/ja053891p} {\bibfield  {journal} {\bibinfo  {journal} {J.
  Am. Chem. Soc.}\ }\textbf {\bibinfo {volume} {127}},\ \bibinfo {pages}
  {13462} (\bibinfo {year} {2005})},\ \bibinfo {note} {pMID:
  16190686}\BibitemShut {NoStop}%
\bibitem [{\citenamefont {Helton}\ \emph {et~al.}(2007)\citenamefont {Helton},
  \citenamefont {Matan}, \citenamefont {Shores}, \citenamefont {Nytko},
  \citenamefont {Bartlett}, \citenamefont {Yoshida}, \citenamefont {Takano},
  \citenamefont {Suslov}, \citenamefont {Qiu}, \citenamefont {Chung},
  \citenamefont {Nocera},\ and\ \citenamefont {Lee}}]{Helton2007}%
  \BibitemOpen
  \bibfield  {author} {\bibinfo {author} {\bibfnamefont {J.~S.}\ \bibnamefont
  {Helton}}, \bibinfo {author} {\bibfnamefont {K.}~\bibnamefont {Matan}},
  \bibinfo {author} {\bibfnamefont {M.~P.}\ \bibnamefont {Shores}}, \bibinfo
  {author} {\bibfnamefont {E.~A.}\ \bibnamefont {Nytko}}, \bibinfo {author}
  {\bibfnamefont {B.~M.}\ \bibnamefont {Bartlett}}, \bibinfo {author}
  {\bibfnamefont {Y.}~\bibnamefont {Yoshida}}, \bibinfo {author} {\bibfnamefont
  {Y.}~\bibnamefont {Takano}}, \bibinfo {author} {\bibfnamefont
  {A.}~\bibnamefont {Suslov}}, \bibinfo {author} {\bibfnamefont
  {Y.}~\bibnamefont {Qiu}}, \bibinfo {author} {\bibfnamefont {J.-H.}\
  \bibnamefont {Chung}}, \bibinfo {author} {\bibfnamefont {D.~G.}\ \bibnamefont
  {Nocera}}, \ and\ \bibinfo {author} {\bibfnamefont {Y.~S.}\ \bibnamefont
  {Lee}},\ }\href {\doibase 10.1103/PhysRevLett.98.107204} {\bibfield
  {journal} {\bibinfo  {journal} {Phys. Rev. Lett.}\ }\textbf {\bibinfo
  {volume} {98}},\ \bibinfo {pages} {107204} (\bibinfo {year}
  {2007})}\BibitemShut {NoStop}%
\bibitem [{\citenamefont {de~Vries}\ \emph {et~al.}(2009)\citenamefont
  {de~Vries}, \citenamefont {Stewart}, \citenamefont {Deen}, \citenamefont
  {Piatek}, \citenamefont {Nilsen}, \citenamefont {R\o{}nnow},\ and\
  \citenamefont {Harrison}}]{deVries2009}%
  \BibitemOpen
  \bibfield  {author} {\bibinfo {author} {\bibfnamefont {M.~A.}\ \bibnamefont
  {de~Vries}}, \bibinfo {author} {\bibfnamefont {J.~R.}\ \bibnamefont
  {Stewart}}, \bibinfo {author} {\bibfnamefont {P.~P.}\ \bibnamefont {Deen}},
  \bibinfo {author} {\bibfnamefont {J.~O.}\ \bibnamefont {Piatek}}, \bibinfo
  {author} {\bibfnamefont {G.~J.}\ \bibnamefont {Nilsen}}, \bibinfo {author}
  {\bibfnamefont {H.~M.}\ \bibnamefont {R\o{}nnow}}, \ and\ \bibinfo {author}
  {\bibfnamefont {A.}~\bibnamefont {Harrison}},\ }\href {\doibase
  10.1103/PhysRevLett.103.237201} {\bibfield  {journal} {\bibinfo  {journal}
  {Phys. Rev. Lett.}\ }\textbf {\bibinfo {volume} {103}},\ \bibinfo {pages}
  {237201} (\bibinfo {year} {2009})}\BibitemShut {NoStop}%
\bibitem [{\citenamefont {Han}\ \emph {et~al.}(2012)\citenamefont {Han},
  \citenamefont {Helton}, \citenamefont {Chu}, \citenamefont {Nocera},
  \citenamefont {Rodriguez-Rivera}, \citenamefont {Broholm},\ and\
  \citenamefont {Lee}}]{Han2012}%
  \BibitemOpen
  \bibfield  {author} {\bibinfo {author} {\bibfnamefont {T.-H.}\ \bibnamefont
  {Han}}, \bibinfo {author} {\bibfnamefont {J.~S.}\ \bibnamefont {Helton}},
  \bibinfo {author} {\bibfnamefont {S.}~\bibnamefont {Chu}}, \bibinfo {author}
  {\bibfnamefont {D.~G.}\ \bibnamefont {Nocera}}, \bibinfo {author}
  {\bibfnamefont {J.~A.}\ \bibnamefont {Rodriguez-Rivera}}, \bibinfo {author}
  {\bibfnamefont {C.}~\bibnamefont {Broholm}}, \ and\ \bibinfo {author}
  {\bibfnamefont {Y.~S.}\ \bibnamefont {Lee}},\ }\href {\doibase
  10.1038/nature11659} {\bibfield  {journal} {\bibinfo  {journal} {Nature}\
  }\textbf {\bibinfo {volume} {492}},\ \bibinfo {pages} {406} (\bibinfo {year}
  {2012})}\BibitemShut {NoStop}%
\bibitem [{\citenamefont {Yan}\ \emph {et~al.}(2011)\citenamefont {Yan},
  \citenamefont {Huse},\ and\ \citenamefont {White}}]{Yan2011}%
  \BibitemOpen
  \bibfield  {author} {\bibinfo {author} {\bibfnamefont {S.}~\bibnamefont
  {Yan}}, \bibinfo {author} {\bibfnamefont {D.~A.}\ \bibnamefont {Huse}}, \
  and\ \bibinfo {author} {\bibfnamefont {S.~R.}\ \bibnamefont {White}},\ }\href
  {http://science.sciencemag.org/content/332/6034/1173} {\bibfield  {journal}
  {\bibinfo  {journal} {Science}\ }\textbf {\bibinfo {volume} {332}},\ \bibinfo
  {pages} {1173} (\bibinfo {year} {2011})}\BibitemShut {NoStop}%
\bibitem [{\citenamefont {Jiang}\ \emph {et~al.}(2012)\citenamefont {Jiang},
  \citenamefont {Wang},\ and\ \citenamefont {Balents}}]{Jiang2012}%
  \BibitemOpen
  \bibfield  {author} {\bibinfo {author} {\bibfnamefont {H.-C.}\ \bibnamefont
  {Jiang}}, \bibinfo {author} {\bibfnamefont {Z.}~\bibnamefont {Wang}}, \ and\
  \bibinfo {author} {\bibfnamefont {L.}~\bibnamefont {Balents}},\ }\href
  {https://doi.org/10.1038/nphys2465} {\bibfield  {journal} {\bibinfo
  {journal} {Nat. Phys.}\ }\textbf {\bibinfo {volume} {8}},\ \bibinfo {pages}
  {902} (\bibinfo {year} {2012})}\BibitemShut {NoStop}%
\bibitem [{\citenamefont {Depenbrock}\ \emph {et~al.}(2012)\citenamefont
  {Depenbrock}, \citenamefont {McCulloch},\ and\ \citenamefont
  {Schollw\"ock}}]{Depenbrock2012}%
  \BibitemOpen
  \bibfield  {author} {\bibinfo {author} {\bibfnamefont {S.}~\bibnamefont
  {Depenbrock}}, \bibinfo {author} {\bibfnamefont {I.~P.}\ \bibnamefont
  {McCulloch}}, \ and\ \bibinfo {author} {\bibfnamefont {U.}~\bibnamefont
  {Schollw\"ock}},\ }\href {\doibase 10.1103/PhysRevLett.109.067201} {\bibfield
   {journal} {\bibinfo  {journal} {Phys. Rev. Lett.}\ }\textbf {\bibinfo
  {volume} {109}},\ \bibinfo {pages} {067201} (\bibinfo {year}
  {2012})}\BibitemShut {NoStop}%
\bibitem [{\citenamefont {Iqbal}\ \emph {et~al.}(2014)\citenamefont {Iqbal},
  \citenamefont {Poilblanc},\ and\ \citenamefont {Becca}}]{Iqbal2014}%
  \BibitemOpen
  \bibfield  {author} {\bibinfo {author} {\bibfnamefont {Y.}~\bibnamefont
  {Iqbal}}, \bibinfo {author} {\bibfnamefont {D.}~\bibnamefont {Poilblanc}}, \
  and\ \bibinfo {author} {\bibfnamefont {F.}~\bibnamefont {Becca}},\ }\href
  {\doibase 10.1103/PhysRevB.89.020407} {\bibfield  {journal} {\bibinfo
  {journal} {Phys. Rev. B}\ }\textbf {\bibinfo {volume} {89}},\ \bibinfo
  {pages} {020407} (\bibinfo {year} {2014})}\BibitemShut {NoStop}%
\bibitem [{\citenamefont {He}\ \emph {et~al.}(2017)\citenamefont {He},
  \citenamefont {Zaletel}, \citenamefont {Oshikawa},\ and\ \citenamefont
  {Pollmann}}]{He2017a}%
  \BibitemOpen
  \bibfield  {author} {\bibinfo {author} {\bibfnamefont {Y.-C.}\ \bibnamefont
  {He}}, \bibinfo {author} {\bibfnamefont {M.~P.}\ \bibnamefont {Zaletel}},
  \bibinfo {author} {\bibfnamefont {M.}~\bibnamefont {Oshikawa}}, \ and\
  \bibinfo {author} {\bibfnamefont {F.}~\bibnamefont {Pollmann}},\ }\href
  {\doibase 10.1103/PhysRevX.7.031020} {\bibfield  {journal} {\bibinfo
  {journal} {Phys. Rev. X}\ }\textbf {\bibinfo {volume} {7}},\ \bibinfo {pages}
  {031020} (\bibinfo {year} {2017})}\BibitemShut {NoStop}%
\bibitem [{\citenamefont {Liao}\ \emph {et~al.}(2017)\citenamefont {Liao},
  \citenamefont {Xie}, \citenamefont {Chen}, \citenamefont {Liu}, \citenamefont
  {Xie}, \citenamefont {Huang}, \citenamefont {Normand},\ and\ \citenamefont
  {Xiang}}]{Liao2017}%
  \BibitemOpen
  \bibfield  {author} {\bibinfo {author} {\bibfnamefont {H.~J.}\ \bibnamefont
  {Liao}}, \bibinfo {author} {\bibfnamefont {Z.~Y.}\ \bibnamefont {Xie}},
  \bibinfo {author} {\bibfnamefont {J.}~\bibnamefont {Chen}}, \bibinfo {author}
  {\bibfnamefont {Z.~Y.}\ \bibnamefont {Liu}}, \bibinfo {author} {\bibfnamefont
  {H.~D.}\ \bibnamefont {Xie}}, \bibinfo {author} {\bibfnamefont {R.~Z.}\
  \bibnamefont {Huang}}, \bibinfo {author} {\bibfnamefont {B.}~\bibnamefont
  {Normand}}, \ and\ \bibinfo {author} {\bibfnamefont {T.}~\bibnamefont
  {Xiang}},\ }\href {\doibase 10.1103/PhysRevLett.118.137202} {\bibfield
  {journal} {\bibinfo  {journal} {Phys. Rev. Lett.}\ }\textbf {\bibinfo
  {volume} {118}},\ \bibinfo {pages} {137202} (\bibinfo {year}
  {2017})}\BibitemShut {NoStop}%
\bibitem [{\citenamefont {Gingras}\ and\ \citenamefont
  {McClarty}(2014)}]{Gingras2014}%
  \BibitemOpen
  \bibfield  {author} {\bibinfo {author} {\bibfnamefont {M.~J.~P.}\
  \bibnamefont {Gingras}}\ and\ \bibinfo {author} {\bibfnamefont {P.~A.}\
  \bibnamefont {McClarty}},\ }\href {\doibase 10.1088/0034-4885/77/5/056501}
  {\bibfield  {journal} {\bibinfo  {journal} {Rep. Prog. Phys.}\ }\textbf
  {\bibinfo {volume} {77}},\ \bibinfo {pages} {056501} (\bibinfo {year}
  {2014})}\BibitemShut {NoStop}%
\bibitem [{\citenamefont {Canals}\ and\ \citenamefont
  {Lacroix}(1998)}]{Canals1998}%
  \BibitemOpen
  \bibfield  {author} {\bibinfo {author} {\bibfnamefont {B.}~\bibnamefont
  {Canals}}\ and\ \bibinfo {author} {\bibfnamefont {C.}~\bibnamefont
  {Lacroix}},\ }\href {\doibase 10.1103/PhysRevLett.80.2933} {\bibfield
  {journal} {\bibinfo  {journal} {Phys. Rev. Lett.}\ }\textbf {\bibinfo
  {volume} {80}},\ \bibinfo {pages} {2933} (\bibinfo {year}
  {1998})}\BibitemShut {NoStop}%
\bibitem [{\citenamefont {Chandra}\ and\ \citenamefont
  {Sahoo}(2018)}]{Chandra2018}%
  \BibitemOpen
  \bibfield  {author} {\bibinfo {author} {\bibfnamefont {V.~R.}\ \bibnamefont
  {Chandra}}\ and\ \bibinfo {author} {\bibfnamefont {J.}~\bibnamefont
  {Sahoo}},\ }\href {\doibase 10.1103/PhysRevB.97.144407} {\bibfield  {journal}
  {\bibinfo  {journal} {Phys. Rev. B}\ }\textbf {\bibinfo {volume} {97}},\
  \bibinfo {pages} {144407} (\bibinfo {year} {2018})}\BibitemShut {NoStop}%
\bibitem [{\citenamefont {Yan}\ \emph {et~al.}(2017)\citenamefont {Yan},
  \citenamefont {Benton}, \citenamefont {Jaubert},\ and\ \citenamefont
  {Shannon}}]{Yan2017}%
  \BibitemOpen
  \bibfield  {author} {\bibinfo {author} {\bibfnamefont {H.}~\bibnamefont
  {Yan}}, \bibinfo {author} {\bibfnamefont {O.}~\bibnamefont {Benton}},
  \bibinfo {author} {\bibfnamefont {L.}~\bibnamefont {Jaubert}}, \ and\
  \bibinfo {author} {\bibfnamefont {N.}~\bibnamefont {Shannon}},\ }\href
  {\doibase 10.1103/PhysRevB.95.094422} {\bibfield  {journal} {\bibinfo
  {journal} {Phys. Rev. B}\ }\textbf {\bibinfo {volume} {95}},\ \bibinfo
  {pages} {094422} (\bibinfo {year} {2017})}\BibitemShut {NoStop}%
\bibitem [{\citenamefont {Polchinski}(1984)}]{Polchinski1984}%
  \BibitemOpen
  \bibfield  {author} {\bibinfo {author} {\bibfnamefont {J.}~\bibnamefont
  {Polchinski}},\ }\href {https://doi.org/10.1016/0550-3213(84)90287-6}
  {\bibfield  {journal} {\bibinfo  {journal} {Nucl. Phys. B}\ }\textbf
  {\bibinfo {volume} {231}},\ \bibinfo {pages} {269} (\bibinfo {year}
  {1984})}\BibitemShut {NoStop}%
\bibitem [{\citenamefont {Wetterich}(1993)}]{Wetterich1993}%
  \BibitemOpen
  \bibfield  {author} {\bibinfo {author} {\bibfnamefont {C.}~\bibnamefont
  {Wetterich}},\ }\href {https://doi.org/10.1016/0370-2693(93)90726-X}
  {\bibfield  {journal} {\bibinfo  {journal} {Phys. Lett. B}\ }\textbf
  {\bibinfo {volume} {301}},\ \bibinfo {pages} {90} (\bibinfo {year}
  {1993})}\BibitemShut {NoStop}%
\bibitem [{\citenamefont {Metzner}\ \emph {et~al.}(2012)\citenamefont
  {Metzner}, \citenamefont {Salmhofer}, \citenamefont {Honerkamp},
  \citenamefont {Meden},\ and\ \citenamefont {Sch\"onhammer}}]{Metzner2012}%
  \BibitemOpen
  \bibfield  {author} {\bibinfo {author} {\bibfnamefont {W.}~\bibnamefont
  {Metzner}}, \bibinfo {author} {\bibfnamefont {M.}~\bibnamefont {Salmhofer}},
  \bibinfo {author} {\bibfnamefont {C.}~\bibnamefont {Honerkamp}}, \bibinfo
  {author} {\bibfnamefont {V.}~\bibnamefont {Meden}}, \ and\ \bibinfo {author}
  {\bibfnamefont {K.}~\bibnamefont {Sch\"onhammer}},\ }\href {\doibase
  10.1103/RevModPhys.84.299} {\bibfield  {journal} {\bibinfo  {journal} {Rev.
  Mod. Phys.}\ }\textbf {\bibinfo {volume} {84}},\ \bibinfo {pages} {299}
  (\bibinfo {year} {2012})}\BibitemShut {NoStop}%
\bibitem [{\citenamefont {Platt}\ \emph {et~al.}(2013)\citenamefont {Platt},
  \citenamefont {Hanke},\ and\ \citenamefont {Thomale}}]{Platt2013}%
  \BibitemOpen
  \bibfield  {author} {\bibinfo {author} {\bibfnamefont {C.}~\bibnamefont
  {Platt}}, \bibinfo {author} {\bibfnamefont {W.}~\bibnamefont {Hanke}}, \ and\
  \bibinfo {author} {\bibfnamefont {R.}~\bibnamefont {Thomale}},\ }\href
  {\doibase 10.1080/00018732.2013.862020} {\bibfield  {journal} {\bibinfo
  {journal} {Adv. Phys.}\ }\textbf {\bibinfo {volume} {62}},\ \bibinfo {pages}
  {453} (\bibinfo {year} {2013})}\BibitemShut {NoStop}%
\bibitem [{\citenamefont {Hedden}\ \emph {et~al.}(2004)\citenamefont {Hedden},
  \citenamefont {Meden}, \citenamefont {Pruschke},\ and\ \citenamefont
  {Schönhammer}}]{Hedden2004}%
  \BibitemOpen
  \bibfield  {author} {\bibinfo {author} {\bibfnamefont {R.}~\bibnamefont
  {Hedden}}, \bibinfo {author} {\bibfnamefont {V.}~\bibnamefont {Meden}},
  \bibinfo {author} {\bibfnamefont {T.}~\bibnamefont {Pruschke}}, \ and\
  \bibinfo {author} {\bibfnamefont {K.}~\bibnamefont {Schönhammer}},\ }\href
  {\doibase 10.1088/0953-8984/16/29/019} {\bibfield  {journal} {\bibinfo
  {journal} {J. Phys.: Condens. Mat.}\ }\textbf {\bibinfo {volume} {16}},\
  \bibinfo {pages} {5279} (\bibinfo {year} {2004})}\BibitemShut {NoStop}%
\bibitem [{\citenamefont {Andergassen}\ \emph {et~al.}(2006)\citenamefont
  {Andergassen}, \citenamefont {Enss}, \citenamefont {Meden}, \citenamefont
  {Metzner}, \citenamefont {Schollw\"{o}ck},\ and\ \citenamefont
  {Sch\"{o}nhammer}}]{Andergassen2006}%
  \BibitemOpen
  \bibfield  {author} {\bibinfo {author} {\bibfnamefont {S.}~\bibnamefont
  {Andergassen}}, \bibinfo {author} {\bibfnamefont {T.}~\bibnamefont {Enss}},
  \bibinfo {author} {\bibfnamefont {V.}~\bibnamefont {Meden}}, \bibinfo
  {author} {\bibfnamefont {W.}~\bibnamefont {Metzner}}, \bibinfo {author}
  {\bibfnamefont {U.}~\bibnamefont {Schollw\"{o}ck}}, \ and\ \bibinfo {author}
  {\bibfnamefont {K.}~\bibnamefont {Sch\"{o}nhammer}},\ }\href {\doibase
  10.1103/PhysRevB.73.045125} {\bibfield  {journal} {\bibinfo  {journal} {Phys.
  Rev. B}\ }\textbf {\bibinfo {volume} {73}},\ \bibinfo {eid} {045125}
  (\bibinfo {year} {2006})}\BibitemShut {NoStop}%
\bibitem [{\citenamefont {Karrasch}\ \emph {et~al.}(2008)\citenamefont
  {Karrasch}, \citenamefont {Hedden}, \citenamefont {Peters}, \citenamefont
  {Pruschke}, \citenamefont {Sch\"onhammer},\ and\ \citenamefont
  {Meden}}]{Karrasch2008a}%
  \BibitemOpen
  \bibfield  {author} {\bibinfo {author} {\bibfnamefont {C.}~\bibnamefont
  {Karrasch}}, \bibinfo {author} {\bibfnamefont {R.}~\bibnamefont {Hedden}},
  \bibinfo {author} {\bibfnamefont {R.}~\bibnamefont {Peters}}, \bibinfo
  {author} {\bibfnamefont {T.}~\bibnamefont {Pruschke}}, \bibinfo {author}
  {\bibfnamefont {K.}~\bibnamefont {Sch\"onhammer}}, \ and\ \bibinfo {author}
  {\bibfnamefont {V.}~\bibnamefont {Meden}},\ }\href {\doibase doi:
  10.1088/0953-8984/20/34/345205} {\bibfield  {journal} {\bibinfo  {journal}
  {J. Phys.: Condens. Mat.}\ }\textbf {\bibinfo {volume} {20}},\ \bibinfo
  {pages} {345205} (\bibinfo {year} {2008})}\BibitemShut {NoStop}%
\bibitem [{\citenamefont {Uebelacker}\ and\ \citenamefont
  {Honerkamp}(2012)}]{Uebelacker2012}%
  \BibitemOpen
  \bibfield  {author} {\bibinfo {author} {\bibfnamefont {S.}~\bibnamefont
  {Uebelacker}}\ and\ \bibinfo {author} {\bibfnamefont {C.}~\bibnamefont
  {Honerkamp}},\ }\href {\doibase 10.1103/PhysRevB.85.155122} {\bibfield
  {journal} {\bibinfo  {journal} {Phys. Rev. B}\ }\textbf {\bibinfo {volume}
  {85}},\ \bibinfo {pages} {155122} (\bibinfo {year} {2012})}\BibitemShut
  {NoStop}%
\bibitem [{\citenamefont {Scherer}\ \emph {et~al.}(2012)\citenamefont
  {Scherer}, \citenamefont {Uebelacker}, \citenamefont {Scherer},\ and\
  \citenamefont {Honerkamp}}]{Scherer2012a}%
  \BibitemOpen
  \bibfield  {author} {\bibinfo {author} {\bibfnamefont {M.~M.}\ \bibnamefont
  {Scherer}}, \bibinfo {author} {\bibfnamefont {S.}~\bibnamefont {Uebelacker}},
  \bibinfo {author} {\bibfnamefont {D.~D.}\ \bibnamefont {Scherer}}, \ and\
  \bibinfo {author} {\bibfnamefont {C.}~\bibnamefont {Honerkamp}},\ }\href
  {\doibase 10.1103/PhysRevB.86.155415} {\bibfield  {journal} {\bibinfo
  {journal} {Phys. Rev. B}\ }\textbf {\bibinfo {volume} {86}},\ \bibinfo
  {pages} {155415} (\bibinfo {year} {2012})}\BibitemShut {NoStop}%
\bibitem [{\citenamefont {Bauer}\ \emph {et~al.}(2013)\citenamefont {Bauer},
  \citenamefont {Heyder}, \citenamefont {Schubert}, \citenamefont {Borowsky},
  \citenamefont {Taubert}, \citenamefont {Bruognolo}, \citenamefont {Schuh},
  \citenamefont {Wegscheider}, \citenamefont {von Delft},\ and\ \citenamefont
  {Ludwig}}]{Bauer2013}%
  \BibitemOpen
  \bibfield  {author} {\bibinfo {author} {\bibfnamefont {F.}~\bibnamefont
  {Bauer}}, \bibinfo {author} {\bibfnamefont {J.}~\bibnamefont {Heyder}},
  \bibinfo {author} {\bibfnamefont {E.}~\bibnamefont {Schubert}}, \bibinfo
  {author} {\bibfnamefont {D.}~\bibnamefont {Borowsky}}, \bibinfo {author}
  {\bibfnamefont {D.}~\bibnamefont {Taubert}}, \bibinfo {author} {\bibfnamefont
  {B.}~\bibnamefont {Bruognolo}}, \bibinfo {author} {\bibfnamefont
  {D.}~\bibnamefont {Schuh}}, \bibinfo {author} {\bibfnamefont
  {W.}~\bibnamefont {Wegscheider}}, \bibinfo {author} {\bibfnamefont
  {J.}~\bibnamefont {von Delft}}, \ and\ \bibinfo {author} {\bibfnamefont
  {S.}~\bibnamefont {Ludwig}},\ }\href {\doibase 10.1038/nature12421}
  {\bibfield  {journal} {\bibinfo  {journal} {Nature}\ }\textbf {\bibinfo
  {volume} {501}},\ \bibinfo {pages} {73} (\bibinfo {year} {2013})}\BibitemShut
  {NoStop}%
\bibitem [{\citenamefont {Eberlein}\ and\ \citenamefont
  {Metzner}(2014)}]{Eberlein2014a}%
  \BibitemOpen
  \bibfield  {author} {\bibinfo {author} {\bibfnamefont {A.}~\bibnamefont
  {Eberlein}}\ and\ \bibinfo {author} {\bibfnamefont {W.}~\bibnamefont
  {Metzner}},\ }\href {\doibase 10.1103/PhysRevB.89.035126} {\bibfield
  {journal} {\bibinfo  {journal} {Phys. Rev. B}\ }\textbf {\bibinfo {volume}
  {89}},\ \bibinfo {pages} {035126} (\bibinfo {year} {2014})}\BibitemShut
  {NoStop}%
\bibitem [{\citenamefont {Bauer}\ \emph {et~al.}(2014)\citenamefont {Bauer},
  \citenamefont {Heyder},\ and\ \citenamefont {von Delft}}]{Bauer2014}%
  \BibitemOpen
  \bibfield  {author} {\bibinfo {author} {\bibfnamefont {F.}~\bibnamefont
  {Bauer}}, \bibinfo {author} {\bibfnamefont {J.}~\bibnamefont {Heyder}}, \
  and\ \bibinfo {author} {\bibfnamefont {J.}~\bibnamefont {von Delft}},\ }\href
  {\doibase 10.1103/PhysRevB.89.045128} {\bibfield  {journal} {\bibinfo
  {journal} {Phys. Rev. B}\ }\textbf {\bibinfo {volume} {89}},\ \bibinfo
  {pages} {045128} (\bibinfo {year} {2014})}\BibitemShut {NoStop}%
\bibitem [{\citenamefont {Schubert}\ \emph {et~al.}(2014)\citenamefont
  {Schubert}, \citenamefont {Heyder}, \citenamefont {Bauer}, \citenamefont
  {Waschneck}, \citenamefont {Stumpf}, \citenamefont {Wegscheider},
  \citenamefont {von Delft}, \citenamefont {Ludwig},\ and\ \citenamefont
  {H\"ogele}}]{Schubert2014}%
  \BibitemOpen
  \bibfield  {author} {\bibinfo {author} {\bibfnamefont {E.}~\bibnamefont
  {Schubert}}, \bibinfo {author} {\bibfnamefont {J.}~\bibnamefont {Heyder}},
  \bibinfo {author} {\bibfnamefont {F.}~\bibnamefont {Bauer}}, \bibinfo
  {author} {\bibfnamefont {B.}~\bibnamefont {Waschneck}}, \bibinfo {author}
  {\bibfnamefont {W.}~\bibnamefont {Stumpf}}, \bibinfo {author} {\bibfnamefont
  {W.}~\bibnamefont {Wegscheider}}, \bibinfo {author} {\bibfnamefont
  {J.}~\bibnamefont {von Delft}}, \bibinfo {author} {\bibfnamefont
  {S.}~\bibnamefont {Ludwig}}, \ and\ \bibinfo {author} {\bibfnamefont
  {A.}~\bibnamefont {H\"ogele}},\ }\href {\doibase DOI 10.1002/pssb.201350218}
  {\bibfield  {journal} {\bibinfo  {journal} {Phys. Status Solidi}\ }\textbf
  {\bibinfo {volume} {B 251}},\ \bibinfo {pages} {1931} (\bibinfo {year}
  {2014})}\BibitemShut {NoStop}%
\bibitem [{\citenamefont {Rentrop}\ \emph {et~al.}(2016)\citenamefont
  {Rentrop}, \citenamefont {Meden},\ and\ \citenamefont
  {Jakobs}}]{Rentrop2016}%
  \BibitemOpen
  \bibfield  {author} {\bibinfo {author} {\bibfnamefont {J.~F.}\ \bibnamefont
  {Rentrop}}, \bibinfo {author} {\bibfnamefont {V.}~\bibnamefont {Meden}}, \
  and\ \bibinfo {author} {\bibfnamefont {S.~G.}\ \bibnamefont {Jakobs}},\
  }\href {\doibase 10.1103/PhysRevB.93.195160} {\bibfield  {journal} {\bibinfo
  {journal} {Phys. Rev. B}\ }\textbf {\bibinfo {volume} {93}},\ \bibinfo
  {pages} {195160} (\bibinfo {year} {2016})}\BibitemShut {NoStop}%
\bibitem [{\citenamefont {Heyder}\ \emph {et~al.}(2015)\citenamefont {Heyder},
  \citenamefont {Bauer}, \citenamefont {Schubert}, \citenamefont {Borowsky},
  \citenamefont {Schuh}, \citenamefont {Wegscheider}, \citenamefont {von
  Delft},\ and\ \citenamefont {Ludwig}}]{Heyder2015}%
  \BibitemOpen
  \bibfield  {author} {\bibinfo {author} {\bibfnamefont {J.}~\bibnamefont
  {Heyder}}, \bibinfo {author} {\bibfnamefont {F.}~\bibnamefont {Bauer}},
  \bibinfo {author} {\bibfnamefont {E.}~\bibnamefont {Schubert}}, \bibinfo
  {author} {\bibfnamefont {D.}~\bibnamefont {Borowsky}}, \bibinfo {author}
  {\bibfnamefont {D.}~\bibnamefont {Schuh}}, \bibinfo {author} {\bibfnamefont
  {W.}~\bibnamefont {Wegscheider}}, \bibinfo {author} {\bibfnamefont
  {J.}~\bibnamefont {von Delft}}, \ and\ \bibinfo {author} {\bibfnamefont
  {S.}~\bibnamefont {Ludwig}},\ }\href {\doibase 10.1103/PhysRevB.92.195401}
  {\bibfield  {journal} {\bibinfo  {journal} {Phys. Rev. B}\ }\textbf {\bibinfo
  {volume} {92}},\ \bibinfo {pages} {195401} (\bibinfo {year}
  {2015})}\BibitemShut {NoStop}%
\bibitem [{\citenamefont {Eberlein}\ \emph {et~al.}(2016)\citenamefont
  {Eberlein}, \citenamefont {Metzner}, \citenamefont {Sachdev},\ and\
  \citenamefont {Yamase}}]{Eberlein2016}%
  \BibitemOpen
  \bibfield  {author} {\bibinfo {author} {\bibfnamefont {A.}~\bibnamefont
  {Eberlein}}, \bibinfo {author} {\bibfnamefont {W.}~\bibnamefont {Metzner}},
  \bibinfo {author} {\bibfnamefont {S.}~\bibnamefont {Sachdev}}, \ and\
  \bibinfo {author} {\bibfnamefont {H.}~\bibnamefont {Yamase}},\ }\href
  {\doibase 10.1103/PhysRevLett.117.187001} {\bibfield  {journal} {\bibinfo
  {journal} {Phys. Rev. Lett.}\ }\textbf {\bibinfo {volume} {117}},\ \bibinfo
  {pages} {187001} (\bibinfo {year} {2016})}\BibitemShut {NoStop}%
\bibitem [{\citenamefont {Wentzell}\ \emph {et~al.}(2020)\citenamefont
  {Wentzell}, \citenamefont {Li}, \citenamefont {Tagliavini}, \citenamefont
  {Taranto}, \citenamefont {Rohringer}, \citenamefont {Held}, \citenamefont
  {Toschi},\ and\ \citenamefont {Andergassen}}]{Wentzell2016}%
  \BibitemOpen
  \bibfield  {author} {\bibinfo {author} {\bibfnamefont {N.}~\bibnamefont
  {Wentzell}}, \bibinfo {author} {\bibfnamefont {G.}~\bibnamefont {Li}},
  \bibinfo {author} {\bibfnamefont {A.}~\bibnamefont {Tagliavini}}, \bibinfo
  {author} {\bibfnamefont {C.}~\bibnamefont {Taranto}}, \bibinfo {author}
  {\bibfnamefont {G.}~\bibnamefont {Rohringer}}, \bibinfo {author}
  {\bibfnamefont {K.}~\bibnamefont {Held}}, \bibinfo {author} {\bibfnamefont
  {A.}~\bibnamefont {Toschi}}, \ and\ \bibinfo {author} {\bibfnamefont
  {S.}~\bibnamefont {Andergassen}},\ }\href {\doibase
  10.1103/PhysRevB.102.085106} {\bibfield  {journal} {\bibinfo  {journal}
  {Phys. Rev. B}\ }\textbf {\bibinfo {volume} {102}},\ \bibinfo {pages}
  {085106} (\bibinfo {year} {2020})}\BibitemShut {NoStop}%
\bibitem [{\citenamefont {Weidinger}\ \emph {et~al.}(2017)\citenamefont
  {Weidinger}, \citenamefont {Bauer},\ and\ \citenamefont {von
  Delft}}]{Weidinger2017}%
  \BibitemOpen
  \bibfield  {author} {\bibinfo {author} {\bibfnamefont {L.}~\bibnamefont
  {Weidinger}}, \bibinfo {author} {\bibfnamefont {F.}~\bibnamefont {Bauer}}, \
  and\ \bibinfo {author} {\bibfnamefont {J.}~\bibnamefont {von Delft}},\ }\href
  {\doibase 10.1103/PhysRevB.95.035122} {\bibfield  {journal} {\bibinfo
  {journal} {Phys. Rev. B}\ }\textbf {\bibinfo {volume} {95}},\ \bibinfo
  {pages} {035122} (\bibinfo {year} {2017})}\BibitemShut {NoStop}%
\bibitem [{\citenamefont {Schimmel}\ \emph {et~al.}(2017)\citenamefont
  {Schimmel}, \citenamefont {Bruognolo},\ and\ \citenamefont {von
  Delft}}]{Schimmel2017}%
  \BibitemOpen
  \bibfield  {author} {\bibinfo {author} {\bibfnamefont {D.~H.}\ \bibnamefont
  {Schimmel}}, \bibinfo {author} {\bibfnamefont {B.}~\bibnamefont {Bruognolo}},
  \ and\ \bibinfo {author} {\bibfnamefont {J.}~\bibnamefont {von Delft}},\
  }\href {\doibase 10.1103/PhysRevLett.119.196401} {\bibfield  {journal}
  {\bibinfo  {journal} {Phys. Rev. Lett.}\ }\textbf {\bibinfo {volume} {119}},\
  \bibinfo {pages} {196401} (\bibinfo {year} {2017})}\BibitemShut {NoStop}%
\bibitem [{\citenamefont {Sbierski}\ and\ \citenamefont
  {Karrasch}(2017)}]{Sbierski2017}%
  \BibitemOpen
  \bibfield  {author} {\bibinfo {author} {\bibfnamefont {B.}~\bibnamefont
  {Sbierski}}\ and\ \bibinfo {author} {\bibfnamefont {C.}~\bibnamefont
  {Karrasch}},\ }\href {\doibase 10.1103/PhysRevB.96.235122} {\bibfield
  {journal} {\bibinfo  {journal} {Phys. Rev. B}\ }\textbf {\bibinfo {volume}
  {96}},\ \bibinfo {pages} {235122} (\bibinfo {year} {2017})}\BibitemShut
  {NoStop}%
\bibitem [{\citenamefont {Vilardi}\ \emph {et~al.}(2017)\citenamefont
  {Vilardi}, \citenamefont {Taranto},\ and\ \citenamefont
  {Metzner}}]{Vilardi2017}%
  \BibitemOpen
  \bibfield  {author} {\bibinfo {author} {\bibfnamefont {D.}~\bibnamefont
  {Vilardi}}, \bibinfo {author} {\bibfnamefont {C.}~\bibnamefont {Taranto}}, \
  and\ \bibinfo {author} {\bibfnamefont {W.}~\bibnamefont {Metzner}},\ }\href
  {\doibase 10.1103/PhysRevB.96.235110} {\bibfield  {journal} {\bibinfo
  {journal} {Phys. Rev. B}\ }\textbf {\bibinfo {volume} {96}},\ \bibinfo
  {pages} {235110} (\bibinfo {year} {2017})}\BibitemShut {NoStop}%
\bibitem [{\citenamefont {Weidinger}\ \emph {et~al.}(2018)\citenamefont
  {Weidinger}, \citenamefont {Schmauder}, \citenamefont {Schimmel},\ and\
  \citenamefont {von Delft}}]{Weidinger2018}%
  \BibitemOpen
  \bibfield  {author} {\bibinfo {author} {\bibfnamefont {L.}~\bibnamefont
  {Weidinger}}, \bibinfo {author} {\bibfnamefont {C.}~\bibnamefont
  {Schmauder}}, \bibinfo {author} {\bibfnamefont {D.~H.}\ \bibnamefont
  {Schimmel}}, \ and\ \bibinfo {author} {\bibfnamefont {J.}~\bibnamefont {von
  Delft}},\ }\href {\doibase 10.1103/PhysRevB.98.115112} {\bibfield  {journal}
  {\bibinfo  {journal} {Phys. Rev. B}\ }\textbf {\bibinfo {volume} {98}},\
  \bibinfo {pages} {115112} (\bibinfo {year} {2018})}\BibitemShut {NoStop}%
\bibitem [{\citenamefont {Weidinger}\ and\ \citenamefont {von
  Delft}()}]{Weidinger2019}%
  \BibitemOpen
  \bibfield  {author} {\bibinfo {author} {\bibfnamefont {L.}~\bibnamefont
  {Weidinger}}\ and\ \bibinfo {author} {\bibfnamefont {J.}~\bibnamefont {von
  Delft}},\ }\href {http://arxiv.org/abs/1912.02700} {\ }\Eprint
  {http://arxiv.org/abs/1912.02700} {arXiv:1912.02700} \BibitemShut {NoStop}%
\bibitem [{\citenamefont {Hille}\ \emph {et~al.}(2020)\citenamefont {Hille},
  \citenamefont {Kugler}, \citenamefont {Eckhardt}, \citenamefont {He},
  \citenamefont {Kauch}, \citenamefont {Honerkamp}, \citenamefont {Toschi},\
  and\ \citenamefont {Andergassen}}]{Hille2020}%
  \BibitemOpen
  \bibfield  {author} {\bibinfo {author} {\bibfnamefont {C.}~\bibnamefont
  {Hille}}, \bibinfo {author} {\bibfnamefont {F.~B.}\ \bibnamefont {Kugler}},
  \bibinfo {author} {\bibfnamefont {C.~J.}\ \bibnamefont {Eckhardt}}, \bibinfo
  {author} {\bibfnamefont {Y.-Y.}\ \bibnamefont {He}}, \bibinfo {author}
  {\bibfnamefont {A.}~\bibnamefont {Kauch}}, \bibinfo {author} {\bibfnamefont
  {C.}~\bibnamefont {Honerkamp}}, \bibinfo {author} {\bibfnamefont
  {A.}~\bibnamefont {Toschi}}, \ and\ \bibinfo {author} {\bibfnamefont
  {S.}~\bibnamefont {Andergassen}},\ }\href {\doibase
  10.1103/PhysRevResearch.2.033372} {\bibfield  {journal} {\bibinfo  {journal}
  {Phys. Rev. Research}\ }\textbf {\bibinfo {volume} {2}},\  (\bibinfo {year}
  {2020})}\BibitemShut {NoStop}%
\bibitem [{\citenamefont {Ehrlich}\ and\ \citenamefont
  {Honerkamp}()}]{Ehrlich2020}%
  \BibitemOpen
  \bibfield  {author} {\bibinfo {author} {\bibfnamefont {J.}~\bibnamefont
  {Ehrlich}}\ and\ \bibinfo {author} {\bibfnamefont {C.}~\bibnamefont
  {Honerkamp}},\ }\href {http://arxiv.org/abs/2004.14711} {\ }\Eprint
  {http://arxiv.org/abs/2004.14711} {arXiv:2004.14711} \BibitemShut {NoStop}%
\bibitem [{\citenamefont {Reuther}\ and\ \citenamefont
  {W\"olfle}(2010)}]{Reuther2010}%
  \BibitemOpen
  \bibfield  {author} {\bibinfo {author} {\bibfnamefont {J.}~\bibnamefont
  {Reuther}}\ and\ \bibinfo {author} {\bibfnamefont {P.}~\bibnamefont
  {W\"olfle}},\ }\href {\doibase 10.1103/PhysRevB.81.144410} {\bibfield
  {journal} {\bibinfo  {journal} {Phys. Rev. B}\ }\textbf {\bibinfo {volume}
  {81}},\ \bibinfo {pages} {144410} (\bibinfo {year} {2010})}\BibitemShut
  {NoStop}%
\bibitem [{\citenamefont {Reuther}\ \emph
  {et~al.}(2011{\natexlab{a}})\citenamefont {Reuther}, \citenamefont
  {Thomale},\ and\ \citenamefont {Trebst}}]{Reuther2011}%
  \BibitemOpen
  \bibfield  {author} {\bibinfo {author} {\bibfnamefont {J.}~\bibnamefont
  {Reuther}}, \bibinfo {author} {\bibfnamefont {R.}~\bibnamefont {Thomale}}, \
  and\ \bibinfo {author} {\bibfnamefont {S.}~\bibnamefont {Trebst}},\ }\href
  {\doibase 10.1103/PhysRevB.84.100406} {\bibfield  {journal} {\bibinfo
  {journal} {Phys. Rev. B}\ }\textbf {\bibinfo {volume} {84}},\ \bibinfo
  {pages} {100406} (\bibinfo {year} {2011}{\natexlab{a}})}\BibitemShut
  {NoStop}%
\bibitem [{\citenamefont {Reuther}\ and\ \citenamefont
  {Thomale}(2011)}]{Reuther2011a}%
  \BibitemOpen
  \bibfield  {author} {\bibinfo {author} {\bibfnamefont {J.}~\bibnamefont
  {Reuther}}\ and\ \bibinfo {author} {\bibfnamefont {R.}~\bibnamefont
  {Thomale}},\ }\href {\doibase 10.1103/PhysRevB.83.024402} {\bibfield
  {journal} {\bibinfo  {journal} {Phys. Rev. B}\ }\textbf {\bibinfo {volume}
  {83}},\ \bibinfo {pages} {024402} (\bibinfo {year} {2011})}\BibitemShut
  {NoStop}%
\bibitem [{\citenamefont {Reuther}\ \emph
  {et~al.}(2011{\natexlab{b}})\citenamefont {Reuther}, \citenamefont {Abanin},\
  and\ \citenamefont {Thomale}}]{Reuther2011b}%
  \BibitemOpen
  \bibfield  {author} {\bibinfo {author} {\bibfnamefont {J.}~\bibnamefont
  {Reuther}}, \bibinfo {author} {\bibfnamefont {D.~A.}\ \bibnamefont {Abanin}},
  \ and\ \bibinfo {author} {\bibfnamefont {R.}~\bibnamefont {Thomale}},\ }\href
  {\doibase 10.1103/PhysRevB.84.014417} {\bibfield  {journal} {\bibinfo
  {journal} {Phys. Rev. B}\ }\textbf {\bibinfo {volume} {84}},\ \bibinfo
  {pages} {014417} (\bibinfo {year} {2011}{\natexlab{b}})}\BibitemShut
  {NoStop}%
\bibitem [{\citenamefont {Singh}\ \emph {et~al.}(2012)\citenamefont {Singh},
  \citenamefont {Manni}, \citenamefont {Reuther}, \citenamefont {Berlijn},
  \citenamefont {Thomale}, \citenamefont {Ku}, \citenamefont {Trebst},\ and\
  \citenamefont {Gegenwart}}]{Singh2012}%
  \BibitemOpen
  \bibfield  {author} {\bibinfo {author} {\bibfnamefont {Y.}~\bibnamefont
  {Singh}}, \bibinfo {author} {\bibfnamefont {S.}~\bibnamefont {Manni}},
  \bibinfo {author} {\bibfnamefont {J.}~\bibnamefont {Reuther}}, \bibinfo
  {author} {\bibfnamefont {T.}~\bibnamefont {Berlijn}}, \bibinfo {author}
  {\bibfnamefont {R.}~\bibnamefont {Thomale}}, \bibinfo {author} {\bibfnamefont
  {W.}~\bibnamefont {Ku}}, \bibinfo {author} {\bibfnamefont {S.}~\bibnamefont
  {Trebst}}, \ and\ \bibinfo {author} {\bibfnamefont {P.}~\bibnamefont
  {Gegenwart}},\ }\href {\doibase 10.1103/PhysRevLett.108.127203} {\bibfield
  {journal} {\bibinfo  {journal} {Phys. Rev. Lett.}\ }\textbf {\bibinfo
  {volume} {108}},\ \bibinfo {pages} {127203} (\bibinfo {year}
  {2012})}\BibitemShut {NoStop}%
\bibitem [{\citenamefont {G\"ottel}\ \emph {et~al.}(2012)\citenamefont
  {G\"ottel}, \citenamefont {Andergassen}, \citenamefont {Honerkamp},
  \citenamefont {Schuricht},\ and\ \citenamefont {Wessel}}]{Goettel2012}%
  \BibitemOpen
  \bibfield  {author} {\bibinfo {author} {\bibfnamefont {S.}~\bibnamefont
  {G\"ottel}}, \bibinfo {author} {\bibfnamefont {S.}~\bibnamefont
  {Andergassen}}, \bibinfo {author} {\bibfnamefont {C.}~\bibnamefont
  {Honerkamp}}, \bibinfo {author} {\bibfnamefont {D.}~\bibnamefont
  {Schuricht}}, \ and\ \bibinfo {author} {\bibfnamefont {S.}~\bibnamefont
  {Wessel}},\ }\href {\doibase 10.1103/PhysRevB.85.214406} {\bibfield
  {journal} {\bibinfo  {journal} {Phys. Rev. B}\ }\textbf {\bibinfo {volume}
  {85}},\ \bibinfo {pages} {214406} (\bibinfo {year} {2012})}\BibitemShut
  {NoStop}%
\bibitem [{\citenamefont {Reuther}\ \emph {et~al.}(2014)\citenamefont
  {Reuther}, \citenamefont {Thomale},\ and\ \citenamefont
  {Rachel}}]{Reuther2014}%
  \BibitemOpen
  \bibfield  {author} {\bibinfo {author} {\bibfnamefont {J.}~\bibnamefont
  {Reuther}}, \bibinfo {author} {\bibfnamefont {R.}~\bibnamefont {Thomale}}, \
  and\ \bibinfo {author} {\bibfnamefont {S.}~\bibnamefont {Rachel}},\ }\href
  {\doibase 10.1103/PhysRevB.90.100405} {\bibfield  {journal} {\bibinfo
  {journal} {Phys. Rev. B}\ }\textbf {\bibinfo {volume} {90}},\ \bibinfo
  {pages} {100405} (\bibinfo {year} {2014})}\BibitemShut {NoStop}%
\bibitem [{\citenamefont {Suttner}\ \emph {et~al.}(2014)\citenamefont
  {Suttner}, \citenamefont {Platt}, \citenamefont {Reuther},\ and\
  \citenamefont {Thomale}}]{Suttner2014}%
  \BibitemOpen
  \bibfield  {author} {\bibinfo {author} {\bibfnamefont {R.}~\bibnamefont
  {Suttner}}, \bibinfo {author} {\bibfnamefont {C.}~\bibnamefont {Platt}},
  \bibinfo {author} {\bibfnamefont {J.}~\bibnamefont {Reuther}}, \ and\
  \bibinfo {author} {\bibfnamefont {R.}~\bibnamefont {Thomale}},\ }\href
  {\doibase 10.1103/PhysRevB.89.020408} {\bibfield  {journal} {\bibinfo
  {journal} {Phys. Rev. B}\ }\textbf {\bibinfo {volume} {89}},\ \bibinfo
  {pages} {020408} (\bibinfo {year} {2014})}\BibitemShut {NoStop}%
\bibitem [{\citenamefont {Rousochatzakis}\ \emph {et~al.}(2015)\citenamefont
  {Rousochatzakis}, \citenamefont {Reuther}, \citenamefont {Thomale},
  \citenamefont {Rachel},\ and\ \citenamefont {Perkins}}]{Rousochatzakis2015}%
  \BibitemOpen
  \bibfield  {author} {\bibinfo {author} {\bibfnamefont {I.}~\bibnamefont
  {Rousochatzakis}}, \bibinfo {author} {\bibfnamefont {J.}~\bibnamefont
  {Reuther}}, \bibinfo {author} {\bibfnamefont {R.}~\bibnamefont {Thomale}},
  \bibinfo {author} {\bibfnamefont {S.}~\bibnamefont {Rachel}}, \ and\ \bibinfo
  {author} {\bibfnamefont {N.~B.}\ \bibnamefont {Perkins}},\ }\href {\doibase
  10.1103/PhysRevX.5.041035} {\bibfield  {journal} {\bibinfo  {journal} {Phys.
  Rev. X}\ }\textbf {\bibinfo {volume} {5}},\ \bibinfo {pages} {041035}
  (\bibinfo {year} {2015})}\BibitemShut {NoStop}%
\bibitem [{\citenamefont {Iqbal}\ \emph {et~al.}(2015)\citenamefont {Iqbal},
  \citenamefont {Jeschke}, \citenamefont {Reuther}, \citenamefont
  {Valent\'{\i}}, \citenamefont {Mazin}, \citenamefont {Greiter},\ and\
  \citenamefont {Thomale}}]{Iqbal2015}%
  \BibitemOpen
  \bibfield  {author} {\bibinfo {author} {\bibfnamefont {Y.}~\bibnamefont
  {Iqbal}}, \bibinfo {author} {\bibfnamefont {H.~O.}\ \bibnamefont {Jeschke}},
  \bibinfo {author} {\bibfnamefont {J.}~\bibnamefont {Reuther}}, \bibinfo
  {author} {\bibfnamefont {R.}~\bibnamefont {Valent\'{\i}}}, \bibinfo {author}
  {\bibfnamefont {I.~I.}\ \bibnamefont {Mazin}}, \bibinfo {author}
  {\bibfnamefont {M.}~\bibnamefont {Greiter}}, \ and\ \bibinfo {author}
  {\bibfnamefont {R.}~\bibnamefont {Thomale}},\ }\href {\doibase
  10.1103/PhysRevB.92.220404} {\bibfield  {journal} {\bibinfo  {journal} {Phys.
  Rev. B}\ }\textbf {\bibinfo {volume} {92}},\ \bibinfo {pages} {220404}
  (\bibinfo {year} {2015})}\BibitemShut {NoStop}%
\bibitem [{\citenamefont {Iqbal}\ \emph
  {et~al.}(2016{\natexlab{a}})\citenamefont {Iqbal}, \citenamefont {Thomale},
  \citenamefont {Parisen~Toldin}, \citenamefont {Rachel},\ and\ \citenamefont
  {Reuther}}]{Iqbal2016}%
  \BibitemOpen
  \bibfield  {author} {\bibinfo {author} {\bibfnamefont {Y.}~\bibnamefont
  {Iqbal}}, \bibinfo {author} {\bibfnamefont {R.}~\bibnamefont {Thomale}},
  \bibinfo {author} {\bibfnamefont {F.}~\bibnamefont {Parisen~Toldin}},
  \bibinfo {author} {\bibfnamefont {S.}~\bibnamefont {Rachel}}, \ and\ \bibinfo
  {author} {\bibfnamefont {J.}~\bibnamefont {Reuther}},\ }\href {\doibase
  10.1103/PhysRevB.94.140408} {\bibfield  {journal} {\bibinfo  {journal} {Phys.
  Rev. B}\ }\textbf {\bibinfo {volume} {94}},\ \bibinfo {pages} {140408}
  (\bibinfo {year} {2016}{\natexlab{a}})}\BibitemShut {NoStop}%
\bibitem [{\citenamefont {Iqbal}\ \emph
  {et~al.}(2016{\natexlab{b}})\citenamefont {Iqbal}, \citenamefont {Ghosh},
  \citenamefont {Narayanan}, \citenamefont {Kumar}, \citenamefont {Reuther},\
  and\ \citenamefont {Thomale}}]{Iqbal2016a}%
  \BibitemOpen
  \bibfield  {author} {\bibinfo {author} {\bibfnamefont {Y.}~\bibnamefont
  {Iqbal}}, \bibinfo {author} {\bibfnamefont {P.}~\bibnamefont {Ghosh}},
  \bibinfo {author} {\bibfnamefont {R.}~\bibnamefont {Narayanan}}, \bibinfo
  {author} {\bibfnamefont {B.}~\bibnamefont {Kumar}}, \bibinfo {author}
  {\bibfnamefont {J.}~\bibnamefont {Reuther}}, \ and\ \bibinfo {author}
  {\bibfnamefont {R.}~\bibnamefont {Thomale}},\ }\href {\doibase
  10.1103/PhysRevB.94.224403} {\bibfield  {journal} {\bibinfo  {journal} {Phys.
  Rev. B}\ }\textbf {\bibinfo {volume} {94}},\ \bibinfo {pages} {224403}
  (\bibinfo {year} {2016}{\natexlab{b}})}\BibitemShut {NoStop}%
\bibitem [{\citenamefont {Iqbal}\ \emph
  {et~al.}(2016{\natexlab{c}})\citenamefont {Iqbal}, \citenamefont {Hu},
  \citenamefont {Thomale}, \citenamefont {Poilblanc},\ and\ \citenamefont
  {Becca}}]{Iqbal2016b}%
  \BibitemOpen
  \bibfield  {author} {\bibinfo {author} {\bibfnamefont {Y.}~\bibnamefont
  {Iqbal}}, \bibinfo {author} {\bibfnamefont {W.-J.}\ \bibnamefont {Hu}},
  \bibinfo {author} {\bibfnamefont {R.}~\bibnamefont {Thomale}}, \bibinfo
  {author} {\bibfnamefont {D.}~\bibnamefont {Poilblanc}}, \ and\ \bibinfo
  {author} {\bibfnamefont {F.}~\bibnamefont {Becca}},\ }\href {\doibase
  10.1103/PhysRevB.93.144411} {\bibfield  {journal} {\bibinfo  {journal} {Phys.
  Rev. B}\ }\textbf {\bibinfo {volume} {93}},\ \bibinfo {pages} {144411}
  (\bibinfo {year} {2016}{\natexlab{c}})}\BibitemShut {NoStop}%
\bibitem [{\citenamefont {Buessen}\ and\ \citenamefont
  {Trebst}(2016)}]{Buessen2016}%
  \BibitemOpen
  \bibfield  {author} {\bibinfo {author} {\bibfnamefont {F.~L.}\ \bibnamefont
  {Buessen}}\ and\ \bibinfo {author} {\bibfnamefont {S.}~\bibnamefont
  {Trebst}},\ }\href {\doibase 10.1103/PhysRevB.94.235138} {\bibfield
  {journal} {\bibinfo  {journal} {Phys. Rev. B}\ }\textbf {\bibinfo {volume}
  {94}},\ \bibinfo {pages} {235138} (\bibinfo {year} {2016})}\BibitemShut
  {NoStop}%
\bibitem [{\citenamefont {Hering}\ and\ \citenamefont
  {Reuther}(2017)}]{Hering2017}%
  \BibitemOpen
  \bibfield  {author} {\bibinfo {author} {\bibfnamefont {M.}~\bibnamefont
  {Hering}}\ and\ \bibinfo {author} {\bibfnamefont {J.}~\bibnamefont
  {Reuther}},\ }\href {\doibase 10.1103/PhysRevB.95.054418} {\bibfield
  {journal} {\bibinfo  {journal} {Phys. Rev. B}\ }\textbf {\bibinfo {volume}
  {95}},\ \bibinfo {pages} {054418} (\bibinfo {year} {2017})}\BibitemShut
  {NoStop}%
\bibitem [{\citenamefont {Buessen}\ \emph
  {et~al.}(2018{\natexlab{a}})\citenamefont {Buessen}, \citenamefont {Hering},
  \citenamefont {Reuther},\ and\ \citenamefont {Trebst}}]{Buessen2018}%
  \BibitemOpen
  \bibfield  {author} {\bibinfo {author} {\bibfnamefont {F.~L.}\ \bibnamefont
  {Buessen}}, \bibinfo {author} {\bibfnamefont {M.}~\bibnamefont {Hering}},
  \bibinfo {author} {\bibfnamefont {J.}~\bibnamefont {Reuther}}, \ and\
  \bibinfo {author} {\bibfnamefont {S.}~\bibnamefont {Trebst}},\ }\href
  {\doibase 10.1103/PhysRevLett.120.057201} {\bibfield  {journal} {\bibinfo
  {journal} {Phys. Rev. Lett.}\ }\textbf {\bibinfo {volume} {120}},\ \bibinfo
  {pages} {057201} (\bibinfo {year} {2018}{\natexlab{a}})}\BibitemShut
  {NoStop}%
\bibitem [{\citenamefont {R\"uck}\ and\ \citenamefont
  {Reuther}(2018)}]{Rueck2018}%
  \BibitemOpen
  \bibfield  {author} {\bibinfo {author} {\bibfnamefont {M.}~\bibnamefont
  {R\"uck}}\ and\ \bibinfo {author} {\bibfnamefont {J.}~\bibnamefont
  {Reuther}},\ }\href {\doibase 10.1103/PhysRevB.97.144404} {\bibfield
  {journal} {\bibinfo  {journal} {Phys. Rev. B}\ }\textbf {\bibinfo {volume}
  {97}},\ \bibinfo {pages} {144404} (\bibinfo {year} {2018})}\BibitemShut
  {NoStop}%
\bibitem [{\citenamefont {Iqbal}\ \emph {et~al.}(2019)\citenamefont {Iqbal},
  \citenamefont {M\"uller}, \citenamefont {Ghosh}, \citenamefont {Gingras},
  \citenamefont {Jeschke}, \citenamefont {Rachel}, \citenamefont {Reuther},\
  and\ \citenamefont {Thomale}}]{Iqbal2019}%
  \BibitemOpen
  \bibfield  {author} {\bibinfo {author} {\bibfnamefont {Y.}~\bibnamefont
  {Iqbal}}, \bibinfo {author} {\bibfnamefont {T.}~\bibnamefont {M\"uller}},
  \bibinfo {author} {\bibfnamefont {P.}~\bibnamefont {Ghosh}}, \bibinfo
  {author} {\bibfnamefont {M.~J.~P.}\ \bibnamefont {Gingras}}, \bibinfo
  {author} {\bibfnamefont {H.~O.}\ \bibnamefont {Jeschke}}, \bibinfo {author}
  {\bibfnamefont {S.}~\bibnamefont {Rachel}}, \bibinfo {author} {\bibfnamefont
  {J.}~\bibnamefont {Reuther}}, \ and\ \bibinfo {author} {\bibfnamefont
  {R.}~\bibnamefont {Thomale}},\ }\href {\doibase 10.1103/PhysRevX.9.011005}
  {\bibfield  {journal} {\bibinfo  {journal} {Phys. Rev. X}\ }\textbf {\bibinfo
  {volume} {9}},\ \bibinfo {pages} {011005} (\bibinfo {year}
  {2019})}\BibitemShut {NoStop}%
\bibitem [{\citenamefont {Buessen}\ \emph {et~al.}(2019)\citenamefont
  {Buessen}, \citenamefont {Noculak}, \citenamefont {Trebst},\ and\
  \citenamefont {Reuther}}]{Buessen2019}%
  \BibitemOpen
  \bibfield  {author} {\bibinfo {author} {\bibfnamefont {F.~L.}\ \bibnamefont
  {Buessen}}, \bibinfo {author} {\bibfnamefont {V.}~\bibnamefont {Noculak}},
  \bibinfo {author} {\bibfnamefont {S.}~\bibnamefont {Trebst}}, \ and\ \bibinfo
  {author} {\bibfnamefont {J.}~\bibnamefont {Reuther}},\ }\href {\doibase
  10.1103/PhysRevB.100.125164} {\bibfield  {journal} {\bibinfo  {journal}
  {Phys. Rev. B}\ }\textbf {\bibinfo {volume} {100}},\ \bibinfo {pages}
  {125164} (\bibinfo {year} {2019})}\BibitemShut {NoStop}%
\bibitem [{\citenamefont {Ghosh}\ \emph
  {et~al.}(2019{\natexlab{a}})\citenamefont {Ghosh}, \citenamefont {M\"uller},
  \citenamefont {Toldin}, \citenamefont {Richter}, \citenamefont {Narayanan},
  \citenamefont {Thomale}, \citenamefont {Reuther},\ and\ \citenamefont
  {Iqbal}}]{Ghosh2019a}%
  \BibitemOpen
  \bibfield  {author} {\bibinfo {author} {\bibfnamefont {P.}~\bibnamefont
  {Ghosh}}, \bibinfo {author} {\bibfnamefont {T.}~\bibnamefont {M\"uller}},
  \bibinfo {author} {\bibfnamefont {F.~P.}\ \bibnamefont {Toldin}}, \bibinfo
  {author} {\bibfnamefont {J.}~\bibnamefont {Richter}}, \bibinfo {author}
  {\bibfnamefont {R.}~\bibnamefont {Narayanan}}, \bibinfo {author}
  {\bibfnamefont {R.}~\bibnamefont {Thomale}}, \bibinfo {author} {\bibfnamefont
  {J.}~\bibnamefont {Reuther}}, \ and\ \bibinfo {author} {\bibfnamefont
  {Y.}~\bibnamefont {Iqbal}},\ }\href {\doibase 10.1103/PhysRevB.100.014420}
  {\bibfield  {journal} {\bibinfo  {journal} {Phys. Rev. B}\ }\textbf {\bibinfo
  {volume} {100}},\ \bibinfo {pages} {014420} (\bibinfo {year}
  {2019}{\natexlab{a}})}\BibitemShut {NoStop}%
\bibitem [{\citenamefont {Ghosh}\ \emph
  {et~al.}(2019{\natexlab{b}})\citenamefont {Ghosh}, \citenamefont {Iqbal},
  \citenamefont {M\"uller}, \citenamefont {Ponnaganti}, \citenamefont
  {Thomale}, \citenamefont {Narayanan}, \citenamefont {Reuther}, \citenamefont
  {Gingras},\ and\ \citenamefont {Jeschke}}]{Ghosh2019}%
  \BibitemOpen
  \bibfield  {author} {\bibinfo {author} {\bibfnamefont {P.}~\bibnamefont
  {Ghosh}}, \bibinfo {author} {\bibfnamefont {Y.}~\bibnamefont {Iqbal}},
  \bibinfo {author} {\bibfnamefont {T.}~\bibnamefont {M\"uller}}, \bibinfo
  {author} {\bibfnamefont {R.~T.}\ \bibnamefont {Ponnaganti}}, \bibinfo
  {author} {\bibfnamefont {R.}~\bibnamefont {Thomale}}, \bibinfo {author}
  {\bibfnamefont {R.}~\bibnamefont {Narayanan}}, \bibinfo {author}
  {\bibfnamefont {J.}~\bibnamefont {Reuther}}, \bibinfo {author} {\bibfnamefont
  {M.~J.~P.}\ \bibnamefont {Gingras}}, \ and\ \bibinfo {author} {\bibfnamefont
  {H.~O.}\ \bibnamefont {Jeschke}},\ }\href {\doibase
  10.1038/s41535-019-0202-z} {\bibfield  {journal} {\bibinfo  {journal} {npj
  Quantum Materials}\ }\textbf {\bibinfo {volume} {4}},\ \bibinfo {pages} {63}
  (\bibinfo {year} {2019}{\natexlab{b}})}\BibitemShut {NoStop}%
\bibitem [{\citenamefont {Kiese}\ \emph
  {et~al.}(2020{\natexlab{a}})\citenamefont {Kiese}, \citenamefont {Buessen},
  \citenamefont {Hickey}, \citenamefont {Trebst},\ and\ \citenamefont
  {Scherer}}]{Kiese2020}%
  \BibitemOpen
  \bibfield  {author} {\bibinfo {author} {\bibfnamefont {D.}~\bibnamefont
  {Kiese}}, \bibinfo {author} {\bibfnamefont {F.~L.}\ \bibnamefont {Buessen}},
  \bibinfo {author} {\bibfnamefont {C.}~\bibnamefont {Hickey}}, \bibinfo
  {author} {\bibfnamefont {S.}~\bibnamefont {Trebst}}, \ and\ \bibinfo {author}
  {\bibfnamefont {M.~M.}\ \bibnamefont {Scherer}},\ }\href {\doibase
  10.1103/PhysRevResearch.2.013370} {\bibfield  {journal} {\bibinfo  {journal}
  {Phys. Rev. Research}\ }\textbf {\bibinfo {volume} {2}},\ \bibinfo {pages}
  {013370} (\bibinfo {year} {2020}{\natexlab{a}})}\BibitemShut {NoStop}%
\bibitem [{\citenamefont {{Diekmann}}\ and\ \citenamefont
  {{Jakobs}}()}]{Diekmann2020}%
  \BibitemOpen
  \bibfield  {author} {\bibinfo {author} {\bibfnamefont {J.}~\bibnamefont
  {{Diekmann}}}\ and\ \bibinfo {author} {\bibfnamefont {S.~G.}\ \bibnamefont
  {{Jakobs}}},\ }\href {http://arxiv.org/abs/2009.04761} {\ }\Eprint
  {http://arxiv.org/abs/2009.04761} {arXiv:2009.04761} \BibitemShut {NoStop}%
\bibitem [{\citenamefont {Katanin}(2004)}]{Katanin2004}%
  \BibitemOpen
  \bibfield  {author} {\bibinfo {author} {\bibfnamefont {A.~A.}\ \bibnamefont
  {Katanin}},\ }\href {\doibase 10.1103/PhysRevB.70.115109} {\bibfield
  {journal} {\bibinfo  {journal} {Phys. Rev. B}\ }\textbf {\bibinfo {volume}
  {70}},\ \bibinfo {pages} {115109} (\bibinfo {year} {2004})}\BibitemShut
  {NoStop}%
\bibitem [{\citenamefont {Eberlein}(2014)}]{Eberlein2014}%
  \BibitemOpen
  \bibfield  {author} {\bibinfo {author} {\bibfnamefont {A.}~\bibnamefont
  {Eberlein}},\ }\href {\doibase 10.1103/PhysRevB.90.115125} {\bibfield
  {journal} {\bibinfo  {journal} {Phys. Rev. B}\ }\textbf {\bibinfo {volume}
  {90}},\ \bibinfo {pages} {115125} (\bibinfo {year} {2014})}\BibitemShut
  {NoStop}%
\bibitem [{\citenamefont {Kugler}\ and\ \citenamefont {von
  Delft}(2018{\natexlab{a}})}]{Kugler2017b}%
  \BibitemOpen
  \bibfield  {author} {\bibinfo {author} {\bibfnamefont {F.~B.}\ \bibnamefont
  {Kugler}}\ and\ \bibinfo {author} {\bibfnamefont {J.}~\bibnamefont {von
  Delft}},\ }\href {\doibase 10.1103/PhysRevB.97.035162} {\bibfield  {journal}
  {\bibinfo  {journal} {Phys. Rev. B}\ }\textbf {\bibinfo {volume} {97}},\
  \bibinfo {pages} {035162} (\bibinfo {year} {2018}{\natexlab{a}})}\BibitemShut
  {NoStop}%
\bibitem [{\citenamefont {Kugler}\ and\ \citenamefont {von
  Delft}(2018{\natexlab{b}})}]{Kugler2018}%
  \BibitemOpen
  \bibfield  {author} {\bibinfo {author} {\bibfnamefont {F.~B.}\ \bibnamefont
  {Kugler}}\ and\ \bibinfo {author} {\bibfnamefont {J.}~\bibnamefont {von
  Delft}},\ }\href
  {https://iopscience.iop.org/article/10.1088/1367-2630/aaf65f} {\bibfield
  {journal} {\bibinfo  {journal} {New J. Phys.}\ }\textbf {\bibinfo {volume}
  {20}},\ \bibinfo {pages} {123029} (\bibinfo {year}
  {2018}{\natexlab{b}})}\BibitemShut {NoStop}%
\bibitem [{\citenamefont {Baez}\ and\ \citenamefont
  {Reuther}(2017)}]{Baez2017}%
  \BibitemOpen
  \bibfield  {author} {\bibinfo {author} {\bibfnamefont {M.~L.}\ \bibnamefont
  {Baez}}\ and\ \bibinfo {author} {\bibfnamefont {J.}~\bibnamefont {Reuther}},\
  }\href {\doibase 10.1103/PhysRevB.96.045144} {\bibfield  {journal} {\bibinfo
  {journal} {Phys. Rev. B}\ }\textbf {\bibinfo {volume} {96}},\ \bibinfo
  {pages} {045144} (\bibinfo {year} {2017})}\BibitemShut {NoStop}%
\bibitem [{\citenamefont {Buessen}\ \emph
  {et~al.}(2018{\natexlab{b}})\citenamefont {Buessen}, \citenamefont {Roscher},
  \citenamefont {Diehl},\ and\ \citenamefont {Trebst}}]{Buessen2018a}%
  \BibitemOpen
  \bibfield  {author} {\bibinfo {author} {\bibfnamefont {F.~L.}\ \bibnamefont
  {Buessen}}, \bibinfo {author} {\bibfnamefont {D.}~\bibnamefont {Roscher}},
  \bibinfo {author} {\bibfnamefont {S.}~\bibnamefont {Diehl}}, \ and\ \bibinfo
  {author} {\bibfnamefont {S.}~\bibnamefont {Trebst}},\ }\href {\doibase
  10.1103/PhysRevB.97.064415} {\bibfield  {journal} {\bibinfo  {journal} {Phys.
  Rev. B}\ }\textbf {\bibinfo {volume} {97}},\ \bibinfo {pages} {064415}
  (\bibinfo {year} {2018}{\natexlab{b}})}\BibitemShut {NoStop}%
\bibitem [{\citenamefont {Kugler}\ and\ \citenamefont {von
  Delft}(2018{\natexlab{c}})}]{Kugler2017}%
  \BibitemOpen
  \bibfield  {author} {\bibinfo {author} {\bibfnamefont {F.~B.}\ \bibnamefont
  {Kugler}}\ and\ \bibinfo {author} {\bibfnamefont {J.}~\bibnamefont {von
  Delft}},\ }\href {\doibase 10.1103/PhysRevLett.120.057403} {\bibfield
  {journal} {\bibinfo  {journal} {Phys. Rev. Lett.}\ }\textbf {\bibinfo
  {volume} {120}},\ \bibinfo {pages} {057403} (\bibinfo {year}
  {2018}{\natexlab{c}})}\BibitemShut {NoStop}%
\bibitem [{\citenamefont {Tagliavini}\ \emph {et~al.}(2019)\citenamefont
  {Tagliavini}, \citenamefont {Hille}, \citenamefont {Kugler}, \citenamefont
  {Andergassen}, \citenamefont {Toschi},\ and\ \citenamefont
  {Honerkamp}}]{Tagliavini2018}%
  \BibitemOpen
  \bibfield  {author} {\bibinfo {author} {\bibfnamefont {A.}~\bibnamefont
  {Tagliavini}}, \bibinfo {author} {\bibfnamefont {C.}~\bibnamefont {Hille}},
  \bibinfo {author} {\bibfnamefont {F.~B.}\ \bibnamefont {Kugler}}, \bibinfo
  {author} {\bibfnamefont {S.}~\bibnamefont {Andergassen}}, \bibinfo {author}
  {\bibfnamefont {A.}~\bibnamefont {Toschi}}, \ and\ \bibinfo {author}
  {\bibfnamefont {C.}~\bibnamefont {Honerkamp}},\ }\href {\doibase
  10.21468/SciPostPhys.6.1.009} {\bibfield  {journal} {\bibinfo  {journal}
  {SciPost Phys.}\ }\textbf {\bibinfo {volume} {6}},\ \bibinfo {pages} {009}
  (\bibinfo {year} {2019})}\BibitemShut {NoStop}%
\bibitem [{\citenamefont {De~Dominicis}\ and\ \citenamefont
  {Martin}(1964)}]{DeDominicis1964a}%
  \BibitemOpen
  \bibfield  {author} {\bibinfo {author} {\bibfnamefont {C.}~\bibnamefont
  {De~Dominicis}}\ and\ \bibinfo {author} {\bibfnamefont {P.~C.}\ \bibnamefont
  {Martin}},\ }\href {\doibase 10.1063/1.1704062} {\bibfield  {journal}
  {\bibinfo  {journal} {J. Math. Phys.}\ }\textbf {\bibinfo {volume} {5}},\
  \bibinfo {pages} {14} (\bibinfo {year} {1964})}\BibitemShut {NoStop}%
\bibitem [{\citenamefont {Bickers}(2004)}]{Bickers2004}%
  \BibitemOpen
  \bibfield  {author} {\bibinfo {author} {\bibfnamefont {N.}~\bibnamefont
  {Bickers}},\ }in\ \href
  {http://link.springer.com/chapter/10.1007%2F0-387-21717-7_6} {\emph {\bibinfo
  {booktitle} {Theoretical Methods for Strongly Correlated Electrons}}},\
  \bibinfo {series and number} {CRM Series in Mathematical Physics},\ \bibinfo
  {editor} {edited by\ \bibinfo {editor} {\bibfnamefont {D.}~\bibnamefont
  {S\'{e}n\'{e}chal}}, \bibinfo {editor} {\bibfnamefont {A.-M.}\ \bibnamefont
  {Tremblay}}, \ and\ \bibinfo {editor} {\bibfnamefont {C.}~\bibnamefont
  {Bourbonnais}}}\ (\bibinfo  {publisher} {Springer New York},\ \bibinfo {year}
  {2004})\ pp.\ \bibinfo {pages} {237--296}\BibitemShut {NoStop}%
\bibitem [{\citenamefont {Abrikosov}(1965)}]{Abrikosov1965}%
  \BibitemOpen
  \bibfield  {author} {\bibinfo {author} {\bibfnamefont {A.}~\bibnamefont
  {Abrikosov}},\ }\href
  {https://journals.aps.org/ppf/abstract/10.1103/PhysicsPhysiqueFizika.2.5}
  {\bibfield  {journal} {\bibinfo  {journal} {Physis}\ }\textbf {\bibinfo
  {volume} {2}},\ \bibinfo {pages} {5} (\bibinfo {year} {1965})}\BibitemShut
  {NoStop}%
\bibitem [{\citenamefont {Roulet}\ \emph {et~al.}(1969)\citenamefont {Roulet},
  \citenamefont {Gavoret},\ and\ \citenamefont {Nozi\`eres}}]{Roulet1969}%
  \BibitemOpen
  \bibfield  {author} {\bibinfo {author} {\bibfnamefont {B.}~\bibnamefont
  {Roulet}}, \bibinfo {author} {\bibfnamefont {J.}~\bibnamefont {Gavoret}}, \
  and\ \bibinfo {author} {\bibfnamefont {P.}~\bibnamefont {Nozi\`eres}},\
  }\href {\doibase 10.1103/PhysRev.178.1072} {\bibfield  {journal} {\bibinfo
  {journal} {Phys. Rev.}\ }\textbf {\bibinfo {volume} {178}},\ \bibinfo {pages}
  {1072} (\bibinfo {year} {1969})}\BibitemShut {NoStop}%
\bibitem [{\citenamefont {Chalupa}\ \emph {et~al.}(2020)\citenamefont
  {Chalupa}, \citenamefont {Hille}, \citenamefont {Kugler}, \citenamefont {von
  Delft}, \citenamefont {Andergassen},\ and\ \citenamefont
  {Toschi}}]{Chalupa2020}%
  \BibitemOpen
  \bibfield  {author} {\bibinfo {author} {\bibfnamefont {P.}~\bibnamefont
  {Chalupa}}, \bibinfo {author} {\bibfnamefont {C.}~\bibnamefont {Hille}},
  \bibinfo {author} {\bibfnamefont {F.~B.}\ \bibnamefont {Kugler}}, \bibinfo
  {author} {\bibfnamefont {J.}~\bibnamefont {von Delft}}, \bibinfo {author}
  {\bibfnamefont {S.}~\bibnamefont {Andergassen}}, \ and\ \bibinfo {author}
  {\bibfnamefont {A.}~\bibnamefont {Toschi}},\ }\href@noop {} {\bibfield
  {journal} {\bibinfo  {journal} {to be published}\ } (\bibinfo {year}
  {2020})}\BibitemShut {NoStop}%
\bibitem [{\citenamefont {Anderson}(1965)}]{Anderson1965}%
  \BibitemOpen
  \bibfield  {author} {\bibinfo {author} {\bibfnamefont {D.~G.}\ \bibnamefont
  {Anderson}},\ }\href {\doibase 10.1145/321296.321305} {\bibfield  {journal}
  {\bibinfo  {journal} {J. ACM}\ }\textbf {\bibinfo {volume} {12}},\ \bibinfo
  {pages} {547–560} (\bibinfo {year} {1965})}\BibitemShut {NoStop}%
\bibitem [{\citenamefont {Walker}\ and\ \citenamefont {Ni}(2011)}]{Walker2011}%
  \BibitemOpen
  \bibfield  {author} {\bibinfo {author} {\bibfnamefont {H.~F.}\ \bibnamefont
  {Walker}}\ and\ \bibinfo {author} {\bibfnamefont {P.}~\bibnamefont {Ni}},\
  }\href {\doibase 10.1137/10078356x} {\bibfield  {journal} {\bibinfo
  {journal} {{SIAM} J. Numer. Anal.}\ }\textbf {\bibinfo {volume} {49}},\
  \bibinfo {pages} {1715} (\bibinfo {year} {2011})}\BibitemShut {NoStop}%
\bibitem [{\citenamefont {Reuther}\ and\ \citenamefont
  {Thomale}(2014)}]{Reuther2014a}%
  \BibitemOpen
  \bibfield  {author} {\bibinfo {author} {\bibfnamefont {J.}~\bibnamefont
  {Reuther}}\ and\ \bibinfo {author} {\bibfnamefont {R.}~\bibnamefont
  {Thomale}},\ }\href {\doibase 10.1103/PhysRevB.89.024412} {\bibfield
  {journal} {\bibinfo  {journal} {Phys. Rev. B}\ }\textbf {\bibinfo {volume}
  {89}},\ \bibinfo {pages} {024412} (\bibinfo {year} {2014})}\BibitemShut
  {NoStop}%
\bibitem [{\citenamefont {Sachdev}(2011)}]{sachdev_QPT}%
  \BibitemOpen
  \bibfield  {author} {\bibinfo {author} {\bibfnamefont {S.}~\bibnamefont
  {Sachdev}},\ }\href {\doibase 10.1017/CBO9780511973765} {\emph {\bibinfo
  {title} {Quantum Phase Transitions}}},\ \bibinfo {edition} {2nd}\ ed.\
  (\bibinfo  {publisher} {Cambridge University Press},\ \bibinfo {year}
  {2011})\BibitemShut {NoStop}%
\bibitem [{\citenamefont {Lichtenstein}\ \emph {et~al.}(2017)\citenamefont
  {Lichtenstein}, \citenamefont {{Sánchez de la Peña}}, \citenamefont {Rohe},
  \citenamefont {{Di Napoli}}, \citenamefont {Honerkamp},\ and\ \citenamefont
  {Maier}}]{Lichtenstein2017a}%
  \BibitemOpen
  \bibfield  {author} {\bibinfo {author} {\bibfnamefont {J.}~\bibnamefont
  {Lichtenstein}}, \bibinfo {author} {\bibfnamefont {D.}~\bibnamefont
  {{Sánchez de la Peña}}}, \bibinfo {author} {\bibfnamefont {D.}~\bibnamefont
  {Rohe}}, \bibinfo {author} {\bibfnamefont {E.}~\bibnamefont {{Di Napoli}}},
  \bibinfo {author} {\bibfnamefont {C.}~\bibnamefont {Honerkamp}}, \ and\
  \bibinfo {author} {\bibfnamefont {S.}~\bibnamefont {Maier}},\ }\href
  {\doibase https://doi.org/10.1016/j.cpc.2016.12.013} {\bibfield  {journal}
  {\bibinfo  {journal} {Comp. Phys. Commun.}\ }\textbf {\bibinfo {volume}
  {213}},\ \bibinfo {pages} {100 } (\bibinfo {year} {2017})}\BibitemShut
  {NoStop}%
\bibitem [{\citenamefont {Galassi}(2009)}]{gnu}%
  \BibitemOpen
  \bibfield  {author} {\bibinfo {author} {\bibfnamefont {M.}~\bibnamefont
  {Galassi}},\ }\href {http://www.gnu.org/software/gsl} {\emph {\bibinfo
  {title} {GNU scientific library reference manual}}}\ (\bibinfo  {publisher}
  {Network Theory},\ \bibinfo {address} {Bristol},\ \bibinfo {year}
  {2009})\BibitemShut {NoStop}%
\bibitem [{\citenamefont {Mermin}\ and\ \citenamefont
  {Wagner}(1966)}]{Mermin1966}%
  \BibitemOpen
  \bibfield  {author} {\bibinfo {author} {\bibfnamefont {N.~D.}\ \bibnamefont
  {Mermin}}\ and\ \bibinfo {author} {\bibfnamefont {H.}~\bibnamefont
  {Wagner}},\ }\href {\doibase 10.1103/PhysRevLett.17.1133} {\bibfield
  {journal} {\bibinfo  {journal} {Phys. Rev. Lett.}\ }\textbf {\bibinfo
  {volume} {17}},\ \bibinfo {pages} {1133} (\bibinfo {year}
  {1966})}\BibitemShut {NoStop}%
\bibitem [{\citenamefont {Popov}\ and\ \citenamefont
  {Fedotov}(1988)}]{Popov1988}%
  \BibitemOpen
  \bibfield  {author} {\bibinfo {author} {\bibfnamefont {V.~N.}\ \bibnamefont
  {Popov}}\ and\ \bibinfo {author} {\bibfnamefont {S.~A.}\ \bibnamefont
  {Fedotov}},\ }\href
  {http://www.jetp.ac.ru/cgi-bin/e/index/e/67/3/p535?a=list} {\bibfield
  {journal} {\bibinfo  {journal} {Sov. Phys. JETP}\ }\textbf {\bibinfo {volume}
  {67}},\ \bibinfo {pages} {535} (\bibinfo {year} {1988})}\BibitemShut
  {NoStop}%
\bibitem [{\citenamefont {Reuther}(2011)}]{Reuther2011c}%
  \BibitemOpen
  \bibfield  {author} {\bibinfo {author} {\bibfnamefont {J.}~\bibnamefont
  {Reuther}},\ }\emph {\bibinfo {title} {Frustrated Quantum {H}eisenberg
  Antiferromagnets: Functional Renormalization-Group Approach in
  Auxiliary-{F}ermion Representation}},\ \href
  {https://publikationen.bibliothek.kit.edu/1000023236} {Ph.D. thesis},\
  \bibinfo  {school} {Karlsruhe Institute of Technology} (\bibinfo {year}
  {2011})\BibitemShut {NoStop}%
\bibitem [{\citenamefont {Roscher}\ \emph {et~al.}(2019)\citenamefont
  {Roscher}, \citenamefont {Gneist}, \citenamefont {Scherer}, \citenamefont
  {Trebst},\ and\ \citenamefont {Diehl}}]{Roscher2019}%
  \BibitemOpen
  \bibfield  {author} {\bibinfo {author} {\bibfnamefont {D.}~\bibnamefont
  {Roscher}}, \bibinfo {author} {\bibfnamefont {N.}~\bibnamefont {Gneist}},
  \bibinfo {author} {\bibfnamefont {M.~M.}\ \bibnamefont {Scherer}}, \bibinfo
  {author} {\bibfnamefont {S.}~\bibnamefont {Trebst}}, \ and\ \bibinfo {author}
  {\bibfnamefont {S.}~\bibnamefont {Diehl}},\ }\href {\doibase
  10.1103/PhysRevB.100.125130} {\bibfield  {journal} {\bibinfo  {journal}
  {Phys. Rev. B}\ }\textbf {\bibinfo {volume} {100}},\ \bibinfo {pages}
  {125130} (\bibinfo {year} {2019})}\BibitemShut {NoStop}%
\bibitem [{\citenamefont {Tsvelik}(2003)}]{Tsvelik2003}%
  \BibitemOpen
  \bibfield  {author} {\bibinfo {author} {\bibfnamefont {A.}~\bibnamefont
  {Tsvelik}},\ }\href {https://www.cambridge.org/9780521529808} {\emph
  {\bibinfo {title} {Quantum field theory in condensed matter physics}}}\
  (\bibinfo  {publisher} {Cambridge University Press},\ \bibinfo {address}
  {Cambridge},\ \bibinfo {year} {2003})\BibitemShut {NoStop}%
\bibitem [{\citenamefont {Niggemann}\ \emph {et~al.}(2020)\citenamefont
  {Niggemann}, \citenamefont {Sbierski},\ and\ \citenamefont
  {Reuther}}]{Niggemann2020}%
  \BibitemOpen
  \bibfield  {author} {\bibinfo {author} {\bibfnamefont {N.}~\bibnamefont
  {Niggemann}}, \bibinfo {author} {\bibfnamefont {B.}~\bibnamefont {Sbierski}},
  \ and\ \bibinfo {author} {\bibfnamefont {J.}~\bibnamefont {Reuther}},\
  }\href@noop {} {}\bibinfo {howpublished} {{in preparation}} (\bibinfo {year}
  {2020})\BibitemShut {NoStop}%
\bibitem [{\citenamefont {Krieg}\ and\ \citenamefont
  {Kopietz}(2019)}]{Krieg2019}%
  \BibitemOpen
  \bibfield  {author} {\bibinfo {author} {\bibfnamefont {J.}~\bibnamefont
  {Krieg}}\ and\ \bibinfo {author} {\bibfnamefont {P.}~\bibnamefont
  {Kopietz}},\ }\href {\doibase 10.1103/PhysRevB.99.060403} {\bibfield
  {journal} {\bibinfo  {journal} {Phys. Rev. B}\ }\textbf {\bibinfo {volume}
  {99}},\ \bibinfo {pages} {060403} (\bibinfo {year} {2019})}\BibitemShut
  {NoStop}%
\bibitem [{\citenamefont {Goll}\ \emph {et~al.}(2019)\citenamefont {Goll},
  \citenamefont {Tarasevych}, \citenamefont {Krieg},\ and\ \citenamefont
  {Kopietz}}]{Goll2019}%
  \BibitemOpen
  \bibfield  {author} {\bibinfo {author} {\bibfnamefont {R.}~\bibnamefont
  {Goll}}, \bibinfo {author} {\bibfnamefont {D.}~\bibnamefont {Tarasevych}},
  \bibinfo {author} {\bibfnamefont {J.}~\bibnamefont {Krieg}}, \ and\ \bibinfo
  {author} {\bibfnamefont {P.}~\bibnamefont {Kopietz}},\ }\href {\doibase
  10.1103/PhysRevB.100.174424} {\bibfield  {journal} {\bibinfo  {journal}
  {Phys. Rev. B}\ }\textbf {\bibinfo {volume} {100}},\ \bibinfo {pages}
  {174424} (\bibinfo {year} {2019})}\BibitemShut {NoStop}%
\bibitem [{\citenamefont {Gardner}\ \emph {et~al.}(1999)\citenamefont
  {Gardner}, \citenamefont {Dunsiger}, \citenamefont {Gaulin}, \citenamefont
  {Gingras}, \citenamefont {Greedan}, \citenamefont {Kiefl}, \citenamefont
  {Lumsden}, \citenamefont {MacFarlane}, \citenamefont {Raju}, \citenamefont
  {Sonier}, \citenamefont {Swainson},\ and\ \citenamefont {Tun}}]{Gardner1999}%
  \BibitemOpen
  \bibfield  {author} {\bibinfo {author} {\bibfnamefont {J.~S.}\ \bibnamefont
  {Gardner}}, \bibinfo {author} {\bibfnamefont {S.~R.}\ \bibnamefont
  {Dunsiger}}, \bibinfo {author} {\bibfnamefont {B.~D.}\ \bibnamefont
  {Gaulin}}, \bibinfo {author} {\bibfnamefont {M.~J.~P.}\ \bibnamefont
  {Gingras}}, \bibinfo {author} {\bibfnamefont {J.~E.}\ \bibnamefont
  {Greedan}}, \bibinfo {author} {\bibfnamefont {R.~F.}\ \bibnamefont {Kiefl}},
  \bibinfo {author} {\bibfnamefont {M.~D.}\ \bibnamefont {Lumsden}}, \bibinfo
  {author} {\bibfnamefont {W.~A.}\ \bibnamefont {MacFarlane}}, \bibinfo
  {author} {\bibfnamefont {N.~P.}\ \bibnamefont {Raju}}, \bibinfo {author}
  {\bibfnamefont {J.~E.}\ \bibnamefont {Sonier}}, \bibinfo {author}
  {\bibfnamefont {I.}~\bibnamefont {Swainson}}, \ and\ \bibinfo {author}
  {\bibfnamefont {Z.}~\bibnamefont {Tun}},\ }\href {\doibase
  10.1103/PhysRevLett.82.1012} {\bibfield  {journal} {\bibinfo  {journal}
  {Phys. Rev. Lett.}\ }\textbf {\bibinfo {volume} {82}},\ \bibinfo {pages}
  {1012} (\bibinfo {year} {1999})}\BibitemShut {NoStop}%
\bibitem [{\citenamefont {Molavian}\ \emph {et~al.}(2007)\citenamefont
  {Molavian}, \citenamefont {Gingras},\ and\ \citenamefont
  {Canals}}]{Molavian2007}%
  \BibitemOpen
  \bibfield  {author} {\bibinfo {author} {\bibfnamefont {H.~R.}\ \bibnamefont
  {Molavian}}, \bibinfo {author} {\bibfnamefont {M.~J.~P.}\ \bibnamefont
  {Gingras}}, \ and\ \bibinfo {author} {\bibfnamefont {B.}~\bibnamefont
  {Canals}},\ }\href {\doibase 10.1103/PhysRevLett.98.157204} {\bibfield
  {journal} {\bibinfo  {journal} {Phys. Rev. Lett.}\ }\textbf {\bibinfo
  {volume} {98}},\ \bibinfo {pages} {157204} (\bibinfo {year}
  {2007})}\BibitemShut {NoStop}%
\bibitem [{\citenamefont {Ross}\ \emph {et~al.}(2011)\citenamefont {Ross},
  \citenamefont {Savary}, \citenamefont {Gaulin},\ and\ \citenamefont
  {Balents}}]{Ross2011}%
  \BibitemOpen
  \bibfield  {author} {\bibinfo {author} {\bibfnamefont {K.~A.}\ \bibnamefont
  {Ross}}, \bibinfo {author} {\bibfnamefont {L.}~\bibnamefont {Savary}},
  \bibinfo {author} {\bibfnamefont {B.~D.}\ \bibnamefont {Gaulin}}, \ and\
  \bibinfo {author} {\bibfnamefont {L.}~\bibnamefont {Balents}},\ }\href
  {\doibase 10.1103/PhysRevX.1.021002} {\bibfield  {journal} {\bibinfo
  {journal} {Phys. Rev. X}\ }\textbf {\bibinfo {volume} {1}},\ \bibinfo {pages}
  {021002} (\bibinfo {year} {2011})}\BibitemShut {NoStop}%
\bibitem [{\citenamefont {Thompson}\ \emph {et~al.}(2011)\citenamefont
  {Thompson}, \citenamefont {McClarty}, \citenamefont {R\o{}nnow},
  \citenamefont {Regnault}, \citenamefont {Sorge},\ and\ \citenamefont
  {Gingras}}]{Thompson2011}%
  \BibitemOpen
  \bibfield  {author} {\bibinfo {author} {\bibfnamefont {J.~D.}\ \bibnamefont
  {Thompson}}, \bibinfo {author} {\bibfnamefont {P.~A.}\ \bibnamefont
  {McClarty}}, \bibinfo {author} {\bibfnamefont {H.~M.}\ \bibnamefont
  {R\o{}nnow}}, \bibinfo {author} {\bibfnamefont {L.~P.}\ \bibnamefont
  {Regnault}}, \bibinfo {author} {\bibfnamefont {A.}~\bibnamefont {Sorge}}, \
  and\ \bibinfo {author} {\bibfnamefont {M.~J.~P.}\ \bibnamefont {Gingras}},\
  }\href {\doibase 10.1103/PhysRevLett.106.187202} {\bibfield  {journal}
  {\bibinfo  {journal} {Phys. Rev. Lett.}\ }\textbf {\bibinfo {volume} {106}},\
  \bibinfo {pages} {187202} (\bibinfo {year} {2011})}\BibitemShut {NoStop}%
\bibitem [{\citenamefont {Fennell}\ \emph {et~al.}(2012)\citenamefont
  {Fennell}, \citenamefont {Kenzelmann}, \citenamefont {Roessli}, \citenamefont
  {Haas},\ and\ \citenamefont {Cava}}]{Fennell2012}%
  \BibitemOpen
  \bibfield  {author} {\bibinfo {author} {\bibfnamefont {T.}~\bibnamefont
  {Fennell}}, \bibinfo {author} {\bibfnamefont {M.}~\bibnamefont {Kenzelmann}},
  \bibinfo {author} {\bibfnamefont {B.}~\bibnamefont {Roessli}}, \bibinfo
  {author} {\bibfnamefont {M.~K.}\ \bibnamefont {Haas}}, \ and\ \bibinfo
  {author} {\bibfnamefont {R.~J.}\ \bibnamefont {Cava}},\ }\href {\doibase
  10.1103/PhysRevLett.109.017201} {\bibfield  {journal} {\bibinfo  {journal}
  {Phys. Rev. Lett.}\ }\textbf {\bibinfo {volume} {109}},\ \bibinfo {pages}
  {017201} (\bibinfo {year} {2012})}\BibitemShut {NoStop}%
\bibitem [{\citenamefont {Kiese}\ \emph
  {et~al.}(2020{\natexlab{b}})\citenamefont {Kiese}, \citenamefont {M\"uller},
  \citenamefont {Iqbal}, \citenamefont {Thomale},\ and\ \citenamefont
  {Trebst}}]{Kiese2020b}%
  \BibitemOpen
  \bibfield  {author} {\bibinfo {author} {\bibfnamefont {D.}~\bibnamefont
  {Kiese}}, \bibinfo {author} {\bibfnamefont {T.}~\bibnamefont {M\"uller}},
  \bibinfo {author} {\bibfnamefont {Y.}~\bibnamefont {Iqbal}}, \bibinfo
  {author} {\bibfnamefont {R.}~\bibnamefont {Thomale}}, \ and\ \bibinfo
  {author} {\bibfnamefont {S.}~\bibnamefont {Trebst}},\ }\href@noop {}
  {\bibfield  {journal} {\bibinfo  {journal} {{in preparation}}\ } (\bibinfo
  {year} {2020}{\natexlab{b}})}\BibitemShut {NoStop}%
\end{thebibliography}%
\bibliographystyle{apsrev4-1}

\end{document}